\documentclass[12pt]{article}
\usepackage{latexsym}
\usepackage{amsmath}
\usepackage{amssymb} 
\usepackage{geometry}

\usepackage{tensor}
\usepackage{physics} 
\usepackage{csquotes} 

\usepackage{graphicx}       	
\graphicspath{{./img/}} 
\usepackage[font=small,labelfont=bf]{caption}
\usepackage{subcaption}
\numberwithin{equation}{section}

\usepackage{mathtools}
\usepackage{hyperref}
\usepackage{xcolor}
\usepackage{tikz-cd}
\usepackage{placeins}
\usepackage[toc,page, title, titletoc]{appendix}
\renewcommand{\vec}[1]{\boldsymbol{#1}}

\newcommand{\calM}{\mathcal{M}}

\newcommand{\calW}{\mathcal{W}}

\newcommand{\Z}{\mathbb{Z}}
\newcommand{\R}{\mathbb{R}}

\newcommand{\e}{\mathrm{e}}
\renewcommand{\i}{\mathrm{i}}

\begin{document}

\hypersetup{%
	linkbordercolor=blue,
}

\begin{titlepage}

\begin{flushright}
FTPI-MINN-24-26 \\ UMN-TH-4406/24
\end{flushright}


\begin{center}
	\Large{{\bf 
		Two types of domain walls in \boldmath{$\mathcal{N}=1$} super-QCD: how they are classified and counted}}

\vspace{5mm}
	
{\large  \bf Shi Chen$^{\,a}$, Evgenii Ievlev$^{\,a,b}$ and Mikhail Shifman$^{\,a,b}$}
\end{center}
\begin{center}
{\it  $^{a}$Department of Physics,
University of Minnesota,
Minneapolis, MN 55455}\\[5pt]
{\it  $^{b}$William I. Fine Theoretical Physics Institute,
University of Minnesota,
Minneapolis, MN 55455}\\
\end{center}

\vspace{5mm}

\begin{center}
{\large\bf Abstract}
\end{center}

We study multiplicities and junctions of BPS domain walls interpolating between different chiral vacua in $\mathcal{N}=1$ supersymmetric QCD (SQCD) with the SU$(N)$ gauge group and a varying number of fundamental quarks. Depending on the number of flavors $F$, two distinct classes of {\em degenerate} domain walls emerge:  (i) locally distinguishable, i.e., those which differ from each other locally, in local experiments; and (ii) those which have identical local structure and are differentiated only topologically, through judiciously chosen compactifications. In the first class, two-wall junctions exist, while in the second class, such junctions do not exist. Acharya and Vafa counted {\em topologically distinguishable} walls in pure super-Yang-Mills. Ritz, Shifman, and Vainshtein counted the {\em locally distinguishable} walls in $F=N$ SQCD. In both cases, the multiplicity of $k$ walls was the same, $\nu_{N,k}^\text{walls}= N!/\big[(N-k)!k!\big]$.

We study the general case $0\leqslant F\leqslant N$, with mixed sets of walls from both classes (i) and (ii) simultaneously, and demonstrate that the above overall multiplicity remains intact. We argue that the growth of the quark masses exhibits no phase transition at any finite mass. The locally distinguishable walls can turn into topologically distinguishable ones only at $m=\infty$. The evolution of the low-energy wall worldsheet theory in the passage from small to large $m$ is briefly discussed. We also propose a candidate for the low-energy description of wall junctions.

The tools used are localization of instantons, supersymmetry enhancement on the walls, and circle compactification.

\end{titlepage}

{
\footnotesize
\tableofcontents
}

\newpage

\section{Introduction}
\label{intro}

In $\mathcal{N}=1$ super-Yang-Mills theory (SYM), the 0-form discrete chiral symmetry is spontaneously broken from $\mathbb{Z}_{2h}$ to $\mathbb{Z}_2$, where $h$ is the dual Coxeter number of the gauge group $G$, equal to the adjoint Casimir invariant $T_G$.
This leads to $h$ degenerate vacua with the gluino condensate as the order parameter \cite{Witten:1982df}.
The value of this condensate in a vacuum number $n$ is given by 
$\expval{\Tr \lambda \lambda}_n \sim e^{2 \pi i n / h} \Lambda^3$ 
\cite{Veneziano:1982ah,Affleck:1983rr,Affleck:1983mk,Novikov:1985ic,Shifman:1987ia}
%
where $\Lambda$ is the strong coupling scale.
The presence of these discrete vacua indicates the existence of domain walls interpolating between them, which turn out to be BPS (Bogomol'nyi-Prasad-Sommerfield) saturated.

Shortly after the discovery of these BPS domain walls in SYM \cite{Dvali:1996xe} (see also \cite{Kovner:1997ca}), Witten identified them as branes \cite{Witten:1997ep}.
In 2001 Acharya and Vafa calculated the multiplicity of degenerate walls interpolating between the vacua $n_i$ and $n_f$ \cite{Acharya:2001dz} in SYM based on a string theory construction. 
Another strategy was used in 1997 in \cite{Kovner:1997ca}.
It was suggested to deform SYM by adding matter fields in such a way that the gauge sector becomes fully Higgsed, allowing one to work in the weak coupling regime.
For $SU(2)$, two solutions with distinct internal structures were found. 
The study was extended to $SU(N)$ SQCD by Ritz et al \cite{Ritz:2002fm}.
With a certain choice of the matter sector, it was shown that the theory on the wall worldvolume is the K\"ahler $\mathcal{N} =2$ sigma model with the Grassmannian target space. 
The Witten index of the wall effective theory determines the multiplicity of the degenerate walls.\footnote{An infrared regularization is needed, see below. In particular, in $SU(2)$ SYM we have only two vacua, the boundary conditions for the wall are unique, and the emerging Chern-Simons is $U(1)$ at level 2 corresponding to multiplicity $\nu^\text{walls}=2$. SYM = SQCD without matter.} 
In particular, for the walls interpolating between the neighboring vacua in SQCD one obtains $\mathbb{CP}(N-1)$ model, with $\nu^\text{walls}=N$ ({\em here and below $\nu^\text{walls}$ stands for the wall multiplicity}). 
It was argued that interpolation in matter mass parameters does not change $\nu^\text{walls}$ as we make the masses large approaching SYM.

One of the questions we address in this paper is the two-wall junctions for degenerate walls, an issue which was not fully resolved previously.  
Acharya and Vafa, who analyzed pure SYM from the very beginning, arrived at the conclusion that a topological field theory --- the Chern-Simons theory (CS) --- emerges on the domain wall worldvolume.
If so, as was noted later \cite{KS,Bashmakov:2018ghn,Benvenuti:2021yqv,Benvenuti:2021com,Delmastro:2020dkz}, the ``degenerate walls'' are in fact {\em locally} identical to each other -- distinctions between them cannot be seen in any local ``experiments.''  
They can be formally revealed only through global properties requiring compactification, e.g. on a torus; we can call the corresponding multiplicity the topological multiplicity.
This implies that two-wall junctions do {\em not} exist in SYM. 

Thus, the crucial question in counting the wall multiplicities is, in fact, how we distinguish degenerate walls connecting a given pair of vacua. 
As will be discussed at length below, some walls are locally distinguishable; therefore, their counting is trivial. Other walls are distinguishable only topologically, being locally identical to each other. If we carry out both countings together, we will get
a fixed number of degenerate walls, independent of the number of flavors, see eq. (\ref{i1}).

We will demonstrate that if the degenerate walls are considered in extended SYM, with fundamental matter added, then some of these walls are distinguishable in local measurements for {\em all values} of mass parameters except at singular points, namely infinite mass parameters, when we return to SYM (the second singular point corresponds to vanishing mass parameters when the bulk theory has no vacuum). 
The corresponding domain wall multiplicity can be called local multiplicity.
We argue that analytically continuing to large masses (strong coupling), we still keep distinctions in the local structure of the degenerate walls as well as two-wall junctions. 
The latter disappear only at the point of infinite matter masses. 
There is no phase transition\footnote{More precisely, we exclude phase transitions associated with the 0-form flavor symmetry directly related to the local multiplicity. We still expect a phase transition for the 1-form center symmetry, which however does not affect the wall multiplicity for finite quark masses. This is explained in Sec.~\ref{sec:phase_trans_result}.} at finite values of masses.

If matter fields are sufficiently heavy, physically we are close to pure SYM. 
However, they show up at short distances inside the two-wall junctions. 
In this aspect, one may note a similarity with the axion walls (see Sec.~4 of \cite{Gabadadze:2002ff}).

Both methods of counting --- through pure SYM with CS on the worldsheet and added matter sector --- produce the same answer,
\begin{equation}
	\nu_{N,k}^\text{walls}=\frac{N!}{k!(N-k)!} \,,
\label{i1}
\end{equation}
but the physics is different. The counting in SYM is a topological counting, and formula \eqref{i1} gives the topological multiplicity in a theory compactified on a torus.
Adding heavy matter fields makes degenerate domain walls locally distinguishable, which manifests itself in two-wall junctions.
Formula \eqref{i1} gives the local multiplicity of domain walls even when the bulk theory is considered on $\mathbb{R}^4$.
In other words, if we consider the effective 2+1d theory on the domain wall, then in the former case of SYM the corresponding Hilbert space just splits into $\nu_{N,k}^\text{walls}$ superselection sectors, while in the latter case of SQCD there really are $\nu_{N,k}^\text{walls}$ ground states and it is possible to have transitions between them.


In this paper we want to argue that the above statement is valid also at strong coupling, i.e. at any value of the matter mass term, except infinity (zero mass gives us a runaway, or no-vacuum theory). 
These two are the only singular points in the ADS (Affleck-Dine-Seiberg) superpotential. 
Moreover, a discrete symmetry of the bulk theory is spontaneously broken inside the wall with the given boundary conditions.
Which particular discrete symmetry is broken depends on the number of flavors $F$.

The main idea that we are going to use in this paper is to constrain the strong coupling dynamics by studying different weak coupling and simplified limits.
To this end, we study the bulk theory compactified on a cylinder of small radius $\mathbb{R}^3 \times \mathbb{S}^1$, in the spirit 
of adiabatic continuity under circle compactification \cite{Davies:1999uw,Davies:2000nw,Unsal:2007vu,Unsal:2007jx}.
In this setup, the gauge field along the compact direction acts as an adjoint scalar in the 3d EFT and Abelianizes the theory.
The ``adiabatic continuity'' refers to the idea that the physics of the theory remains smoothly connected (i.e., without a phase transition) as the compactification radius is varied from large to small values. 
This concept is particularly useful for understanding the strongly coupled dynamics of non-abelian gauge theories by relating them to a more controllable regime, and it has been widely used, e.g. for pure SYM.
In this paper, we focus on SQCD with fundamental quark hypermultiplets.
We will eventually arrive at a certain phase diagram in Sec.~\ref{sec:would-be-phase-diagram}, arguing that on the domain wall there is no phase transition (with respect to the spontaneously broken flavor symmetry) at finite compactification scale $L$ or quark mass $m$.

By counting the BPS wall trajectories, we identify the ``local'' and ``topological'' contributions to the wall multiplicities.
The resulting formula in eq.~(\ref{multiplicity_cylinder_full}) agrees with \eqref{i1}, but in a non-trivial way: one has to sum over different sectors.

As a byproduct, we obtain effective theories on the domain wall worldvolume.
We also consider the junctions of several domain walls in SQCD and obtain the worldsheet effective theory (see eq.~\eqref{multiwall_junc_theory_N_F}), thereby extending Gaiotto's proposal \cite{Gaiotto:2013gwa} to the case with matter fields.

We start this paper from Sec.~\ref{sec:review} introducing notation and presenting the simplest example of SU(2) theory.
In Sec.~\ref{sec:semicl} we review the superpotential of the circle compactified bulk theory, making a few additional checks in the case with fundamental quarks.
These superpotentials are applied in Sec.~\ref{sec:walls} to analyze the BPS domain wall trajectories and derive the effective theory on the wall worldvolume.
Next, we discuss junctions of two or more domain walls in Sec.~\ref{sec:junc_2} and Sec.~\ref{sec:junc_multi}, arriving at the effective theory on the junction worldsheet.
Sec.~\ref{sec:phase_trans_result} presents our main results concerning the domain wall physics. Here we deliver our punchline and argue that the phase transition between the ``local'' and ``topological'' domain wall multiplicities (related to spontaneous breaking of flavor symmetry on the wall) happens right at the endpoint of the infinite quark mass, $m = \infty$.
Sec.~\ref{sec:concl} summarizes our findings.
Appendices present some additional discussion and review material.

\section{Warm-up example. Formulation and notation}
\label{sec:review}

In this example, the bulk theory has two supersymmetric vacua. Since the $SU(2)$ group is quasi-real (unlike $SU(N)$ with $N>2$), each quark flavor consists of two chiral superfields in the fundamental representation of the gauge group $SU(2)$ (i.e., two color doublets $Q^f$). 
We introduce a subflavor index $f$ to distinguish the subflavors, $f=1,2$. The Lorentz and color indices are omitted unless stated otherwise. One can consider matter sectors with either one or two flavors. We will start with the former.

After that, we will move to a more general case with $N$ colors (gauge group $SU(N)$) and $F$ flavors of matter in the fundamental representation of the gauge group (quarks), assuming $0 \leqslant F \leqslant N$.

\subsection{\boldmath{$SU(2)$} with one flavor. Why the ADS superpotential is valid at small and large $m$}
\label{sec:review_su2}

The gauge part of the model under consideration is 
\begin{equation}
{\mathcal L}_\text{g}=-\frac{1}{4g^{2}}\int d^{2}\theta\,
W^{a\alpha}W_{\alpha}^{a}+{\rm H.c.}= 
-\frac{1}{4g^{2}}G_{\mu \nu}^{a}G^{a\mu \nu} 
+\frac{1}{2g^{2}}D^{a}D^{a}+\frac{i}{g^{2}}\lambda^{a}\sigma^{\mu}\mathcal{D}_{\mu}\bar{\lambda}^{a},
\end{equation}
while its matter kinetic term has the form
\begin{equation}
{\mathcal L}_\text{kin}= \int d^{2}\theta d^{2}\bar{\theta}\,\bar{Q}^{\,f}e^{V}Q_{f}\,
\end{equation}
where $f=1,2$ is a subflavor index for SU(2)$_{\rm color}$ and one flavor. 
We also add a superpotential (mass) term at the classical level,
%
\begin{equation}
	{\cal L}_{\rm cl} =\frac{m_0}{2}\int d^{2}\theta\,
	Q_{\alpha}^{f}Q_{\,f}^{\alpha}+{\rm H.c.}
\label{cl_superpotential_intro}
\end{equation}
where $\alpha=1,2$ is the color index in the fundamental representation. 
In what follows, it will be convenient to introduce the ``meson'' chiral superfield
%
\begin{equation}
	M = \frac{1}{2} Q_{\alpha}^{f}Q_{\,f}^{\alpha} \,.
\label{fourp}
\end{equation}
%

The quantum (non-perturbative) part of the superpotential is generated by one-instanton contribution,
%
\begin{equation}
\mathcal{W}_{\mathrm{inst}}(M)=\frac{\Lambda_{F=1}^{5}}{M} \,.
\label{four}
\end{equation}
The subscript $F=1$ of the strong coupling scale (defined generally in eq.~\eqref{Lambda_def}) marks the theory with one-flavor matter sector.
In the present case it is given by\footnote{At an arbitrary renormalization point $M_0$, eq.~\eqref{Lambda_N=2_F=1} would also contain the renormalization factor $Z$ for the matter field. At the UV point $M_0 = M_\text{uv}$ we have $Z=1$, which is why it does not appear in eq.~\eqref{Lambda_N=2_F=1} (and in eq.~\eqref{Lambda_def}, for that matter).}
\begin{equation}
	\Lambda_{F=1}^{5} \stackrel{\rm def}{=} 
	\frac{e^{-8\pi^{2}/g^{2}_0}}{g^{4}_0} (M_\text{uv})^5\,.
\label{Lambda_N=2_F=1}
\end{equation}
The full superpotential term is
\begin{equation}
{\cal L}_{\rm sp}={\cal L}_{\rm cl} + \left( \int\, d^2 \theta\, {\mathcal{W}}_{\rm{inst}} +{\rm H.c.}\right) \,.
\label{six}
\end{equation}

Why Eq. (\ref{four}) is valid even at large $m_0$ when the vacua lie at strong coupling? This is a crucial question.

The answer is as follows.  Equation (\ref{four}) is {\it exact}.
We do {\em not} have to integrate out the Higgsed gluon fields to obtain (\ref{four}). Increasing $m_0$ toward $\Lambda_{F=1}$
eventually we come to a point at which the matter mass becomes $\sim \Lambda_{F=1}$.
No fields can be integrated out. Further increase of $m_0$ makes the matter field heavy.
Still we do {\em not} integrate them out, keeping all fields in the Lagrangian.

The functional dependence in (\ref{four}) is unambiguously determined by the conserved $\tilde{R}$ charge. The instanton-induced superpotential  cannot depend on $m_0$ since the $\tilde{R}$ charge of $m_0$  is non-vanishing,
$\tilde R =4$. The overall coefficient is determined by a single instanton.  
The one-instanton calculation is {\em saturated} by {\em zero-size} instantons. 
This observation was made in \cite{Novikov:1985ic} (see Sec. 4 there) long before the invention of the Nekrasov localization \cite{Nekrasov:2002qd}. 
It will also echo in Sec.~\ref{sec:semicl} of this paper.

Indeed, the instanton measure in the model under consideration has the form 
\cite{Novikov:1985ic,Shifman:1999mv}
(see also Sec.~10.19-20 of the textbook \cite{Shifman:2012zz})
\begin{equation}
	d\mu = \frac{1}{2^{5}} \frac{\Lambda_{F=1}^{5}}{M(x_0,\theta_{0})}
		\exp\left(-4\pi^{2} |M| \rho_{\mathrm{inv}}^{2}\right)
		\frac{\mathrm{d}\rho^{2}}{\rho^{2}}\,d^{4}x_{0}d^{2}\theta_{0}\,d^{2}\bar{\beta}\,d^{2}\bar{\theta}_{0}
\label{seven}
\end{equation}
where
\begin{equation}
	\bar{\beta}_{\mathrm{inv}}=\frac{\bar{\beta}}{1-4i(\bar{\beta}\bar{\theta}_{0})}, \qquad
	\rho_{\mathrm{inv}}^{2}=\frac{\rho^{2}}{1-4i(\bar{\beta}\bar{\theta}_{0})}\,,
\label{eight}
\end{equation}
while $\bar\theta_0$ and $\bar\zeta$ are collective coordinates defined in Appendix~\ref{sec:inst_loc}.
In that Appendix we show that the integrated instanton measure \eqref{seven} produces
\begin{equation}
\begin{aligned}
	\int d\mu 
		&\propto  \int\,\frac{d\rho^2}{\rho^2} d^2\bar\beta_{\rm inv}  \, d^2\bar\theta_0 \,\exp\left( -\frac{8\pi^2}{g^2}-4\pi^2|v|^2\rho^2_{\rm inv} \right) \\
		&= 16 \exp\left( -\frac{8\pi^2}{g^2_0 }\right)
\end{aligned}
\label{inst_measure_integrated}
\end{equation}
where $g_0^2$ is the bare coupling constant corresponding to $\rho^2 \to 0$.
This confirms that only the zero-size instantons contribute to the superpotential \eqref{four} with \eqref{Lambda_N=2_F=1}.

\subsubsection{Vacua}

Superpotential in \eqref{six} dictates the vacua as 
%
%
\begin{equation}
	\expval{M} 
		= \pm \frac{1}{m_0} \sqrt{m_0 \Lambda_{F=1}^{5}}
		= \pm\frac{1}{m_0} \sqrt{m_0 \frac{e^{-8\pi^{2}/g^{2}_0}}{g^{4}_0} M_\text{uv}^5}
\label{intro_quark_condensate}
\end{equation}
where we substituted the definition of $\Lambda_{F=1}$ from \eqref{Lambda_N=2_F=1}.
The gluino condensate can then be inferred from the Konishi anomaly as
\begin{equation}
	\langle { \rm Tr}\lambda\lambda \rangle
		= \pm16\pi^2 \left[m_0 \frac{e^{-8\pi^{2}/g^{2}_0}}{g^{4}_0} M_\text{uv}^5\right]^{1/2} \,.
\end{equation}
Because the superpotential is exact, we can take the limit $m_0\to M_\text{uv}$ when the matter decouples and we recover pure SYM.
This yields
\begin{equation}
	\expval{  \frac{ \Tr \lambda\lambda }{16\pi^2} } 
		= \pm \left[\frac{e^{-8\pi^{2}/g^{2}_0}}{g^{4}_0} M_\text{uv}^6 \right]^{1/2}
		= \Lambda_\text{SYM}^3
\label{intro_lambda_condensate}
\end{equation}
where we used the definition of the strong coupling scale in pure SYM (cf. eq.~\eqref{Lambda_def})
\begin{equation}
	\Lambda_\text{SYM}^3 = \frac{e^{-4\pi^{2}/g^{2}_0}}{g^{2}_0} M_\text{uv}^{ 3} \,.
\label{Lambda_N=2_F=0}
\end{equation}

\subsubsection{Two domain walls at weak coupling}

\begin{figure}[t]
    \centering
    \includegraphics[width=0.4\textwidth]{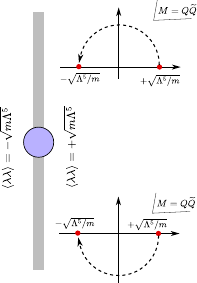}
    \caption{
    	Junction of two domain walls in $SU(2)$ SQCD with 1 flavor.
    	Gluino condensate has different values on the two sides of the walls.
    	On the right we show the domain wall trajectories eq.~\eqref{1six} on the complex plane of the flavor modulus $M$ defined in eq.~\eqref{fourp}.
    	}
    \label{fig:junc}
\end{figure}

Now let us return to the weakly coupled SQCD (small quark mass $m_0$).
Consider a domain wall interpolating between the vacua \eqref{intro_quark_condensate}.
The BPS wall equations are 
\begin{equation}
	\frac{\partial M}{\partial z}
		= \pm \frac{1}{2} |M| \frac{\partial \Big(\bar{\cal{W}}_{\rm cl}+ \bar{\cal{W}}_{\rm inst} \Big) }{\partial\bar {M}} \,.
\label{1five}
\end{equation}
The BPS wall solutions take the form (see Sec. 5 in \cite{Kovner:1997ca})
\begin{equation}
\begin{aligned}
	M_{\rm wall} &= \abs{\expval{M}} \, e^{\pm i\alpha(z)}\,,\\
	i\alpha(z) &= 2\log \left(\frac{1+i\,e^{m_0(z-z_0)}}{\sqrt{1+e^{2m_0(z-z_0)}}}\right)
\end{aligned}
\label{1six}
\end{equation}
with $|M_{\rm wall}| = |\expval{M}| ={\rm const}$ on the wall trajectory.
See Fig.~\ref{fig:junc} for a visual representation.

The solution (\ref{1six}) is obtained in the quasiclassical limit of small $m_0$.
Unlike the BPS protected quantities, the trajectory is subject to corrections; however, these corrections do not qualitatively change the wall profile even if $m_0$ becomes large. 
In equation (\ref{1six}) the wall thickness is $\sim m_0^{-1}$. We do not see the core of the wall built of heavy fields --- Higgsed gauge bosons.

In the case at hand, we do not have moduli of the wall solution other than its translational moduli.
Therefore, there is no field theory on the wall worldvolume. 
{\em  If}  we had extra moduli, we would have a continuum of degenerate walls, in much the same way as in \cite{Shifman:1997wg}. 
To detect a discrete degeneracy on the worldvolume in local low-energy experiments, one must have light ``quasi-moduli.'' 

Needless to say, to observe differences in the internal wall structure, one must have {em probe} fields coupled to matter $M$ with an arbitrary weak interaction.

\subsubsection{From small to large \boldmath{$m$}}

Now let us move to larger values of the quark mass $m_0$.
As it approaches $\Lambda_{F=1}$, the wall's internal structure evolves. 
The inside of the wall 
now has both the matter fields and the gauge bosons.  
With further increase of $m_0$, the matter fields become heavier than the gauge fields. Correspondingly, the core of the wall will be made mostly of the matter fields, while its tails mostly of the gauge fields. Further growth of $m_0$ will lead to a 
thinner core, with the thickness $\Lambda_{F=1}^{-1}$ and even $\ll\sim  \Lambda_{\rm SYM}^{-1}$ (cf. \eqref{Lambda_N=2_F=1} and \eqref{Lambda_N=2_F=0}) if $m_0$ is sufficiently large, $m_0 \gg \Lambda_{\rm SYM}$.

A gedanken ``experimenter'' must go to higher energies
to be able to probe the wall's internal structure. Since the internal structure is different, a two-wall junction must exist.

The crucial feature is that for any finite $m_0$, there are no moduli on the wall, except translational. 
Therefore, in low-energy experiments, the two solutions in (\ref{1six}) cannot be distinguished from each other. However, 
if the experimenter could measure inside the wall, say, at the center of the wall profile in the $z$ direction, the distinction between them would become observable. 

On the wall trajectory \eqref{1six}, the full superpotential \eqref{six} becomes
\begin{equation}
\mathcal{W} = \
\frac{1}{2} \sqrt{m_0\Lambda^5_{\rm 1fl}}\,\cos 2\alpha\,,\qquad 2\alpha \in [0, \pi].
\end{equation}
i.e., Im$\,\mathcal{W}=0$, despite the complex nature of the solution (\ref{1six}), as required for any BPS wall with the given boundary condition by the BPS saturation. 

A few words on the wall thickness $\ell$. 
There are a few possible scales.

If  $m_0$ is small the appropriate scale is $\ell= m_0^{-1}$. If $m_0$ is large, the relevant scale is $\ell= \Lambda_{\rm SYM}^{-1}$. A transitional domain lies at $\ell\sim \Lambda_{F=1}^{-1}$. As $m_0$ approaches $M_\text{uv}$,
the 4d vacuum energy density (as measured by the energy-momentum tensor of the theory) approaches $\Lambda_{\rm SYM}^4$, and the wall quark core width  
shrinks to $1/M_\text{uv}$, i.e., to zero.
In this limit, the wall is built from gluon fields since we find ourselves in pure SYM. Now, two walls are indistinguishable in any local measurement. They are counted by a topological theory on the wall --- the Chern-Simons (CS) theory.
The two-wall junction disappears.

\subsection{Other number of colors \boldmath{$N$} and flavors \boldmath{$F$}}

If we increase the number of flavors and consider, say, SQCD with gauge group $SU(2)$ and two matter flavors, we have a non-trivial flavor symmetry.
A domain wall would break some part of this flavor symmetry, which then leads to the emergence of flavor quasi-moduli.

Let us continue this discussion directly in a more general case of $SU(N)$ SQCD with an arbitrary number of colors $N$ and a number $F < N$ of matter multiplets $Q_f$ and $\tilde{Q}^{\bar{g}}$ in the (anti-)fundamental representation; these are quark flavors, with flavor indices $f, \bar{g} = 1, \ldots, F$.
For massless quarks, the theory has runaway vacua, which are stabilized if we introduce the quark mass $m$ (we take masses of all flavors to be the same).
In the same way as in the $SU(2)$ SQCD case, the quark superpotential gets a contribution from instantons \cite{Affleck:1983rr,Affleck:1983mk}.
It is convenient to pass to the meson superfield
\begin{equation}
	M_f^{\bar{g}} = Q_f \tilde{Q}^{\bar{g}} \,, \quad
	f, \bar{g} = 1, \ldots, F \,.
\label{meson_field_def}
\end{equation}
The full superpotential for the theory with $N$ colors and $F$ flavors (for $0<F<N$) reads
\begin{equation}
	\mathcal{W} = m \Tr M + (N - F) \left( \frac{ \Lambda_{F}^{3N - F} }{ \det M } \right)^{\frac{1}{N-F}} \,.
\label{quark_superpotential}
\end{equation}
Here, $\Lambda_{F}$ is the strong coupling scale for the theory with $F$ quark flavors.
The theory has $N$ gapped vacua with vacuum expectation values (VEV)
\begin{eqnarray}
	&&\expval{M}_n = \mathtt{M}_n \cdot \mathbb{I}_{F \times F} \,, \quad
	\mathtt{M}_n =  e^{\frac{2 \pi i n}{N}} \cdot \left( \frac{ \Lambda_{F}^{3N - F} }{ m^{N - F} } \right)^{1/N}  \,, \quad n=1,2,...N,
	\nonumber\\[2mm]
	&&\expval{  \frac{ \Tr \lambda\lambda }{16\pi^2} }_n = m \mathtt{M}_n \,,
\label{quark_vacua}
\end{eqnarray}
where $\mathbb{I}_{F \times F}$ is an identity matrix.
The last formula follows from the Konishi relation.
The bare quark mass in these vacua is still $m$ (the physical mass is corrected by the $Z$-factors coming from the kinetic $D$-term, which is not protected against renormalization\footnote{For a discussion of possible normalization choices and implications, see p. 542-544 of \cite{Shifman:2012zz}.}).

When the number of flavors reaches the number of colors, $F=N$, the ADS superpotential is replaced by a constraint, and the superpotential can be written as
\begin{equation}
	\mathcal{W} = m \Tr M + \lambda \left[ \det M - \Lambda_{F=N}^{2N} \right]  \,.
\label{quark_superpotential_F=N}
\end{equation}
Here, $\lambda$ is a Lagrange multiplier, $m$ is again a common value of the quark masses, and we focus on the non-baryonic branch.
Vacua that follow from the superpotential \eqref{quark_superpotential_F=N} are given by the same eq.~\eqref{quark_vacua} with the substitution $F = N$.
In a more general case when quark masses are distinct and given by an $N \times N$ mass matrix $\hat{m}$, the vacua become
\begin{eqnarray}
	&&\expval{M}_n = \mathtt{M}_n \cdot \mathbb{I}_{F \times F} \,, \quad
	\mathtt{M}_n =  e^{\frac{2 \pi i n}{N}} \cdot \hat{m}^{-1}  \left( \mathrm{det} \hat{m} \right)^{1/N} \Lambda_{F=N}^2  \,, \quad n=1,2,...N,
	\nonumber\\[2mm]
	&&\expval{  \frac{ \Tr \lambda\lambda }{16\pi^2} }_n = m \mathtt{M}_n \,.
\label{quark_vacua_mhat}
\end{eqnarray}

\begin{figure}[t]
    \centering
    \includegraphics[width=0.4\textwidth]{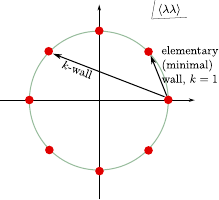}
    \caption{
    	Complex plane of $\expval{\lambda\lambda}$.
        Red dots represent the vacua \eqref{quark_vacua}.
        Arrows schematically represent domain walls
    	}
    \label{fig:k_walls_lambdalambda}
\end{figure}

In the remainder of this paper, we will study domain walls between different chiral vacua \eqref{quark_vacua}, as well as junctions between these walls.
For definiteness, we will consider a wall stretching in space along the $x,y$ directions with the $z$ axis perpendicular to the wall.
If the spacetime is a cylinder, we will choose to compactify the time direction (note that we will deal with Euclidean spacetimes, so we use the term ``time direction'' only for convenience).
A wall interpolating from a vacuum number $n$ at $z \to -\infty$ to the vacuum number $n+k$ at $z \to \infty$ is referred to as a $k$-wall, see Fig.~\ref{fig:k_walls_lambdalambda}.

Domain walls in $SU(N)$ SQCD with $F=N$ flavors were studied in \cite{Ritz:2002fm,Ritz:2004mp,Benvenuti:2021com}.
The $k$-wall multiplicity in this case turns out to be given by
\begin{equation}
	\nu_{N,k}^\text{walls} = \frac{N!}{ k! (N-k)! } \,,
\label{wall_index}
\end{equation}
while the worldvolume theory living on the wall is an $\mathcal{N}=2$ sigma model with its target space given by the complex Grassmannian
\begin{equation}
	\widetilde{\mathcal{M}}_k = G(k,N) = \frac{ U(N) }{ U(k) \times U(N-k) } \,.
\label{grassmanian}
\end{equation}
The Witten index for this sigma model, which is given by the Euler characteristic of the target space, coincides with the multiplicity formula \eqref{wall_index}.

We will reproduce the multiplicity \eqref{wall_index} and the wall worldvolume theory \eqref{grassmanian} below in Sec.~\ref{sec:walls_F=N}.
To this end, we are going to adopt the counting method of \cite{Ritz:2002fm,Ritz:2004mp} to the case when the bulk theory is compactified on a cylinder $\mathbb{R}^3 \times \mathbb{S}^1$ (sometimes referred to as ``circle compactification''); the derivation is very similar, so we will not review it here.
After that, we will extend these results to SQCD with other number of flavors $F<N$.

\section{Superpotentials of SQCD on \boldmath{$\mathbb{R}^3 \times \mathbb{S}^1$} }
\label{sec:semicl}

In this section, we are going to analyze the bulk SQCD theory compactified on a cylinder $\mathbb{R}^3 \times \mathbb{S}^1$, also referred to as ``circle compactification''; we will denote the $\mathbb{S}^1$ circumference by $L$.
The material in this section is largely a review, with some additional checks and comments.

\subsection{Pure SYM and the monopole moduli \boldmath{$Y_j$}}
\label{sec:SYM_review}

To start and introduce notation, we briefly review pure SYM (gauge group $SU(N)$) without quarks under circle compactification~\cite{Davies:1999uw}.
Consider the semiclassical regime where the size of the circle is small (compared to $1/\Lambda_\text{SYM}$).
The low-energy theory and the values of the gluino condensate were studied in~\cite{Davies:2000nw}.
On the Coulomb branch, the non-Abelian gauge group is broken down to an Abelian subgroup; we focus on the case
\begin{equation}
    SU(N) \to U(1)^{N-1}
\label{coulomb_branch_breaking_pattern}
\end{equation}
The 3d infrared effective field is an $(N\!-\!1)$-component dimensionless chiral superfield in a Cartan subalgebra of $\mathfrak{su}(N)$,
\begin{equation}
    \vec{X}=\vec{x}+\sqrt{2}\boldsymbol{\theta}\vec{\lambda}+\boldsymbol{\theta}^2\vec{F} \,.
\end{equation}
The scalar field $\vec{x}$ is the 3d reduction of the gauge field, i.e.,
\begin{equation}
    \vec{x}=\frac{8\pi^2}{Ng^2 }\,\vec{\rho} - \left(\frac{4\pi}{g^2}\, \vec{\sigma}  - \i\vec{\gamma}\right),\qquad \vec{\rho}\equiv\sum_{j=1}^{N-1}\vec{\mu}_j
\label{x_moduli_def}
\end{equation}
where $\vec{\mu}_j$'s are the simple weights and $\vec{\rho}$ is usually called the Weyl vector. Moreover,  $g$ is the coupling of the 3d theory (cf. Table~\ref{tab:3d-4d}).
The real part $\vec{\sigma}$ is the component of the gauge field along the compact direction $\mathbb{S}^1$, and the imaginary part $\vec{\gamma}$ is the dual gauge field, i.e. the dual scalar of the spatial gauge field, see Appendix~\ref{sec:3d_susy} for an elementary introduction.
They take values in an orbifold,
\begin{equation}
    (\vec{\sigma},\vec{\gamma})\in\left(\frac{\R^{N-1}}{2\pi\Lambda_r}\times\frac{\R^{N-1}}{2\pi\Lambda_w}  \right)\Bigg/W_{\mathfrak{su}(N)}
\label{monopole_moduli_space}
\end{equation}
where $\Lambda_{r}$, $\Lambda_{w}$, and $W_{\mathfrak{su}(N)}$ denote the root lattice, the weight lattice, and the Weyl group, respectively.
The Coulomb branch is realized as long as $\vec{\sigma}$ is not far from $\frac{2\pi}{N}\vec{\rho}$ and stays away from the boundary of the orbifold \eqref{monopole_moduli_space}, i.e. $\vec{\sigma}$ should stay away from the values where some non-Abelian subgroup of the gauge group remains unbroken.

In 3d reduction, the theory has 0-form $\Z_{2N}$ chiral symmetry, a 0-form $\Z_N$ center symmetry, and a 1-form $\Z_N$ center symmetry.
The two 0-form symmetries act on $\vec{X}$ as follows,
\begin{subequations}
\begin{gather}
   \! \!\text{center: }\quad\vec{X}\to\vec{X}+2\pi\vec{\mu}\,,\quad \vec{\mu}\in\Lambda_w\!\!\!\mod\Lambda_r \,,\\
    \text{chiral: }\,\, \quad \vec{X}\to\vec{X}+\i\frac{2\pi k}{N}\vec{\rho}\,,\quad k\in\Z\!\!\!\mod 2N \,.
\end{gather}
\label{center_chiral_X}
\end{subequations}
Recall that $\Lambda_w/\Lambda_r=\Z_N$.
The Coulomb branch is conveniently parameterized by operators (sometimes called the monopole moduli)
\begin{equation}
    Y_j\equiv\exp{\vec{\alpha}_j \cdot\vec{X}}\,,\qquad j\in\Z\mod N
\label{Y_def}
\end{equation}
where $\vec{\alpha}_{j=1,\cdots,N-1}$'s are simple roots\footnote{Note that in $\mathfrak{su}(N)$ the root vectors coincide with the co-root vectors.}, $$\vec{\alpha}_0=-\sum_{j=1}^{N-1}\vec{\alpha}_j$$ is the affine root.
Thus these $N$ operators are subject to a constraint
\begin{equation}
    \prod_{j\in\Z_N} Y_j = 1\,.
\label{Y_constraint}
\end{equation}
The monopoles corresponding to $Y_j$ with $j=1,\cdots,N-1$ are the usual (fundamental) monopoles, while the one corresponding to $Y_0$ with the affine root is the Kaluza-Klein (KK) monopole.

The actions of symmetry generators \eqref{center_chiral_X} now take a very simple form:
\begin{subequations}
\begin{gather}
    \label{center_chiral_Y_a} \!\!\!\text{center: }\quad Y_j\to Y_{j+1}\,,\\[1mm]
    \label{center_chiral_Y_b} \text{chiral: }\quad Y_j\to Y_j\e^{\i\frac{2\pi}{N}}\,.
\end{gather}
\label{center_chiral_Y}
\end{subequations}
The second line here follows directly from the second shift in eq.~\eqref{center_chiral_X}.
To understand the first line in eq.~\eqref{center_chiral_Y}, recall that the first shift in eq.~\eqref{center_chiral_X} gets us outside of the fundamental domain \eqref{monopole_moduli_space}, and to get back we need to perform a Weyl reflection; the resulting transformation amounts to a rotation.

The 1-form symmetry \eqref{center_chiral_Y_a} is realized as a solitonic symmetry acting on line defects (and flux tubes, if the latter exist) with nontrivial winding number for $\vec{\gamma}$.
At the energies well below the $W$ boson mass the gauge group is Abelian \eqref{coulomb_branch_breaking_pattern}, and the stable line defects are classified by windings $\pi_1(U(1)^{N-1})=\mathbb{Z}^{N-1}$.
It is known that, say in the case with the gauge group $U(1)$, the 1-form symmetry acting on Wilson lines is actually $U(1)$, see e.g. the discussion in Sec.~4.1 of \cite{Gaiotto:2014kfa}.
In our case, the 1-form symmetry is enhanced from $\Z_N$ to $U(1)^{N-1}$ on the Coulomb branch \eqref{coulomb_branch_breaking_pattern}.
Such an IR symmetry enhancement has nothing to do with supersymmetry and always happens on the Coulomb branch.

As pointed out by the authors of~\cite{Davies:2000nw}, monopole-instantons generate the superpotential in the 3d EFT
\begin{equation}
	\calW = L\Lambda_\text{SYM}^3 \sum_{j\in\Z_N} Y_j \,.
\label{W_SYM_1}
\end{equation}
See eq.~\eqref{Lambda_def} defining our convention for $\Lambda_\text{SYM}$ (note that the beta function coefficient $b=3N$ in the present case).
Formula \eqref{W_SYM_1} clearly respects the 0-form center symmetry and has a definite charge under the chiral symmetry.
Solving $\partial_{\vec{X}}\calW_{3d}=0$, we can find $N$ vacua parameterized by $n=1,\cdots,N$,
\begin{equation}
    \expval{\vec{X}}_n = -\i\frac{2\pi n}{N}\vec{\rho}\quad (\mathrm{mod}\ \i2\pi\Lambda_w)\,,\qquad \expval{Y_j}_n = \e^{-\i\frac{2\pi n}{N}}\,.
\end{equation}
The values of the superpotential in these vacua are given by
\begin{equation}
    (\calW_{3d})_n = L \Lambda_\text{SYM}^3 \e^{-\i\frac{2\pi n}{N}}\,.
\end{equation}
Clearly, chiral symmetry \eqref{center_chiral_Y} acts non-trivially on these vacua and center symmetry leaves them invariant, which signifies the chiral symmetry breaking $\Z_{2N}\to\Z_2$ and the confinement of the Polyakov loops (unbroken 0-form center symmetry).
The Wilson loops are also confined (unbroken 1-form center symmetry).

\begin{figure}[h]
    \centering
    \includegraphics[width=0.3\textwidth]{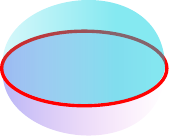}
    \caption{
    	Wall-strings bounded by a Wilson line in 3d
    	}
    \label{fig:wall_string}
\end{figure}

Solving the classical equation of motion in the presence of a $\vec{\gamma}$ defect loop, we can find a ``pancake'' bounded by the defect loop, which generates an area law for the defect loop~\cite{Anber:2015kea}; see Fig.~\ref{fig:wall_string}.

\begin{figure}
     \centering
     \begin{subfigure}[b]{0.3\textwidth}
         \centering
         \includegraphics[width=\textwidth]{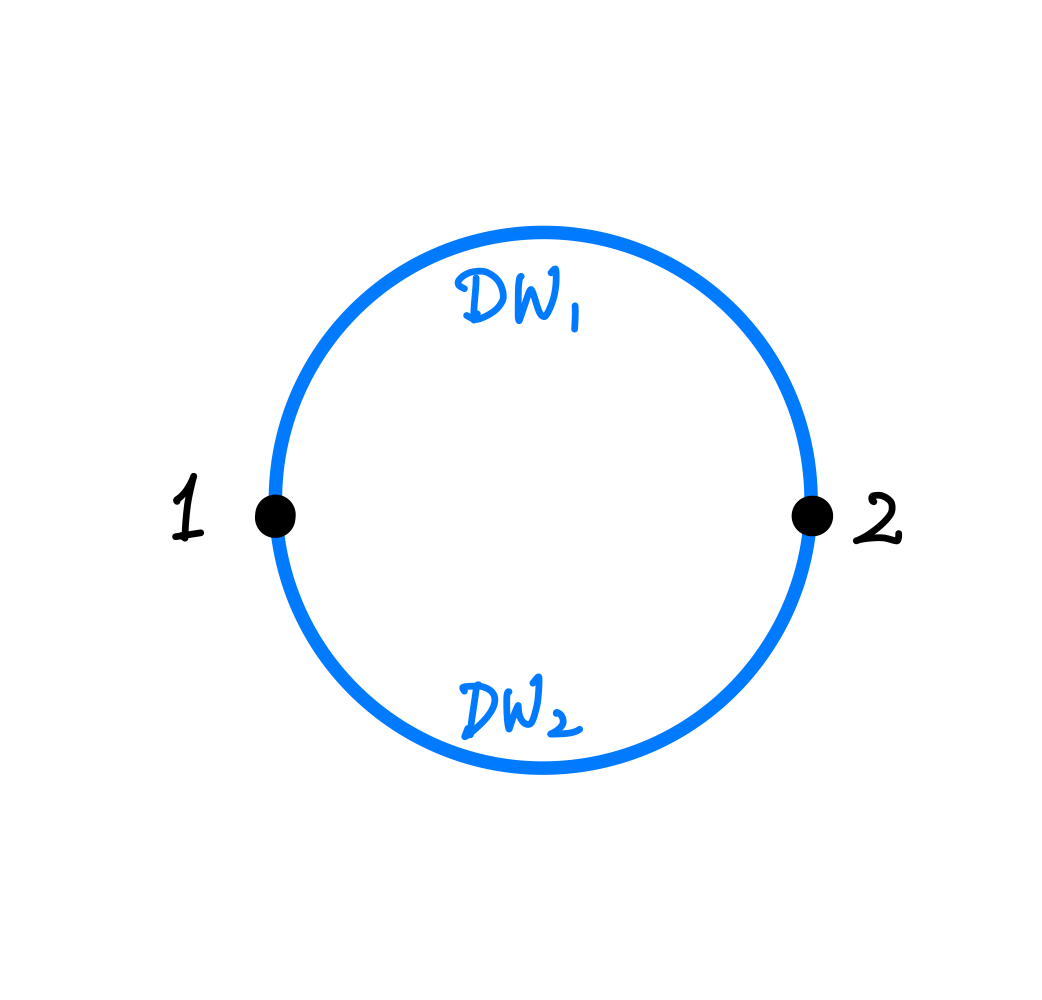}
         \subcaption{$N=2$}
     \end{subfigure}
     $\qquad$
     \begin{subfigure}[b]{0.3\textwidth}
         \centering
         \includegraphics[width=1.5\textwidth]{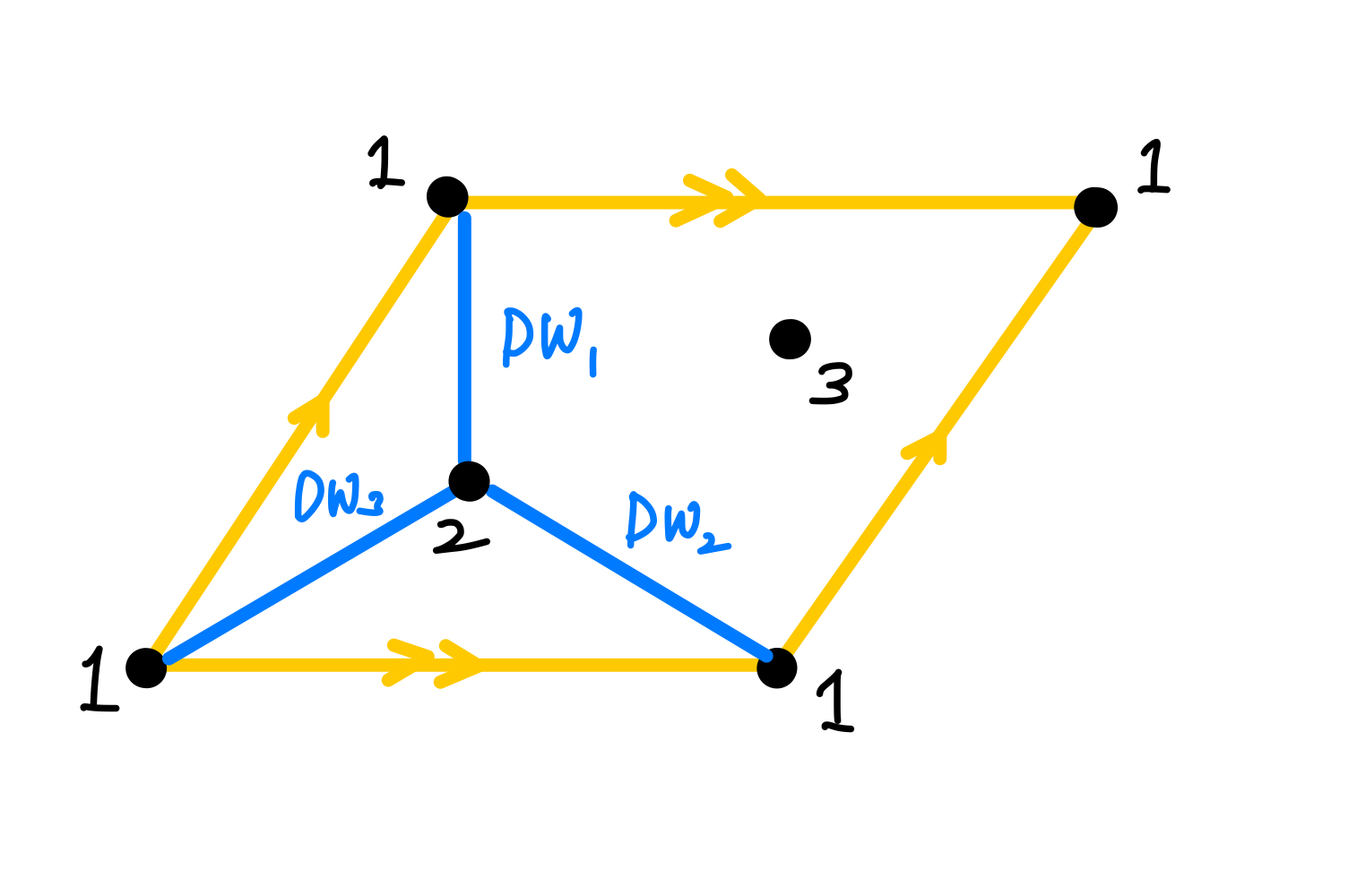}
         \subcaption{$N=3$}
     \end{subfigure}
        \caption{The compact moduli of $\vec{\gamma}=\Im\vec{x}$. 
        For $N=2$ the moduli space is a one-dimensional circle $\mathbb{S}^1$, while for $N=2$ it is a two-dimensional torus $\mathbb{T}^2$.
        The black dots indicate the vacua.
        The blue line segments indicate the domain walls between vacua 1 and 2.}
        \label{fig:YM}
\end{figure}

Figure~\ref{fig:YM} sketches the moduli space of $\vec{\gamma}=\Im\vec{x}$, as well as vacua and domain walls, for $N=2,3$.
In the case of $N=2$, the moduli space is $\mathbb{S}^1$.
There are two vacua, and there are two inequivalent domain walls interpolating between these vacua.
Because the moduli space is not simply connected, there can be no physical junction between these two domain walls; such a junction can only be a singularity.
This singularity is exactly a line defect with nontrivial $\vec{\gamma}$ winding, i.e., a Wilson line.
Therefore, the junction singularity is protected by the 1-form center symmetry.

In other words, a confining string consists of domain walls, and domain walls are fractional confining strings. 
Indeed, since a domain wall in 4d separates two different vacua, for a consistent picture we must demand that in this domain wall wraps the compact dimension $\mathbb{S}^1$.
In the 3d compactified theory this domain wall is a two-dimensional object.
If two different domain walls meet in a junction, for consistency we must require that this junction carries quantum numbers of a Wilson loop.
Thus a domain wall effectively becomes a worldsheet of the confining string.

This picture remains qualitatively valid also for $N>2$, where the moduli space has more dimensions.

\vspace{10pt}

Definition of $Y_j$ above is very convenient in the pure SYM case, as it makes some symmetries among $Y$'s manifest.
However, fundamental matter explicitly breaks some of these symmetries.
Below we are going to consider theories with fundamental quark multiplets.
These are known to enjoy the non-perturbative ADS superpotential in 4d.
In order to make the connection between 4d and 3d more transparent, it is useful to rescale the monopole moduli (e.g. shift the definition of $\vec{x}$ in \eqref{x_moduli_def}) so that the superpotential becomes  \cite{Seiberg:1996nz,Aharony:1997bx,Davies:1999uw}
\begin{equation}
    \calW = \eta Y_0 + \sum_{j=1}^{N-1} Y_j - \lambda \left( \prod_{j=0}^{N-1} Y_j - 1 \right) \,, \quad
    \eta \equiv \left( L \Lambda^3 \right)^N \,.
\label{W_SYM_2}
\end{equation}
Here $\lambda$ is the Lagrange multiplier implementing the constraint (\ref{Y_constraint}).

Essentially, this amounts to dropping the $\vec{\rho}$ term in the definition of $\vec{x}$ in \eqref{x_moduli_def}.
The constraint \eqref{Y_constraint} is not modified; we included it in the superpotential with the help of a Lagrange multiplier..
Note that $\eta \sim \Lambda^{b}$, where $b = 3N$ is the beta function coefficient for pure SYM, see eq.~\eqref{eta_def}.
Here, $\eta$ is a 3d analog of the 4d one-instanton factor.

\vspace{10pt}

We end this subsection with a note on a subtlety about the relation of chiral symmetry breaking in 4d and in the compactified theory.
When the bulk theory lives on $\mathbb{R}^4$, the order parameter for the chiral symmetry breaking is $\expval{ \lambda \lambda }$. 
Since $\lambda \lambda$ is the lowest component of the field strength superfield, this does not contradict supersymmetry.

However after the circle compactification the order parameter is the VEVs of the scalars $\vec{x}$ (or $Y_j$).
After the dualization of the 3d gauge field into $\vec{\gamma}$ the relevant gauge field strength superfield is the linear superfield whose lowest component is actually the scalar $\vec{x}$, see eq.~\eqref{dualized_susy_scalar}.
Although this does not forbid the non-zero value of the bi-fermion condensate $\expval{ \lambda \lambda }$, the calculation of this condensate now involves an additional step.
For more details see Appendix~\ref{sec:lamlam_3d}.


\subsection{Quarks and confinement of monopoles}
\label{sec:mon_conf}

The superpotential of the compactified theory, \eqref{W_SYM_1} or \eqref{W_SYM_2}, is generated by the instanton-monopoles.
When we introduce matter multiplets in the (anti-)fundamental representations (quarks), their scalar components can develop VEVs.
Generally speaking, this gives mass to the gauge fields and leads to confinement of the monopoles.

To understand which monopoles are affected when we introduce quarks, it is useful to consider a generic scenario.
Let us start from a theory with a gauge group $G$.
After the circle compactification we are left with a 3d EFT where an Abelian gauge group $U(1)^r$ is unbroken by the adjoint VEVs (the Cartan torus of $G$, $r = \text{rank}(G)$).
Monopole charges $\vec{g}$ lie on the co-root lattice $\Lambda_R^*$; fundamental monopoles correspond to the simple roots, the KK monopole corresponds to the affine root.
Quarks that were in the fundamental reps of $G$ now acquire electric charges $\vec{Q}^\text{el}_j = \vec{w}_j$ that are in the weight lattice,  $\vec{w}_j \in \Lambda_W$.

Now suppose that a single quark $q$ with charge $\vec{Q}^\text{el}_j = \vec{w}_j$ acquires a scalar VEV $\expval{q} = v_j$.
All the monopoles with charges that have non-zero scalar product with $\vec{w}_j$ become confined, and there are exactly two such monopoles --- the one with magnetic charge $\vec{Q}^\text{magn}_j = \vec{\alpha}_j$ such that $\vec{w}_j \cdot \vec{\alpha}_j = 1$, and the KK monopole with charge $\vec{Q}^\text{magn}_0 = \vec{\alpha}_0 = - \sum \vec{\alpha}_j$ such that $\vec{w}_j \cdot \vec{\alpha}_0 = -1$.
To see that these monopoles are indeed confined, note that the potential from a free monopole of charge $\vec{Q}^\text{magn}$ is $A \sim \vec{Q}^\text{magn} / |x|$.
Because of the term $  |A q|^2$ in the Lagrangian ($q$ is the squark with a VEV $\expval{q} \neq 0$) the monopole action $S_\text{mon}$ diverges. 
The monopole weight in the path integral vanishes,
\begin{equation}
	e^{ - S_\text{mon} } = 0  \,.
\end{equation}
In a more general case when there are $F$ different quark flavors that develop VEVs, $F+1$ monopoles become confined.

Below we will see that the VEV $\expval{q}$ is indeed non-zero for quarks of finite mass, and vanishes only at the endpoint of infinite mass.
The corresponding U(1) gauge field mass is
$m_{U(1)} = g_{3d} \expval{q_{3d}}$, see Table~\ref{tab:3d-4d}.

A pair of such monopoles with charges $\vec{\alpha}_j$ and $\vec{\alpha}_0$ form a confined object with a non-zero net magnetic charge 
\begin{equation}
	\vec{Q}_{0,j}^\text{magn} = \vec{\alpha}_j + \vec{\alpha}_0 = - \sum_{i \neq j}  \vec{\alpha}_i \,.
\end{equation}
Note that this $\vec{Q}_{0,j}^\text{magn}$ is orthogonal to the quark electric charge $\vec{Q}^\text{el}_j = \vec{w}_j$, which renders the action of this composite monopole finite.

In a more general scenario when we introduce $F$ flavors of quarks, the gauge symmetry is broken as
\begin{equation}
	SU(N) \xrightarrow{\text{ adj } \sigma \text{ } } U(1)^{N-1} \xrightarrow{\text{ fund } q \text{ } } U(1)^{N - 1 - F} \,.
\end{equation}
Correspondingly, $F+1$ out of $N$ monopoles become confined.
All these $F+1$ monopoles can form a free ``molecule'' with a non-zero net magnetic charge, generalizing the confined pair above.
This picture remains valid for $F < N-1$.
When the number of flavors reaches $F = N - 1$ or $F = N$, the gauge group is completely broken, and all the monopoles become confined.
The net magnetic charge of the ``molecule'' is now zero\footnote{The difference between $F = N - 1$ and $F = N$ in this context manifests itself in SQCD with gauge group $U(N)$ rather than $SU(N)$. In the former theory, at $F=N-1$ the monopole ``molecule'' resembles a chain with a non-zero leftover magnetic charge that is still unconfined, while at $F=N$ that leftover magnetic flux is also confined, and the ``monopole chain'' closes to a ``monopole necklace''.}.
However, it still can be beneficial to keep track of the Coulomb branch (which is now completely lifted) by keeping one monopole modulus in the superpotential \cite{Aharony:2013dha,Aharony:1997bx}.

\subsection{\boldmath{$SU(2)$} SQCD with one flavor}

As a flavor warm-up, we start from the theory with gauge group $SU(2)$ and $F = 1$ quark flavor (two subflavors in the fundamental representation, see Sec.~\ref{sec:review_su2}).
We start with the superpotential that includes monopole as well as quark degrees of freedom.

In pure SYM, because of the constraint \eqref{Y_constraint}, there is only one independent Coulomb branch modulus.
In SQCD, the quark is in the fundamental representation of the gauge $SU(2)$.
In the 3d EFT, it will be electrically charged under the unbroken $U(1)$, which will lead to confinement of the 3d monopoles; however, it will be instructive to keep the (globally defined) monopole modulus $Y_0$.
The quark is described by a single meson operator $M$ (the meson operator in 3d is defined in the same way as in 4d, see eq.~\eqref{meson_field_def}).
Classically, the scalar potential is given by $V \sim |\sigma q|^2$ ($q$ is the scalar component of a quark multiplet and $\sigma$ is the 3d adjoint scalar), which leads to a classical moduli space constraint $ M Y_0 = 0$.
On the quantum level, this constraint is modified as $M Y_0 = 1$ \cite{Aharony:1997bx,Aharony:2013dha} (see also \cite{Poppitz:2013zqa}), and the superpotential can be written in the form 
\begin{equation}
    \mathcal{W} = \eta_{1} Y_0 + \lambda (M Y_0 - 1) + m M \,. 
\label{W_circle_su2}
\end{equation}
Here $\lambda$ is a Lagrange multiplier, and $\eta_{1} = L^2 \Lambda_{1}^5$; the power 5 comes from the beta function $b = 3 N - F$, see Appendix~\ref{sec:betafunc}; the subscript reminds us that $F=1$.
We have also added a bare quark mass term.

When the quark mass $m$ is small, it is natural to use the meson modulus $M$ for describing the low-energy theory.
Using the constraint enforced by $\lambda$, we can re-express the superpotential in terms of the meson operator only:
\begin{equation}
    \mathcal{W}  = \frac{\eta_{1}}{M} + m M \,.
\label{W_circle_su2_meson}
\end{equation}
This directly corresponds\footnote{Recall that 3d and 4d quantities are related as $\mathcal{W}_{3d} = L \cdot \mathcal{W}_{4d}$, $M_{3d} = L \cdot M_{4d}$, see Table~\ref{tab:3d-4d}.} to the ADS superpotential coming from 4d.

With a heavy quark, it makes sense to integrate out $M$; we obtain:
\begin{equation}
	\mathcal{W} = \eta_1 Y_0 + \frac{m}{Y_0} \,.
\label{W_circle_su2_monopole}
\end{equation}
To match the scales, one has to redefine
\begin{equation}
	Y_0 \to \tilde{Y}_0 = \frac{Y_0}{m} \,, \quad
	\eta_{1} \to \eta_0 = m \, \eta_{1} \,,
\label{SU2_decouple_rescale}
\end{equation}
cf. eq.~\eqref{scales_matching_F-1} and eq.~\eqref{scales_matching_F-1_eta}.
After this rescaling, we recover the superpotential of pure SYM, see eq.~\eqref{W_SYM_2}.

Supersymmetric vacua can be easily found as extrema of the superpotential.
From eq.~\eqref{W_circle_su2_meson}, we find two vacua
\begin{equation}
	M = \pm \sqrt{\eta_1 / m} = \pm L \sqrt{\Lambda_{1}^5 / m} \,.
\end{equation}
which precisely agrees with the 4d result \eqref{quark_vacua}.
In terms of the monopole modulus, from eq.~\eqref{W_circle_su2_monopole}, we see that 
$Y_0 = \pm \sqrt{ m / \eta_1 }$.
In either formulation, we find two vacua that correspond to the two chiral vacua of the 4d theory.

This theory also enjoys domain walls interpolating between these two vacua.
Specifically, consider a domain wall with $x, y$ coordinates stretching along the wall and $z$ being the orthogonal direction, and consider a wall interpolating from the vacuum 
$M = 1/Y_0 = \sqrt{\eta_1 / m}$ at $z \to - \infty$ to the vacuum $M = 1/Y_0 = - \sqrt{\eta_1 / m}$ at $z \to + \infty$.
Just as in the 4d case from Sec.~\ref{sec:review_su2}, there are two walls satisfying these boundary conditions.
On one wall, the meson $M$ winds clockwise, while on the other wall, $M$ winds counter-clockwise, see Fig.~\ref{fig:junc}.
Because of the constraint $M Y_0 = 1$, the monopole modulus $Y_0$ is forced to wind in the opposite direction.
Note that when we decouple the quarks by sending $m \to \infty$, the trajectory of $M$ shrinks, but the rescaled monopole modulus \eqref{SU2_decouple_rescale} stays finite.

\subsection{\boldmath{$SU(N)$} SQCD with \boldmath{$F = N$} and \boldmath{$F = N - 1$} on cylinder}
\label{sec:F=N}

Now let us move on to a more general case.
Consider SQCD with gauge group $SU(N)$ and $F=N$ quarks.
The superpotential of the 3d theory can be written as \cite{Aharony:2013dha,Aharony:1997bx}
\begin{equation}
    \mathcal{W} = Y_0 ( \eta_N - \det(M) + B \tilde{B} ) + \Tr(\hat{m} M) \,. 
\label{superpotential_3d_Nf=Nc}
\end{equation}
Here, $\eta_N = (L \Lambda_{N}^2)^N = L^N \, \Lambda_{N}^b$ ($b$ is the beta function coefficient, see Appendix~\ref{sec:betafunc}), and $\hat{m}$ is the mass matrix.
As explained in Sec. 8.2 of \cite{Aharony:1997bx}, at least part of the Coulomb branch is lifted in the presence of massless quarks, and we can use $Y_0$ for describing the remaining part of the branch.
$B$ and $\tilde{B}$ are the baryon operators; however, here we will focus on the non-baryonic branch where $B \tilde{B} = 0$.

One can see from the superpotential \eqref{superpotential_3d_Nf=Nc} that the Coulomb branch is in fact completely lifted, and the corresponding coordinate $Y_0$ can be integrated out.
The low-energy dynamics in this case is described by the meson moduli with the constraint 
\begin{equation}
	\det(M) = \eta_N\,,
\label{M_constraint_F=N}
\end{equation}
in complete analogy with the 4d theory.
We find $N$ chiral vacua (cf. eq.~\eqref{quark_vacua_mhat})
\begin{equation}
	\expval{M}_n = \hat{m}^{-1} \left( \det(\hat{m}) \eta_N \right)^{1/N} \, e^{\frac{2 \pi i n}{N}} \,.
\end{equation}
In the simplest case when the masses of all quarks are the same, this formula reduces to
\begin{equation}
	\expval{M}_n = \mathtt{M}_n \cdot \mathbb{I}_{N \times N} \,, \quad
	\mathtt{M}_n =  e^{\frac{2 \pi i n}{N}} \cdot \left( \eta_N \right)^{1/N}  \,, \quad
	\hat{m} \equiv m \, \mathbb{I}_{N \times N} \,.
\label{quark_vacua_3d}
\end{equation}
At first sight, it might seem that the quark VEVs do not depend on $m$.
However, when the quarks are heavy, one should measure the energy with respect to $\eta_0 \equiv \eta_\text{SYM} = m^N \eta_N$ (see eq.~\eqref{scales_matching_F-1_eta}), so that \eqref{quark_vacua_3d} actually gives $$\expval{M} \sim \left( \eta_\text{SYM} \right)^{1/N} / m\,.$$ This vanishes in the infinite mass limit.

In order to pass on to the case with $F = N - 1$, we may decouple one flavor by sending its mass to infinity.
As follows from the discussion in Sec.~\ref{sec:mon_conf}, all the monopoles are still confined in this theory.
In this case, the Coulomb and Higgs branches of the 3d theory merge in a deformed moduli space, which can be implemented with a Lagrange multiplier.
The superpotential in this case is a generalization of the $SU(2)$ case \eqref{W_circle_su2} and reads
\begin{equation}
    \mathcal{W} = \lambda (Y_0 \det(M') - 1) + \eta_{N-1} Y_0 + \Tr(\hat{m}' M') 
\label{W_SUN_F=N-1}
\end{equation}
where $M^\prime$ and $\hat{m}'$ are now $(N-1) \times (N-1)$ matrices.
This superpotential was derived in \cite{Aharony:1997bx,Aharony:2013dha} based on holomorphy and global symmetries.
Here we note that it can also be derived by integrating out one flavor from \eqref{superpotential_3d_Nf=Nc}. 
Indeed, let us make the $N^\text{th}$ quark flavor heavy and take the mass matrix in the block-diagonal form
\begin{equation}
	 \hat{m} = \left(\begin{array}{ccc|c}
           &  &  & 0 \\ 
           & \hat{m}' & & \vdots \\ 
           & & & 0 \\ \hline
          0 & \cdots & 0 & \hat{m}_{NN}
        \end{array}\right), \;\;\;\;\; \hat{m}_N^N \gg (\hat{m}')_{fg},\, \quad
        f,g=1,\ldots, N-1 \,.
\end{equation}
Then we have
 \begin{equation}
    {\mathcal W} = Y_0 \big( \eta_{N-1} - \det M' \big) + \hat{m}_{NN} \, M_{NN} + \Tr \big(\hat{m}' \, M'\big) \,.
\label{superpotential_3d_F=N-1_derivation}
\end{equation}
Extremizing this superpotential with respect to $$Y_0\,,M_{NN}\,,M_{Nf}\,, M_{fN}\,, \qquad (f=1,\ldots,N-1)$$ 
then gives
\begin{equation}
	M_{NN} = \frac{\eta_{N-1}}{ \det M' } \,, \quad
	Y_0 = \frac{ \hat{m}_{NN} }{ \det M' } \,, \quad
	M_{Nf} = M_{fN} = 0 \,.
\label{YMNNonstr}
\end{equation}
Using the constraints \eqref{YMNNonstr} and rescaling $Y_0$ and $\eta$ as in \eqref{SU2_decouple_rescale}, we can bring the superpotential \eqref{superpotential_3d_F=N-1_derivation} to the form \eqref{W_SUN_F=N-1}.

Note that integrating out $Y$ from \eqref{W_SUN_F=N-1} yields 
\begin{equation}
    \mathcal{W} = \frac{\eta_{N-1} }{ \det(M') } + \Tr(\hat{m}' M') \,.
\end{equation}
This form corresponds to the familiar ADS superpotential in 4d.

Vacua that follow from the superpotential \eqref{W_SUN_F=N-1} are given by
\begin{equation}
	\expval{M'}_n = (\hat{m}')^{-1} \left( \det(\hat{m}') \, \eta_{N-1} \right)^{1/N} \, e^{\frac{2 \pi i n}{N}} \,.
\label{vacua_3d_F=N-1_matrix-mass}
\end{equation}
In the simplest case when the masses of all quarks are the same, $\hat{m}' \equiv m \, \mathbb{I}_{(N-1) \times (N-1)}$, this formula reduces to
\begin{equation}
	\expval{M}_n = \mathtt{M}_n \cdot \mathbb{I}_{(N-1) \times (N-1)} \,, \quad
	\mathtt{M}_n =  e^{\frac{2 \pi i n}{N}} \cdot \left( \frac{\eta_{N-1}}{m} \right)^{1/N}  \,.
\label{vacua_3d_F=N-1_eqmass}
\end{equation}
Thus for $F=N$ and $F=N-1$ flavors, we see $N$ different vacua, much as in the uncompactified case.

\subsection{\boldmath{$SU(N)$} SQCD with \boldmath{$0< F < N-1$} on \boldmath{$\mathbb{R}^3 \times \mathbb{S}^1$} }

When the number of quark flavors $F$ is less than $N-1$, one has to be a bit more careful.
As follows from the discussion in Sec.~\ref{sec:mon_conf}, in this case we have $N-F-1$ deconfined monopoles.
As we will see below, accounting for all the relevant monopole moduli is important for understanding the domain wall physics.

Naively, we can start by integrating out a number of quark flavors from the superpotential \eqref{superpotential_3d_Nf=Nc} or \eqref{W_SUN_F=N-1} and obtain
\begin{equation}
    \mathcal{W} = \eta_F Y_0 + (N - F - 1) \frac{1}{ \left( Y_0 \det(M) \right)^{ \frac{1}{N - F - 1} } }  + \Tr(m M) \,. 
\label{superpotential_3d_Nf}
\end{equation}
Here $\eta_F = L^N \Lambda_{F}^{3N-F}$ (see Appendix~\ref{sec:betafunc}), and we take the masses of all quark flavors to be the same $m$.
This procedure is completely analogous to the one familiar from studies of the ADS superpotential.
Note that integrating out $Y_0$ we recover the familiar form of the ADS superpotential,
\begin{equation}
    \mathcal{W} = (N - F) \left[ \frac{ \eta }{ \det(M) } \right]^{ \frac{1}{N - F} } + \Tr(m M) \,.
\label{W_recover_ADS}
\end{equation}
Let us pause here briefly for a comment.
In all the examples above, we have always recovered the correct 4d quark physics (ADS superpotential or the deformed moduli space constraint) from the physics of the compactified theory.
However, for that we did not have to take the decompactification limit.
This is not surprising.
As was explained in Sec.~\ref{sec:review}, the relevant quantities in 4d are saturated by one instanton localized at zero size.
Because of the zero size, the instanton does not ``see'' the compactness in the present $\mathbb{R}^3 \times \mathbb{S}^1$ setup.
Such an instanton is governed by the microscopic physics only, but at the micro-level our theory is 4d.

\vspace{10pt}

Formula \eqref{superpotential_3d_Nf} was derived in \cite{Aharony:1997bx} based on the consideration of global symmetries and holomorphy (see Sec.~8 there).
The authors used $Y_0$ because it is a quantity globally defined on the Coulomb branch.
The form of the superpotential \eqref{superpotential_3d_Nf} can be used for finding the vacua of the 3d SQCD and matching the $SU(N)$ gauge theory without quarks.
Indeed, integrating out all the meson moduli from \eqref{superpotential_3d_Nf} yields
\begin{equation}
    M = x \, \mathbb{I}_{F \times F} \,, \quad
    x = \frac{1}{ Y_0^{ \frac{1}{N - 1} } m^{ \frac{N - F - 1}{N - 1} } }
\end{equation}
and
\begin{equation}
    \mathcal{W} = \eta Y_0 + (N - 1) \frac{ 1 }{ Y_0^{ \frac{1}{N - 1} } }
\label{W_recover_SYM}
\end{equation}
where we also rescaled $Y_0$ and the $\eta$-factor according to eq.~\eqref{scales_matching_F-1_eta}.
It is a trivial check that eq.~\eqref{W_recover_SYM} reproduces the pure SYM superpotential \eqref{W_SYM_2} with all moduli integrated out except $Y_0$.
Extrema of \eqref{W_recover_SYM} also give $N$ correct vacua of the pure SYM.

However, superpotential \eqref{W_recover_SYM} does not help us much in studying domain walls interpolating between these vacua.
Roughly speaking, this happens because more than one monopole might be unconfined, and we cannot fully describe the associated moduli with just one coordinate.
We need to reinstate all the relevant coordinates on the moduli space.

The full superpotential can be derived from the path integral; the computation carried out in \cite{Csaki:2017mik,Shirman:2019mqv} accounts for the fact that with nonvanishing quark VEVs, (some of the) monopoles become confined.
The result in its simplified form matches \eqref{superpotential_3d_Nf}, while the full superpotential reads (here we also added a bare mass term):
\begin{equation}
    \mathcal{W} = \eta Y_0 + \sum_{i = F + 2}^{N - 1} Y_i + \frac{ [ \prod_{i=1}^{F+1} Y_i ] }{\det M} - \lambda \left( Y_0 \prod_{j=1}^{N - 1} Y_j - 1 \right) + \Tr( \hat{m} M) \,.
\label{superpotential_3d_Nf_Yj}
\end{equation}
The product $[ \prod_{i=1}^{F+1} Y_i ]$ represents a molecule of $F+1$ confined monopoles forming a bound state (see Sec.~\ref{sec:mon_conf}) and should be treated as a single variable.
We checked that, upon integrating out $Y_{j}$ $(j=F+2, \ldots, N-1)$ and $[ \prod_{i=1}^{F+1} Y_i ]$ we recover eq.~\eqref{superpotential_3d_Nf}
from \eqref{superpotential_3d_Nf_Yj}.

For studying domain walls, it is actually more convenient to change variables so that the superpotential and formula for the vacua would look more homogeneous.
Let us define
\begin{equation}
	Y_{\text{conf}} \equiv \frac{ [ \prod_{i=1}^{F+1} Y_i ] }{\det M} \,.
\label{Y_conf_def}
\end{equation}
The superpotential \eqref{superpotential_3d_Nf_Yj} then becomes
\begin{equation}
    \mathcal{W} = \eta Y_0 + \sum_{i = F + 2}^{N - 1} Y_i + Y_{\text{conf}} - \lambda \left( \det(M) \cdot Y_0 \cdot Y_{\text{conf}} \cdot \prod_{i=F+2}^{N - 1} Y_i - 1 \right) + \Tr( \hat{m} M) \,.
\label{superpotential_3d_Nf_Yconf}
\end{equation}
%
We again find $N$ vacua:
\begin{equation}
\begin{aligned}
    \expval{Y_0}_n &= \frac{ (m^F \eta)^{1/N} }{\eta} e^{2 \pi i n / N} \,, \quad \expval{Y_{\text{conf}}}_n = (m^F \eta)^{1/N}  e^{2 \pi i n / N} \,, \\
    \expval{Y_j}_n &=  (m^F \eta)^{1/N}  e^{2 \pi i n / N} \quad (j = F+2, \ldots, N-1) \,, \\
    \expval{M}_n &= \mathtt{M}_n \cdot \mathbb{I}_{F \times F} \,, \quad \mathtt{M}_n = \frac{ (m^F \eta)^{1/N} }{m} e^{2 \pi i n / N} \,, \\
    \expval{\lambda}_n &= (m^F \eta)^{1/N}  e^{2 \pi i n / N} \,.
\end{aligned}
\label{vacua_3d_Nf_Yj}
\end{equation}
Note that we can view the case with $F=N-1$ flavors as a limiting case when we also include $Y_0$ into the product \eqref{Y_conf_def}.
In this case, the superpotential \eqref{superpotential_3d_Nf_Yconf} and vacua \eqref{vacua_3d_Nf_Yj} reduce to eq.~\eqref{W_SUN_F=N-1} and eq.~\eqref{vacua_3d_F=N-1_eqmass}, respectively.
Below we will not consider the $F=N-1$ case separately, as the results for general $F$ will reproduce this case as well.

The $N$ vacua from eq.~\eqref{vacua_3d_Nf_Yj} are reminiscent of the chiral vacua of the uncompactified 4d theory.
In the next Section, we are going to study domain walls interpolating between these vacua.

\section{Domain walls}
\label{sec:walls}

In this Section, we are going to apply our knowledge of the superpotentials of bulk SQCD compactified on a circle.
We are going to study the BPS domain wall trajectories, which will enable us to count the wall multiplicities and make a proposal for an effective theory living on the wall.

When the bulk theory is compactified on $\mathbb{R}^3 \times \mathbb{S}^1$ with circumference $L$, for self-consistency, we require that the domain wall wraps the compact dimension.
Otherwise, we would only be able to consider the wall-antiwall pairs, because in our setup, the vacuum structure has to be periodic on $\mathbb{S}^1$, see \cite{Ritz:2006zz} for a related discussion.

\subsection{Domain walls at \boldmath{$F=N$}: fully Higgsed SQCD}
\label{sec:walls_F=N}

The task of domain wall counting in the 4d SQCD with $N$ quark flavors was carried out in \cite{Ritz:2002fm}.
The method is trivially adopted to the 3d EFT at hand.

We start by rewriting the superpotential \eqref{superpotential_3d_Nf=Nc} in more convenient variables.
Let us introduce a dimensionless superfield
\begin{equation}
	X = \frac{1}{m \, \eta^{1/N}} \hat{m} M \,, \quad
	m \equiv ( \det \hat{m} )^{1/N} \,.
\end{equation}
To demonstrate the generality of this approach, we are going to keep the mass matrix $\hat{m}$ to be general in this Subsection, but later for simplicity we will take the equal masses for all quark flavors.

Let $x_i$, $i=1,\ldots,N$ be the eigenvalues of the matrix $X$.
In terms of these eigenvalues, the superpotential \eqref{superpotential_3d_Nf=Nc} (on the non-baryonic branch) can be rewritten as
\begin{equation}
    \mathcal{W} = m \eta_N^{1/N} \left\{ Y_0 \cdot \left( 1 - \prod_{i=1}^{N} x_i \right) + \sum_{i=1}^{N} x_i \right\} \,.
\label{superpotential_3d_Nf=Nc_xi}
\end{equation}
The vacua are obtained as minima of the superpotential.
There are $N$ of them, and in each of the vacua the matrix $X$ is proportional to the identity and has all eigenvalues equal,
\begin{equation}
	\expval{x_j}_{n } = e^{\frac{2 \pi i n}{N}} \,, \quad
	n = 0, \ldots, N-1 \,.
\label{SUN_F=N_vac}
\end{equation}

\begin{figure}[h]
    \centering
    \includegraphics[width=0.45\textwidth]{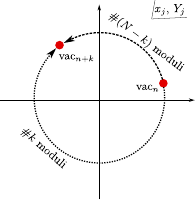}
    \caption{
    	Winding of the bulk moduli inside a BPS domain wall.
    	The bulk moduli are the eigenvalues of the meson matrix $x_j$ ($F$ of them, where $F$ is the number of quark flavors) and/or the monopole moduli $Y_j$ ($N-F$ of them for the gauge group $SU(N)$).
    	The red dots represent two chiral vacua, see eq.~\eqref{SUN_F=N_vac} or eq.~\eqref{vacua_3d_Nf_Yj}
    	}
    \label{fig:moduli_windings}
\end{figure}

Now, consider a domain wall interpolating between two vacua, say from $n=n_0$ to $n= n_0 + k$ (remember, we can call it $k$-wall).
We will focus on the case $0 < k \leqslant N/2$ (the walls with $k > N$ are related to these by parity transformation).
One can characterize the eigenvalue trajectories by their winding numbers,
\begin{equation}
	w_i = \frac{1}{2 \pi i} \int_{\Gamma_i} \frac{d x_i}{x_i}
\label{winding_def}
\end{equation}
where $\Gamma_i$ is the eigenvalue trajectory in the complex plane (upon traversing the wall from $z \to -\infty$ to $z \to \infty$, $z$ being the direction orthogonal to the wall).
Suppose that $N_L$ of the eigenvalues wind clockwise along a trajectory $\Gamma_i \equiv x_L$ with windings $w_L = k/N - 1$, while the rest $N_R = N - N_L$ wind counter-clockwise along a trajectory $\Gamma_i \equiv x_R$ with windings $w_R = k/N$.
Superpotential \eqref{superpotential_3d_Nf=Nc_xi} encodes a constraint 
\begin{equation}
	\prod_{i=1}^{N} x_i = 1
\label{F=Nonstraint}
\end{equation}
which implies that the condition
\begin{equation}
	N_R \, \arg (x_R) + N_L \, \arg (x_L) = 0 \ (\text{mod } 2\pi)
\end{equation}
has to be fulfilled continuously along the entire domain wall trajectory, including the endpoints (i.e. the vacua \eqref{SUN_F=N_vac} with $n=n_0$ and $n= n_0 + k$).
Under the assumption that the eigenvalues do not wind more than once \cite{Ritz:2002fm,Ritz:2003qq}, this fixes uniquely
\begin{equation}
	N_L = k \,, \quad
	N_R = N - k \,.
\label{windings_1}
\end{equation}
Note that this result is valid only when we consider $0 < k \leqslant N/2$ (which is easy to understand, because for the constraint \eqref{F=Nonstraint} to hold we need a small number of eigenvalues to wind the long way and compensate it by a relatively large number of eigenvalues winding the short way in the opposite direction).
This conclusion, eq.~\eqref{windings_1}, has also been confirmed numerically \cite{Benvenuti:2021com}.
For a visual representation see Fig.~\ref{fig:moduli_windings}.

To count the number of possible $k$-walls note that there is an ambiguity in choosing which of the $N$ eigenvalues $x_j$ wind clockwise and which wind counter-clockwise.
Different possible choices yield inequivalent (but degenerate) domain walls.
The number of such choices counts the domain wall multiplicity and is given by the binomial coefficient
\begin{equation}
	\nu^\text{wall}_{N,k} = \begin{pmatrix} N \\ k \end{pmatrix} = \frac{N!}{ k! (N-k)! } \,.
\label{wall_multiplicity_F=N}
\end{equation}

The superpotential \eqref{superpotential_3d_Nf=Nc_xi} enjoys $U(N)$ flavor symmetry\footnote{In the full SQCD, the flavor symmetry that actually acts faithfully is $U(N)/\mathbb{Z}_N$; however, this discrete factor will not matter for the considerations in this paper. For example, as the elements of this $\mathbb{Z}_N$ are proportional to the identity, they act on all the quarks with the same phase, so that this $\mathbb{Z}_N$ is not broken by the wall, and this discrete factor simply ``cancels'' between the numerator and the denominator of eq. \eqref{F=N_moduli_space}.}.
Each domain wall solution with windings \eqref{windings_1} breaks this flavor symmetry down to $U(k) \times U(N-k)$.
While a particular wall solution is unchanged by the action of this $U(k) \times U(N-k)$, the transformations parametrized by the quotient
\begin{equation}
	\mathcal{M}_k^{\text{wall, } F=N} = \mathbb{G}\mathrm{r}(k,N) = \frac{ U(N) }{ U(k) \times U(N-k) } \,.
\label{F=N_moduli_space}
\end{equation}
transform a wall solution into a different wall solution exactly degenerate with the initial one. 
Thus the quotient \eqref{F=N_moduli_space} effectively parametrizes different degenerate walls (with fixed vacua spacing $k$, number of colors $N$, and number of flavors chosen as $F=N$), at least classically.
One can expect that the Grassmannian in eq. \eqref{F=N_moduli_space} is the true moduli space of the domain wall solutions (excluding the trivial translational sector).
For the case $F=N$ this has been checked in \cite{Ritz:2002fm}; below we will generalize this to $F<N$ as well.

For the minimal (elementary)  $k=1$ domain wall interpolating between the neighboring vacua, this moduli space reduces to $\mathbb{CP}(N-1)$.
Classically, this continuous moduli space would lead to massless modes living on the wall worldvolume.
However, quantum mechanically, these modes are lifted and become quasimoduli; the set of degenerate ground states becomes discrete, see below.

This leads us to propose that the effective theory living on the domain wall is a supersymmetric 2d sigma model with the target space\footnote{In this paper we omit the trivial translational moduli, as they correspond to free and decoupled fields in the wall effective theory. Moreover, recall that our domain wall wraps the compact $\mathbb{S}^1$ dimension, so that in the bulk 3d EFT the wall is a 2d worldsheet. Normally, when the bulk is $\mathbb{R}^4$, the worldvolume of the domain wall is three-dimensional, but here the (Euclidean) time direction of the wall worldvolume wraps the compact dimension, so that the wall worldvolume becomes a two-dimensional worldsheet.} 
\eqref{F=N_moduli_space}.
One can naturally expect that the domain wall is a 1/2 BPS state.
Together with $\mathcal{N}=2$ supersymmetry of the bulk 3d theory (i.e. four real supercharges), this would imply $\mathcal{N}=(1,1)$ supersymmetry on the 2d worldsheet.
However, because the target space \eqref{F=N_moduli_space} is a K\"ahler manifold, the worldsheet supersymmetry is automatically enhanced to $\mathcal{N}=(2,2)$ \cite{Ritz:2004mp}, see also Sec.~4.2 of \cite{Bashmakov:2018ghn}.

When we take the decompactification limit $L \to \infty$ and return to the $\mathbb{R}^4$ bulk theory, nothing dramatic happens to the flavor moduli.
The physics is still IR regularized by the mass $m$.
Therefore, we can expect that the wall counting \eqref{wall_multiplicity_F=N} is still valid.
Moreover, the domain wall effective theory should still be given by a sigma model with the same target space \eqref{F=N_moduli_space}; of course, it is now a 3d worldvolume theory.
The amount of supersymmetry in the compactified wall theory, namely $\mathcal{N}=(2,2)$, suggests that the decompactified 3d wall worldvolume theory should have $\mathcal{N}=2$ supersymmetry.

These expectations have been confirmed in \cite{Ritz:2002fm,Ritz:2004mp} by a direct study of the 4d theory (without compactification).
It survives as a non-trivial test of the 3d EFT approach for inferring the 4d physics.
Later we are going to extend it to the case with fewer quark flavors.

\subsection{Wall multiplicity for \boldmath{$0 \leqslant F \leqslant N$}}
\label{sec:wall_multiplicity_FlessN}

Now let us consider a theory with fewer flavors.
We will take $m$ to be the mass of all the quark flavors.
The superpotential in this case is given by eq.~\eqref{superpotential_3d_Nf_Yconf}.

\subsubsection{Preliminary considerations}

Take a $k$-wall interpolating between vacua $n=n_0$ and $n= n_0 + k$ of eq.~\eqref{vacua_3d_Nf_Yj} (we will again focus on $0< k \leqslant N/2$).
We denote eigenvalues of the meson matrix $M$ by $x_i$, $i=1,\ldots,F$.
The constraint encoded by the superpotential \eqref{superpotential_3d_Nf_Yconf} reads
\begin{equation}
	\left( \prod_{i=1}^F x_i \right) \cdot Y \cdot Y_{\text{conf}} \cdot \left( \prod_{i=F+2}^{N - 1} Y_i \right) = 1 \,.
\label{F_constraint}
\end{equation}
Thus, for any $F$, we have a product of $N$ complex variables, constrained to be 1.

One can immediately see that equation \eqref{F_constraint} is basically the same as eq.~\eqref{F=Nonstraint} analyzed for SQCD with $F=N$ flavors.
Thus the whole domain wall counting here essentially repeats the argument of Sec.~\ref{sec:F=N} with the caveat that it gives us the \textit{total} number of domain walls.
The \textit{total} domain wall multiplicity is still given by the same $\nu_{N,k}^\text{walls}$ from eq.~\eqref{wall_multiplicity_F=N}.
However, we need to pay more attention to where the different contributions are coming from, as there are \textit{local} and \textit{topological} contributions to the domain wall multiplicity.

Indeed, consider these two examples.
\begin{itemize}
\item 
In SQCD with $F=N$ flavors, there are effectively no dynamical monopole moduli $Y$ (all the monopoles are confined).
The domain wall multiplicity $\nu_{N,k}^\text{walls}$ is coming purely from $N$ flavor moduli.
The effective theory living on the wall is a sigma model \eqref{F=N_moduli_space}, even in the decompactification limit $L \to \infty$ when we return to 4d.
Different domain walls are locally distinguishable\footnote{At least under an appropriate IR regularization, i.e., finite compactification $L$ or unequal masses for quark hypermultiplets.}, and junctions between them are possible (the latter will be discussed in Sec.~\ref{sec:junc_2}).

\item
In pure SYM, there are no flavor moduli.
Instead, the superpotential \eqref{W_SYM_2} is formulated in terms of $N$ monopole moduli $Y_i$.
This superpotential has nominally the same form as in the $F=N$ SQCD, cf. eq.~\eqref{superpotential_3d_Nf=Nc_xi}, and gives the same domain wall multiplicity.
However, as reviewed in Sec.~\ref{sec:SYM_review}, there can be no junctions between (topologically) different domain walls.
We know in this case that in the limit $L \to \infty$ when we decompactify to 4d, the effective theory on the wall is a $U(1)_N$ CS TQFT \cite{Acharya:2001dz}.
Different domain walls are not distinguishable locally.
\end{itemize}

The two approaches, introducing massive flavors \cite{Ritz:2002fm} or considering the theory on a torus or a circle \cite{Acharya:2001dz}, yield the same result for the multiplicity $\nu_{N,k}^\text{walls}$,
\begin{equation}
	\nu_{N,k}^\text{walls} = \begin{pmatrix} N \\ k \end{pmatrix} = \frac{N!}{ k! (N-k)! } \,.
\label{wall_multiplicity_F=N_discuss}
\end{equation}

This suggests the following interpretation of different contributions to the domain wall multiplicity.
The part of the multiplicity that comes from the flavor moduli $x_i$ persists upon decompactifying the bulk to $\mathbb{R}^4$; junctions between the corresponding walls are physical.
At the same time, the counting coming from the monopole moduli $Y_j$ turns in the limit of large $L$ into counting of ground states of a TQFT; such domain walls are locally indistinguishable, and there can be no junctions between them.
Let us now elaborate on this statement.

\vspace{10pt}

To conclude, let us remark on the domain walls in pure SYM on $\mathbb{R}^4$.
When the domain wall is non-compact and its worldvolume spans $\mathbb{R}^3$, for fixed $k$ one should think of the domain wall as a single wall.
One might worry that there might be a contradiction with the Witten index of the worldvolume theory that reproduces the multiplicity formula \eqref{wall_multiplicity_F=N_discuss}. 

However, in this limit the Witten index of the wall effective theory is not well-defined.
Indeed, despite the fact that the Chern-Simons theory renders the gauge field massive, the zero mode survives.
This makes the spectrum gapless and continuous, and the partition function is ill-defined.
In the current context, this has been discussed in \cite{Witten:1999ds} and \cite{Acharya:2001dz} (the last passage of Sec.~3 in the latter).

A similar situation occurs in SQCD with $F=N$ flavors when the quark hypermultiplets have exactly equal masses.
In this case, the Goldstones of the broken flavor symmetry on the wall do not pick up masses.
A proper computation of the Witten index requires the theory to be regularized in the IR.
For example, unequal quark masses would render these modes as pseudo-Goldstones; a compactification on a circle would turn on IR physics that makes the effective theory gapped.

\subsubsection{Analysis of the BPS wall trajectories}

We want to study more closely the problem of domain wall counting based on the superpotential \eqref{superpotential_3d_Nf_Yconf} and the corresponding constraint \eqref{F_constraint}.
In total, we have $N$ complex variables: $F$ of them are flavor moduli $x_i$, and the rest $N-F$ are the monopole moduli $Y_0, Y_\text{conf}, Y_i$.
Algebraically, this problem is no different from the counting in Sec.~\ref{sec:walls_F=N}.
Repeating the argument between eq.~\eqref{winding_def} and eq.~\eqref{windings_1}, we conclude that for a $k$-wall
exactly $k$ out of these $N$ variables have to wind clockwise, while the rest $N-k$ wind counter-clockwise, see Fig.~\ref{fig:moduli_windings}.

To distinguish between local and topological contributions to the multiplicity, we actually need to take a closer look at the windings of the monopole and flavor moduli separately.
{\it Let $J$ be the number of flavor moduli that wind clockwise.}

Obviously, $J$ cannot be larger than the total number of flavors $F$, and because exactly $k$ out of $N$ complex variables wind clockwise, $J$ cannot exceed $k$; thus $J \leqslant \min(k,F)$.
On the other hand, $J$ must be non-negative, and because exactly $N-k$ variables wind counter-clockwise ($F-J$ of those are the flavor moduli $x_i$), we have $N-k \geqslant F-J$; thus $J \geqslant \max(0,F+k-N)$.
All in all, the allowed values of $J$ lie in the interval
\begin{equation}
	\max(0,F + k - N) \leqslant J \leqslant \min(k,F) \,.
\label{J_interval}
\end{equation}

Now we are ready to write down the formula for domain wall multiplicity.
In a sector with a given $J$, we have $\begin{pmatrix} F \\ J \end{pmatrix}$ choices to pick which of the flavor moduli wind clockwise, and on top of that we also have $\begin{pmatrix} N-F \\ k-J \end{pmatrix}$ choices to pick which of the monopole moduli wind clockwise.
Once these choices are made, the rest of the moduli are bound to wind in the opposite direction.
Summing up over the range of possible values of $J$ gives the total multiplicity of $k$-walls
\begin{equation}
    \nu_{N,k}^\text{walls} \Big|_{\mathbb{R}^3 \times \mathbb{S}^1} = \sum_{J=\max(0,F + k - N)}^{ \min(k,F) } 
        \begin{pmatrix} N-F \\ k-J \end{pmatrix}
        \begin{pmatrix} F \\ J \end{pmatrix}
        = \begin{pmatrix} N \\ k \end{pmatrix}
\label{multiplicity_cylinder_full}
\end{equation}
which of course reproduces\footnote{As was noted in footnote 13 of \cite{Bashmakov:2018ghn}, the last equality sign in \eqref{multiplicity_cylinder_full} can be proven by applying the binomial expansion formula on the two sides of the identity $(1+t)^{N-F} (1+t)^F = (1+t)^{N}$.} 
\eqref{i1} and \eqref{wall_multiplicity_F=N} protected by the Witten index.
Multiplicity formula \eqref{multiplicity_cylinder_full} was derived previously in \cite{Bashmakov:2018ghn} from an entirely different argument.

We stress that the multiplicity formula \eqref{multiplicity_cylinder_full} is valid for any number of quark flavors in the interval $0 \leqslant F \leqslant N$.
In particular, for $F=N$ there are no monopole moduli so that $J \equiv k$ for all walls, while in the pure SYM case there are no flavor moduli and $J \equiv 0$.
Below we will see how the total multiplicity \eqref{multiplicity_cylinder_full} can be derived from the wall effective theory perspective, and after that we will be able to identify the local and topological contributions to this formula, see Sec.~\ref{sec:local_top_mult_discuss}.

\begin{figure}[h]
    \centering
    \includegraphics[width=0.6\textwidth]{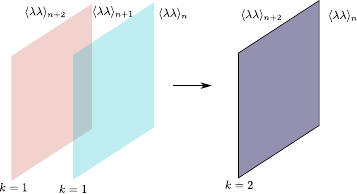}
    \caption{
    	Two BPS domain walls can merge to form a new wall, which may or may not be BPS
    	}
    \label{fig:two_wall_merging}
\end{figure}

We can perform an interesting check of our construction.
Consider pure SYM compactified on a cylinder, and take two parallel $k=1$ domain walls, see Fig.~\ref{fig:two_wall_merging} 
(These requirements are not essential; one can consider two $k_1$- and $k_2$-walls in a theory with a number $F$ of quark flavors; this would only slightly complicate the argument below.)
Since these walls have $k=1$ and are BPS, exactly one monopole modulus winds clockwise in each wall; let us denote the corresponding moduli $Y_{j_1}$ and $Y_{j_2}$ respectively.
The multiplicity of each domain wall is $N$ (corresponding to different choices for $j_1$ and $j_2$), so there are a total of $N^2$ such pairs of domain walls.
Now let these two walls come together to form a single $k=2$ domain wall.
Not all resulting configurations are BPS.
Indeed, in a BPS $k=2$ wall, we need to have two different monopole moduli winding clockwise.
This is the case when $j_1 \neq j_2$, and then the ordering of the pair $(j_1,j_2)$ does not matter for the final state of the wall.
However, when $j_1 = j_2$, we obtain a wall in which a single monopole modulus winds more than a full circle; according to the argument above, such a wall is not BPS.
Therefore, the total number of BPS $k=2$ domain walls equals the total number of the unordered pairs of numbers $(j_1,j_2)$ with $j_1 \neq j_2$, which is precisely $\begin{pmatrix} N \\ 2 \end{pmatrix}$ and agrees with the general multiplicity formulas \eqref{i1} and \eqref{multiplicity_cylinder_full}.

\subsection{Effective theory on the domain wall for \boldmath{$0 \leqslant F \leqslant N$}}
\label{sec:wall_effective}

This approach from the circle compactification allows us to go one step further and make a proposal for the effective theory on the domain wall.
In each of the $N$ vacua from eq.~\eqref{vacua_3d_Nf_Yj}, the symmetry of the bulk 3d theory is 
\begin{equation}
	U(N-F)_\text{mon} \times U(F)_\text{quark}
\label{sym-wall-1}
\end{equation}
where the first factor comes from the symmetry between the monopole moduli, while the second comes from the quark flavor symmetry.
Now, each domain wall solution with given $k$ and $J$ breaks the symmetry \eqref{sym-wall-1} down to
\begin{equation}
	[U(N-F)]_\text{mon} \times [U(F)]_\text{quark}
	\to
	[U(k-J) \times U(N-F-k+J)]_\text{mon} \times [U(J) \times U(F-J)]_\text{quark} \,.
\label{sym-wall-2}
\end{equation}
Recall that $k$ labels the closeness of the vacua on the two sides of the wall (i.e., the wall interpolates from vacuum $n_0$ to $n_0+k$ for some $n_0$), while $J$ is an additional discrete parameter describing the details of the flavor dynamics inside the wall (semiclassically, $J$ is the number of flavor moduli that wind clockwise as we traverse the wall).
Factoring \eqref{sym-wall-1} over \eqref{sym-wall-2} gives the moduli space of the wall; we get a product of two Grassmannians (the subscript $\mathbb{R}^3\times\mathbb{S}^1$ reminds us that we are in a compactified 3d theory)
\begin{equation}
	\mathcal{M}^{k,J}_{\mathbb{R}^3\times\mathbb{S}^1} = \mathbb{G}\mathrm{r}(N-F,k-J)_\text{mon} \times \mathbb{G}\mathrm{r}(F,J)_\text{quark} \,.
\label{wall_moduli_space_3d}
\end{equation}
The Witten index\footnote{For Witten index computation in a more general 2d sigma model, see Sec.~\ref{sec:multi_wall_junc_F=N} below.} of a sigma model with this target space reproduces the corresponding term in the sum \eqref{multiplicity_cylinder_full}.

However, we are actually interested in gaining insights into the physics of domain walls of the decompactified theory on $\mathbb{R}^4$ or close to that limit.
Let us consider what theory may emerge after taking the limit $L \to \infty$.
From the discussion in Sec.~\ref{sec:walls_F=N}, it follows that nothing dramatic should happen to the flavor moduli.
The quarks are IR-regularized by their masses, and taking $L \to \infty$ does not interfere with that.
Moreover, as we repeatedly saw above, the 3d superpotential always recovers the ADS superpotential in the decompactification limit.
Thus, we expect that the quark sector will still give the $\mathbb{G}\mathrm{r}(F,J)$ NLSM living on the domain wall in this limit.
Therefore, we expect that the last factor in the moduli space \eqref{wall_moduli_space_3d} survives.

It is a bit more difficult to answer what happens to the contribution of the monopole moduli.
Strictly speaking, when we pass from 3d to 4d, there are no monopole moduli anymore, and the corresponding sector of the gauge theory becomes strongly coupled.
We can understand this from the Lagrangian point of view as follows.
When we want to take the limit of large $L$ and rescale the fields and parameters accordingly (see Table~\ref{tab:3d-4d}), the 3d superpotential \eqref{superpotential_3d_Nf_Yj} for both the monopole and the flavor moduli is naturally proportional to $L$, which corresponds to integration over the compact dimension.
The 3d K\"ahler potential for quarks  
\begin{equation}
	\mathcal{K}^Q_{3d} = L \bar{Q} Q
\end{equation}
is also proportional to $L$ (cf. Table~\ref{tab:3d-4d}).
However, the story is different for the monopole moduli.
As follows from the definition \eqref{x_moduli_def} and \eqref{Y_def}, the K\"ahler potential for the monopole moduli has the form 
\begin{equation}
	\mathcal{K}^Y_{3d} = \frac{g_{4d}^2}{L} \ln( \bar{Y} ) \, \ln( Y )
\end{equation}
where $g_{4d}^2$ is the dimensionless coupling of the 4d gauge theory.
We conclude that in the decompactification limit $L \to \infty$ the dynamical NLSM associated with quarks survives ($\mathbb{G}\mathrm{r}(F,J)_\text{quark}$ above), while instead of the monopole NLSM, we are left with its far IR counterpart.
We would like to argue that it is a TQFT.
This is a natural guess, since we know the answer for pure SYM \cite{Acharya:2001dz} (see also e.g. Sec. 2 of \cite{Bashmakov:2018ghn} and Sec. 4.2 of \cite{Cox:2024tgo} for a review): in that case, the effective theory living on the 3d domain wall worldvolume is given by\footnote{Note that eq.~\eqref{TQFT_SYM} is written in conventions for the non-supersymmetric CS theory. It is obtained from $\mathcal{N}=1$ $U(k)_N$ CS with an adjoint scalar \cite{Acharya:2001dz} by integrating out the scalar and the gauginos.}
\begin{equation}
	U(k)_{N-k,N} \equiv \frac{ SU(k)_{N-k} \times U(1)_{N} }{ \mathbb{Z}_k } \text{ CS} \,, \quad 
	0 < k \leqslant \frac{N}{2} \,.
\label{TQFT_SYM}
\end{equation}
Equation (\ref{TQFT_SYM}) presents the definition of $U(a)_{b,c}$ repeatedly used below.

Here we would like to stress that we focus on $k$-walls with $0 < k \leqslant N/2$.
The walls with other values of $k$ are related to these by parity and $k \sim k + N$ periodicity.
Normalization of the $U(1)$ level in \eqref{TQFT_SYM} is such that $U(k)_{N,N}$ is a theory that one would get by considering a CS term with a trace in the fundamental representation of $U(N)$.
In the simplest case of the $k=1$ wall, eq.~\eqref{TQFT_SYM} reduces to $U(1)_N$.

In our case, the sigma model of interest is a 2d theory with a Grassmannian target space,
\begin{equation}
    \mathbb{G}\mathrm{r}(N-F,k-J)_\text{mon}\,.
\label{grassmanian_mon_discuss_1}
\end{equation}
This is part of the domain wall effective theory when the bulk (and the domain wall) is compactified on a small circle.
We are going to argue that, upon decompactification, this part of the domain wall theory actually flows to a 3d TQFT.
Admittedly, our argument is not very rigorous, but it provides us with a good guess.

What TQFT can naturally emerge from $\mathbb{G}\mathrm{r}(N-F,k-J)_\text{mon}$ in the decompactification limit?
To answer this question, recall that the non-linear sigma model with the target space \eqref{grassmanian_mon_discuss_1} can be described as a gauged linear sigma model either with an auxiliary $U(k-J)$ gauge field or with an auxiliary $U(N-F-k+J)$ gauge field (appropriately supersymmetrized, of course).
If we start with the formulation of the model with the $U(k-J)$ gauge field, we might expect that the resulting TQFT should involve factors $SU(k-J)$ and $U(1)$ at some levels.
However, we could just as well use the auxiliary $U(N-F-k+J)$ gauge field, as it does not matter for the Grassmannian; in this case, we would expect factors $SU(N-F-k+J)$ and $U(1)$ at some levels in the resulting TQFT.
One can expect that this arbitrariness of description should be respected by the level-rank duality $U(n)_{m-n,m} \leftrightarrow U(m-n)_{-n,-m}$ \cite{Hsin:2016blu,Aharony:2015mjs}.
These simple observations already suggest the TQFT to be
\begin{equation}
	U(k-J)_{(N-F) - (k-J), N-F}  \text{ CS} \,.
\label{TQFT_FkJ}
\end{equation}
The level-rank duality actually corresponds here to a parity transformation that relates a $k,J$-wall to an $(N-k),(F-J)$-wall with the effective theory 
\begin{equation}
    U \big( (N-F) - (k-J) \big)_{-(k-J), -(N-F)} \,.
\end{equation}
For $F=0$ we have $J=0$ by definition, and formula \eqref{TQFT_FkJ} recovers the SYM case \eqref{TQFT_SYM}.
The level of the non-Abelian factor in \eqref{TQFT_SYM} and \eqref{TQFT_FkJ} reflects the total number of monopole or quark moduli winding counter-clockwise, while the second index shows the total number of the monopole moduli.

We can conclude that, in the decompactified bulk SQCD on $\mathbb{R}^4$ (or on $\mathbb{R}^3 \times \mathbb{S}^1$ with a circle of large radius), the domain wall effective theory is a product
\begin{equation}
\boxed{	\text{ CS TQFT defined in \eqref{TQFT_FkJ} } \times  \frac{ U(F) }{ U(J) \times U(F-J) } \text{ NLSM} \,.  }
\label{domain_wall_theory}
\end{equation}
This is our proposal for the effective theory on the domain wall for \textit{any} finite value of the quark mass.
The degrees of freedom of the TQFT and of the NLSM in \eqref{domain_wall_theory} are decoupled from each other.

We stress that while our argument is not rigorous and should be viewed as a justification of a plausible guess rather than a derivation, it is consistent with itself and with the known result for pure SYM \eqref{TQFT_SYM}. Moreover, the only conjectural step is really the assumption that the monopole moduli theory flows to a TQFT; if one accepts this, then it is more or less uniquely fixed what that TQFT should be.

Here we would like to take a pause and discuss an important point.
In \cite{Bashmakov:2018ghn}, the authors propose another 3d theory living on the domain wall worldvolume.
That theory has its own effective mass parameter, and, depending on the sign of that mass, the theory can be in one of two phases.
Our proposed theory in eq.~\eqref{domain_wall_theory} agrees with the result of \cite{Bashmakov:2018ghn} for negative values of that effective mass parameter\footnote{See eq. (4.12) of \cite{Bashmakov:2018ghn}. Note that in passing from $\mathcal{N}=1$ formulation to the usual formulation of CS, the level of the non-Abelian factor is shifted as $U(n)_k \to U(n)_{k-\frac{n}{2}}$ after integrating out the chiral adjoint fermions at one loop.}.
However, for positive effective mass parameter, the theory of \cite{Bashmakov:2018ghn} is in a different phase, with the phase transition happening at zero effective mass.
The authors conjecture that the zero effective mass in that model corresponds to some finite value of the quark mass in 4d.
Our analysis suggests that there is no such phase transition at any finite value of the quark masses.
Therefore, we can conclude that the point of zero effective mass in the model of \cite{Bashmakov:2018ghn} actually corresponds to an infinite mass of the 4d quarks.

\subsection{Local and topological multiplicity}
\label{sec:local_top_mult_discuss}

Having learned about the nature of the domain wall effective theory \eqref{domain_wall_theory}, we can now take a step back and have another look at the multiplicity formula \eqref{multiplicity_cylinder_full}, which we reprint here:
\begin{equation}
    \nu_{N,k}^\text{walls} \Big|_{\mathbb{R}^3 \times \mathbb{S}^1} = \sum_{J=\max(0,F + k - N)}^{ \min(k,F) } 
        \begin{pmatrix} N-F \\ k-J \end{pmatrix}
        \begin{pmatrix} F \\ J \end{pmatrix}
        = \begin{pmatrix} N \\ k \end{pmatrix} \,.
\label{multiplicity_cylinder_full_duplicate_1}
\end{equation}
This formula was derived above by counting solutions of the BPS wall equations.
Let us now compare it to the number of ground states of the effective theory \eqref{domain_wall_theory}.

First of all, the Witten index for the $\mathbb{G}\mathrm{r}(J,F)$ sigma model in \eqref{domain_wall_theory} is\footnote{\label{ft:witten_IR} In order to make the Witten index well-defined, we need to IR regularize the massless Goldstone modes of the spontaneously broken continuous flavor symmetry. This can be done, e.g., by keeping the compactification circle radius finite; in this case, the effective theory becomes two-dimensional in the long-wavelength limit. For Witten index computation in a more general 2d sigma model, see Sec.~\ref{sec:multi_wall_junc_F=N} below. An alternative method of regularization is to take different masses for the quark flavors, thus changing the would-be Goldstone modes to massive pseudo-Goldstone excitations.} 
\begin{equation}
I_W= \begin{pmatrix} F \\ J \end{pmatrix}
\end{equation}
which corresponds to the second factor under the sum in eq.~\eqref{multiplicity_cylinder_full_duplicate_1}.
This part of the multiplicity is due to the flavor degrees of freedom.
Different domain walls in this sector can be distinguished locally, and junctions between them are possible (see more on that below).
On the wall effective theory, these junctions correspond to solitons of the sigma model interpolating between different ground states.
This motivates us to call this contribution \textit{local} multiplicity.

On the contrary, the first factor under the sum in eq.~\eqref{multiplicity_cylinder_full_duplicate_1} is due to a very different physics of the gauge fields.
The Witten index for the $U(k-J)_{(N-k) - (F-J), N-F}$ CS TQFT from eq.~\eqref{domain_wall_theory} is well-defined only when the theory is compactified on a torus or a cylinder \cite{Acharya:2001dz} (the size of the compact dimension(s) may be large but finite).
Only in that case do we see the corresponding contribution $\begin{pmatrix} N-F \\ k-J \end{pmatrix}$ to the wall multiplicity\footnote{
	For computation of the Witten index of the CS theory in the current context, see \cite{Acharya:2001dz,Ohta:1999iv,Bergman:1999na}; the original paper is \cite{Witten:1999ds}.
}.
On the wall worldvolume, this index counts the number of ground states of this TQFT; however, there can be no junctions between them.
Moreover, if we decompactify completely and return to the $\mathbb{R}^4$ bulk, this multiplicity disappears; the Witten index is not well-defined in this case.
This motivates us to call this contribution \textit{topological} multiplicity.

Note that formulas \eqref{domain_wall_theory} and \eqref{multiplicity_cylinder_full_duplicate_1} work even for the borderline cases of $F=N-1$ and $F=N$.
The latter is pretty obvious: the TQFT in \eqref{domain_wall_theory} has level zero and is therefore trivial, while the sum in \eqref{multiplicity_cylinder_full_duplicate_1} contains only one term.
The case of $F=N-1$ is a bit more tricky.
As an example, take $N=2$ and $F=1$; this corresponds to the SQCD discussed in Sec.~\ref{sec:review_su2}, and we expect to find two domain walls with no non-trivial moduli\footnote{Except the translational moduli, which we do not discuss here, since they are free and decoupled.} and with a possible two-wall junction.
Indeed, the multiplicity formula \eqref{multiplicity_cylinder_full_duplicate_1} gives
\begin{equation}
	  \begin{pmatrix} 1 \\ 1 \end{pmatrix}  \begin{pmatrix} 1 \\ 0 \end{pmatrix} 
	+ \begin{pmatrix} 1 \\ 0 \end{pmatrix}  \begin{pmatrix} 1 \\ 1 \end{pmatrix} 
	= 2 \,.
\end{equation}
The two walls in Fig.~\ref{fig:junc} correspond to $J=0$ for the upper one, while $J=1$ for the lower wall. 
The formula \eqref{domain_wall_theory} gives a trivial theory for each of these walls. 
The junction in this case is not describable within the domain wall theory; see more on that in the next Section.

\section{Two-wall junctions}
\label{sec:junc_2}

In this Section, we are going to discuss the junctions of two domain walls.
The junctions might arise when the multiplicity of domain walls interpolating between a given pair of vacua is greater than one.

Such junctions are absent when the bulk theory is pure SYM.
However, when we add matter flavors, the question of whether or not such junctions exist becomes non-trivial.
When they do exist, there are strong indications that these junctions are 1/4 BPS with respect to the 4d $\mathcal{N}=1$ supersymmetry algebra 
\cite{Gorsky:1999hk,Ritz:2004mp,Carroll:1999wr}.


\subsection{Two-wall junctions in \boldmath{$SU(2)$ $F=2$} SQCD}
\label{sec:two_wall_junc_N=2_F=2}

As a warm-up, we start from SQCD with gauge group SU(2) on the $\mathbb{R}^4$ bulk. 
We will include $F=2$ quark flavors with masses $m_1$, $m_2$.
In this theory, there are two chiral vacua.
As we discussed above, in this case, there can be two distinct domain walls interpolating between these two vacua.
The two walls can form a junction, see Fig.~\ref{fig:junc}.
The effective theory on the wall worldvolume is the 3d $\mathbb{CP}(1)$ model (this was also derived in \cite{Ritz:2004mp}), and the junction represents a domain line of this theory.

In the hierarchical limit of quark masses $m_2 \gg m_1$, the effective $\mathbb{CP}(1)$ sigma model is weakly coupled, and the junction tension computed in \cite{Ritz:2004mp} reads
\begin{equation}
	T_{\text{junc}} = \pi\left|\frac{\left|m_1\right|-\left|m_2\right|}{\sqrt{\left|m_1 m_2\right|}}\right| \Lambda_{F=2}^2
	\xrightarrow{m_2 \to \infty}
	\pi \sqrt{ \frac{\Lambda_{F=1}^5}{|m_1|} } \,.
\label{junc_tension_weak_1}
\end{equation}
When taking $m_2$ to infinity, one should keep $\Lambda_{F=1}^5 = m_2 \Lambda_{F=2}^4$ fixed, see Appendix~\ref{sec:betafunc}.

When we compactify the bulk to a cylinder $\mathbb{R}^3 \times \mathbb{S}^1$, the domain wall also wraps the compact dimension.
Consider the case when the junction also wraps the compact dimension.
In this case, the domain wall theory reduces to the 2d $\mathbb{CP}(1)$ model, while the junction becomes a kink interpolating between two vacua (i.e., the two distinct domain walls).
The dynamical scale generated in the 2d effective theory is
\begin{equation}
	\Lambda_{2d} = \mu \, \exp( - \frac{2\pi}{g_{2d}^2(\mu)} )	\,, \quad
	\frac{1}{g_{2d}^2(\mu)} = L \, \frac{\Lambda_{F=2}^2}{ \sqrt{m_1 m_2} } \,,
\label{2d_scale}
\end{equation}
at least when the quark masses $m_1$ and $m_2$ are small compared to the bulk SQCD scale $\Lambda_{F=2}$,
see e.g. \cite{Ritz:2004mp} (Sec.~III.B and Sec.~IV.B there).
This scale becomes relevant in the strong coupling regime $m_1 \sim m_2$, and we can make only some qualitative statements.
Here, $\mu$ is the UV scale of the effective theory set by the bulk quark masses.

At weak coupling, the kink mass in the 2d $\mathbb{CP}(1)$ model is given by
\begin{equation}
	M_{kink} \approx \frac{ |\Delta m| }{ g_{2d}^2 }
\label{2d_kink_mass_weak}
\end{equation}
where $\Delta m = m_2 - m_1$ is the mass scale of the theory.
At small $\Delta m$, the theory becomes strongly coupled, and the kink mass is given by the central charge \cite{Dorey:1998yh}, which at $\Delta m = 0$ reads
\begin{equation}
	M_{kink} = \frac{1}{ 2 \pi e }  \Lambda_{2d} \,.
\label{2d_kink_mass_strong}
\end{equation}
where $e = 2.718 \ldots$ is Euler's number.

To see the consistency between these results, note that when the junction world line wraps the compact dimension, the junction tension and the mass of the corresponding kink in the 2d effective theory on the domain wall are related as
\begin{equation}
	M_{kink} = L \, T_{\text{junc}} \,.
\label{kink_junc_mass_relation}
\end{equation}
Then, at weak coupling, \eqref{junc_tension_weak_1} and \eqref{kink_junc_mass_relation} together give \eqref{2d_kink_mass_weak}.

At strong coupling, $g_{2d}^2$ becomes large, and \eqref{2d_scale} and \eqref{2d_kink_mass_strong} imply that the 2d kinks become heavy, with masses of the order of the bulk quark mass.
When $ \Delta m  = m_1 - m_2 \ll m_1, m_2$, one can naively expect that the strong coupling sets in; this is indeed the case when the quark masses are not large compared to the bulk scale $\Lambda_{F=2}$.
However, below in Sec.~\ref{sec:junc_tension_various_m}, we are going to argue that $g_{2d}^2$ actually becomes small in the limit of very large quark masses.

\vspace{10pt}

We end this subsection with the following note.
The supersymmetric 2d $\mathbb{CP}(1)$ sigma model at $\Delta m = 0$ has two degenerate kink supermultiplets corresponding to the CFIV index = 2 \cite{Cecotti:1992qh}. 
The corresponding kinks have the same topological charge 1 and form a doublet with respect to the SU(2) flavor symmetry of this sigma model (see e.g. \cite{Witten:1978bc,Hori:2000kt} and references therein).
This suggests that the two-wall junctions in the $\mathbb{R}^3 \times \mathbb{S}^1$ setup also form a representation with respect to the flavor symmetry.

Away from the point $\Delta m = 0$, flavor symmetry is explicitly broken, and the junction tensions are no longer degenerate.
In particular, at large $\Delta m$, the bulk theory flows to the SU(2) SQCD with $F=1$ flavor.
In this case, the two kinks mentioned above have very different masses, only one of them being relatively light.
Precisely this light kink is seen as the junction in the 4d SQCD \cite{Ritz:2004mp}; in this case, the one flavor that we have interpolates between going clockwise in the first wall and going counter-clockwise in the second wall, see Fig.~\eqref{fig:junc} for the wall trajectories and the corresponding junction.

\subsection{Two-wall junctions in \boldmath{$SU(N)$} SQCD}

Now let us discuss the general case of $SU(N)$ SQCD with $F$ quark flavors.
As we saw above, for generic $0 \leqslant F \leqslant N$, domain walls are characterized by two numbers, $k$ 
(the number of the walls interpolating between bulk vacua number $n$ and $n+k$, see eq. \eqref{quark_vacua}, \eqref{quark_vacua_3d}, \eqref{vacua_3d_Nf_Yj}) 
and an additional parameter $J$ characterizing the inner flavor structure of the wall,
\vspace{1mm}

 \noindent
{\it i.e., $J$ of the eigenvalues of the meson matrix $M$ wind clockwise; the remaining $F-J$ wind counter-clockwise,}
\vspace{1mm}

\noindent
see Sec.~\ref{sec:wall_multiplicity_FlessN}).
The domain wall effective theory is then, generally speaking, a product of a TQFT and a sigma model, see eq.~\eqref{domain_wall_theory}.

When it comes to considering two-wall junctions, the question is meaningful only when the two domain walls interpolate between the same pair of vacua, i.e., the two walls should have the same value of $k_1 = k_2 = k$ (the indices 1 and 2 refer to the two walls).
After that, there are a few possibilities, depending on $J$ and the number of flavors $F$:
\begin{enumerate}
\item Junctions between two walls with the same $J_1 = J_2 = J$ (in particular, this is always the case if $F = N$)
\item Junctions between two walls with different $J_1 \neq J_2$ in the theory with $F = N-1$
\item Junctions between two walls with different $J_1 \neq J_2$ in the theory with $F < N-1$
\end{enumerate}
Now let us discuss these possibilities one by one.

\subsubsection{Two walls with \boldmath{$J_1 = J_2$}}
\label{sec:junc_2_SUN_J1=J2}

In this case, the effective theories \eqref{domain_wall_theory} are identical on the two walls.
In pure SYM, i.e., for $F=0$, there are no such junctions, but they become possible when we introduce matter flavors.

To gain some insights about the junctions, let us first discuss the properties of the domain wall effective theory.
The supersymmetric Grassmannian sigma model $\mathbb{G}\mathrm{r}(F,J)$ in 2d is known to have $\begin{pmatrix} F \\ J \end{pmatrix}$ vacua; this model is relevant to our domain wall if we keep the compactification scale $L$ finite.
The low-lying spectrum around a given vacuum consists of an $F$-plet of kinks transforming in the fundamental representation of the flavor $SU(F)$, with mass given by a formula similar to eq.~\eqref{2d_kink_mass_weak} and eq.~\eqref{2d_kink_mass_strong} (at weak and strong coupling, respectively). 
This picture is very similar to the $\mathbb{CP}(F-1)$ case.
These low-lying kinks interpolate between ``neighboring'' vacua, but there are also heavier kinks interpolating between more ``distant'' vacua \cite{Ireson:2019huc}.

\begin{figure}[h]
    \centering
    \includegraphics[width=0.2\textwidth]{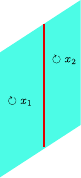}
    \caption{
    	Junction of two $k=1$ walls.
    	In each wall exactly one meson matrix eigenvalue winds counter-clockwise
    	}
    \label{fig:junction_2wall_simple}
\end{figure}

Now let us interpret these results in terms of the domain wall junctions.
When the masses of the bulk SQCD quarks are hierarchical, the domain wall theory becomes weakly coupled.
The $\mathbb{G}\mathrm{r}(F,J)$ kinks are not degenerate.
The lightest kinks correspond to junctions between domain walls that have the same $k$ and $J$ but which differ by a pair of meson eigenvalue trajectories.
For example, if $x_j$, $j=1,\ldots,F$ are the meson matrix eigenvalues, then we can consider two walls for which the trajectories of all the meson eigenvalues are the same in the two walls, except that in the first wall $x_{1}$ winds clockwise and the rest counter-clockwise, while for the second wall $x_{2}$ winds clockwise and the rest counter-clockwise, see Fig.~\ref{fig:junction_2wall_simple}.

When the mass differences of all the SQCD quarks tend to zero, the Grassmannian sigma model $\mathbb{G}\mathrm{r}(F,J)$ living on the domain wall is at strong coupling.
To stabilize the junctions, we need to render the would-be Goldstone modes massive; this can be achieved by keeping the compactification scale $L$ finite.
In this case, the low-lying kinks correspond to an $F$-plet of two-wall junctions with degenerate tensions.
The analysis of Sec.~\ref{sec:two_wall_junc_N=2_F=2} above suggests that the tensions of these junctions grow with the bulk quark masses.


To reiterate, two different domain walls with the same $k$ and $J$ can come together in a junction describable within the effective sigma model living on the wall (the possible TQFT factor is irrelevant in this case).
When the bulk quark masses are equal, these domain wall junctions have degenerate tensions and form an $F$-plet of the flavor $SU(F)$ symmetry (of the domain wall theory).
Large unequal quark masses lift this degeneracy.

Note that in SQCD with $F = N$ flavors, we always have $J_1 = J_2 = k$.

\subsubsection{\boldmath{$F = N-1$}, two walls with \boldmath{$J_1 \neq J_2$}}

When there are $F = N-1$ quark flavors in the bulk SQCD, there are no low-lying gauge degrees of freedom living on the wall.
This is seen on $\mathbb{R}^4$ as full Higgsing of the gauge group, and in the circle compactified theory as the fact that all monopole moduli become confined, see Sec.~\ref{sec:mon_conf}.
The TQFT factor in the effective theory on the wall \eqref{domain_wall_theory} is trivial.

In this case, for a $k$-wall we have $J = k$ or $J = k-1$ (see eq.~\eqref{J_interval}).
When the two walls with $J_1 = J_2$ meet in a junction, such a junction is describable within the effective sigma model living on the wall, see above.

However, when $J_1 \neq J_2$, the sigma models of \eqref{domain_wall_theory} living on the two walls are different (unless $k = N/2$, which is a special case).
Because the effective theories differ on the two sides, the junction is not describable within the domain wall; the bulk degrees of freedom become involved.
This complicates the study of such junctions; we leave this task for future work.
The junction in $N=2$, $F=1$ SQCD mentioned in Sec.~\ref{sec:review_su2}, Sec.~\ref{sec:local_top_mult_discuss}, and Sec.~\ref{sec:two_wall_junc_N=2_F=2} is exactly of this type.

\subsubsection{\boldmath{$0 < F < N-1$}, two walls with \boldmath{$J_1 \neq J_2$}}

Much as in the previous case, the effective theories on the two walls, eq.~\eqref{domain_wall_theory}, are different on the two walls.
In particular, the sigma models for the flavor moduli are different, and we do not expect this junction to be describable within the wall effective theory.

Apart from the flavor contribution to the junction dynamics, we also have two different TQFTs living on the two walls.
This suggests that, if such a junction exists, it might support gapless degrees of freedom, possibly a chiral CFT.
However, the analysis of the next section seems to suggest that this is not the case.
We leave detailed investigation of this puzzle for future work.

\section{Multi-wall junctions}
\label{sec:junc_multi}

\begin{figure}[ht]
    \centering
    \includegraphics[width=0.5\textwidth]{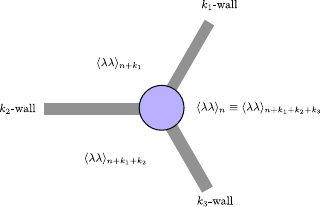}
    \caption{
    	3-wall junction
    	}
    \label{fig:junc_multi}
\end{figure}

In the previous Section we briefly discussed junctions that can be formed by two domain walls; they are made possible by the presence of the flavor degrees of freedom on the wall.
Here we consider junctions formed by a number $N_\text{walls} \geqslant 3$ of domain walls meeting along a line, see Fig.~\ref{fig:junc_multi}; these are possible even in pure SYM.
These walls divide the space into $N_\text{walls}$ regions with different values of the gluino condensate $\expval{ \lambda \lambda }$. 
Take the sequence of vacua around the junction as ($n$, $n+k_1$, $n + k_1 + k_2$, $\ldots$), then the domain walls between these vacua are characterized by the positive integers $k_a$ subject to the condition
\begin{equation}
	\sum_a k_a = N \,; \quad
	k_a > 0 \,, \quad
	a = 1,\ldots,N_\text{walls} \,.
\label{ka_condition}
\end{equation}
In principle, we could have picked a different condition of the form $\sum k_a = p N$ with any $p \in \mathbb{Z}$, but eq.~\eqref{ka_condition} is the simplest nontrivial case.
We want to understand what low-lying degrees of freedom can be supported on this junction.
The answer can be expected to depend on the number of quark flavors $F$ present in the theory.

We will not give here a detailed analysis of the multi-wall junctions' tension in this case, constraining ourselves to a few comments.

\subsection{Pure SYM and Gaiotto's proposal}
\label{sec:gaiotto_cft_junc}

As a warm-up, we start with the pure SYM case, i.e. $F=0$.
The effective theory living on the junction worldsheet in this case was proposed by Gaiotto in \cite{Gaiotto:2013gwa}: it is a supersymmetric chiral WZW coset CFT
\begin{equation}
	T_{\left(k_a\right)}=\left[\frac{u(N)_N}{\prod_a u\left(k_a\right)_N}\right] \,.
\label{gaiotto_junc_scft}
\end{equation}
The naive amount of supersymmetry respected by this 1/4 BPS junction is $\mathcal{N}=(1,0)$, but actually the model \eqref{gaiotto_junc_scft} possesses $\mathcal{N}=(2,0)$ ``hidden'' supersymmetry \cite{Gaiotto:2013gwa}.

When the bulk theory is considered on $\mathbb{R}^4$, the theory \eqref{gaiotto_junc_scft} lives in 2d.
We want to understand whether we can say something about this theory by considering the compactified bulk theory.
As in Sec.~\ref{sec:wall_effective}, we can start from the moduli space describing the object under consideration --- in this case, the junction.
The main guiding principle for us will be that the domain wall configuration should be BPS.
In the vacuum $n$ (first in our sequence), all the monopole moduli $Y_j$ have the same value.
When we pass through the first $k_1$-wall, we have $k_1$ of the monopole moduli winding clockwise and $N - k_1$ winding counter-clockwise (see Sec.~\ref{sec:wall_multiplicity_FlessN}), and we pass to the vacuum number $n + k_1$.
On the second $k_2$-wall, there are $k_2$ monopole moduli winding clockwise and $N - k_2$ winding counter-clockwise, and so on.

As we trace a closed path around the junction, each monopole modulus $Y_j$ winds in total $2 \pi w(Y_j)$ around the origin, with $w(Y_j)$ being the winding number.
Since we return to the same vacuum number $n$ that we initially started from, each $w(Y_j)$ is an integer.
If a given $w(Y_j)$ is non-zero, then the junction must carry an appropriate Wilson line (otherwise the corresponding $Y_j$ inside the junction would interpolate through the origin, which is forbidden; see Sec.~\ref{sec:SYM_review}).

In the simplest case, when we do not have a Wilson line, all the windings $w(Y_j)$ vanish.
Let us understand which junctions satisfy this condition.
To this end, recall that when a modulus winds clockwise, it picks up a phase $2\pi (k_a/N - 1)$, while winding counter-clockwise inside a $k_a$-wall, it picks up a phase $2 \pi k_a/N$.
If a given modulus $Y_j$ winds clockwise inside a number $N_L^{(j)}$ of walls, then this modulus winds counter-clockwise inside $N_R^{(j)} = N_\text{walls}-N_L^{(j)}$ walls.
The total winding number is then
\begin{equation}
	w(Y_j) =   \sum_{a \in \text{clockwise} } \left(\frac{k_{a}}{N} - 1 \right)   + \sum_{a \in \text{counter-clockwise} }  \frac{k_{a}}{N}
	= 1 - N_L^{(j)}
\end{equation}
where we used eq.~\eqref{ka_condition}.
Thus, to have zero winding, we require 
\begin{equation}
	N_L^{(j)} = 1 \,, \quad
	\text{for all } j = 0,\ldots,N-1 \,.
\label{1_wind}
\end{equation}
To check that this requirement is consistent, note that on one hand this implies
\begin{equation}
	\sum_{j=0}^{N-1} N_L^{(j)} = N \,.
\label{SYM_NR_constr_1}
\end{equation}
On the other hand, since in a $k_a$-wall there must be exactly $k_a$ moduli winding clockwise (see Sec.~\ref{sec:wall_multiplicity_FlessN}), we have an additional constraint, namely
\begin{equation}
	\sum_{j=0}^{N-1} N_L^{(j)} = \sum_{a=1}^{N_\text{walls}} k_a = N
\label{SYM_NR_constr_2}
\end{equation}
where we used eq.~\eqref{ka_condition} again.
One can see that \eqref{SYM_NR_constr_1} and \eqref{SYM_NR_constr_2} are indeed consistent.

To reiterate, in order to have zero winding for each monopole modulus, $w(Y_j) \equiv 0$, we require that each of these moduli winds clockwise inside exactly one wall (and cannot do it twice).

It follows then that the inner structure of each junction can be characterized semiclassically by a cyclically ordered partition of $N$ monopole moduli labels into subsets of cardinalities $k_a$ (the $a^\text{th}$ subset specifies which of $Y_j$ wind clockwise in the $a^\text{th}$ wall).
This breaks the underlying $U(N)$ symmetry into a product $\prod U(k_a)$, and the junction moduli space becomes
\begin{equation}
	\mathcal{M}^\text{junc, SYM}_{\left(k_a\right)} = \frac{ U(N) }{ \prod_a U(k_a) } \,.
\label{semiclassical_junc_moduli_space}
\end{equation}
One can see that this simple argument reproduces the correct symmetry of the junction effective theory \eqref{gaiotto_junc_scft}.
This suggests that when we decompactify the bulk, the 1d quantum mechanics on the moduli space \eqref{semiclassical_junc_moduli_space} becomes the CFT \eqref{gaiotto_junc_scft}.

The analysis above can be extended to the case when the junction does carry one or more Wilson lines.
Let us suppose that we have inserted into the junction core a combination of Wilson lines corresponding to a vector from a weight lattice, $\vec{w}_\text{line} \in \Lambda_W$.
Then, as we make a full circle around the junction, the monopole moduli $Y_j$ pick up windings $w(Y_j) = \vec{w}_\text{line} \cdot \vec{\alpha}_j$.
Because for the affine root we have $\vec{\alpha}_0 = -\sum_{j=1}^{N-1}\vec{\alpha}_j$, the condition \eqref{SYM_NR_constr_2} is still satisfied.

\begin{figure}[t]
    \centering
    \begin{subfigure}[t]{0.49\textwidth}
        \centering
        \includegraphics[width=0.9\textwidth]{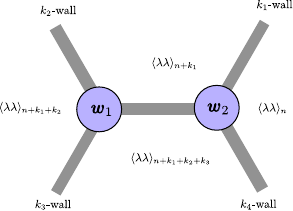}
        \label{fig:junction_3wall_3wall}
    \end{subfigure}%
    ~ 
    \begin{subfigure}[t]{0.49\textwidth}
        \centering
        \includegraphics[width=0.65\textwidth]{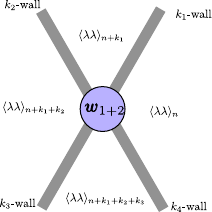}
        \label{fig:junction_4wall}
    \end{subfigure}%
	\caption{
		Two 3-wall junctions can merge to form one 4-wall junction.
		If the initial junctions carried Wilson lines with charges $\vec{w}_1$ and $\vec{w}_2$, the resulting junction will carry a charge $\vec{w}_{1+2} = \vec{w}_1 + \vec{w}_2$ Wilson line
	}
\label{fig:junction_manipulation}
\end{figure}

Because the winding numbers and Wilson line charges are additive, this analysis appears to be consistent under junction manipulation.
For an example, see Fig.~\ref{fig:junction_manipulation}.

\subsection{Multi-wall junctions in SQCD}

By extending the above argument, we can make a guess about the effective theory on the junction worldsheet in the presence of quarks.
In this case, each wall in a junction is characterized by not just one but two numbers, $k_a$ and $J_a$.
We will still impose the condition \eqref{ka_condition}.

The main difference between the flavor moduli and the monopole moduli is that the origin of the moduli space is no longer protected by any symmetry.
A junction can interpolate through the origin.
This will, generally speaking, lift the restriction on the total winding of the flavor moduli (the eigenvalues of the meson matrix $M$).
It is not completely clear what constraint the BPS-ness condition puts on these windings.
One can expect that the configurations with the lowest possible windings are BPS; however, there are indications that other junctions (not satisfying the lowest winding condition) might also be BPS, see below.

This motivates us to introduce the term ``minimal junction''.
We will call a junction ``minimal'' if the flavor moduli windings around it satisfy the same condition as the windings of the monopole moduli in pure SYM (with no Wilson lines inserted), namely that each modulus has to wind clockwise inside precisely one domain wall and counter-clockwise inside all other walls.
In this case, the condition \eqref{ka_condition} is supplemented by $\sum J_a = F$, so that the resulting constraint reads
\begin{equation}
	\sum_a k_a = N \,, \quad
	\sum_a J_a = F \,; \quad
	k_a, J_a > 0 \,, \quad
	a = 1,\ldots,N_\text{walls} \,.
\label{kJa_condition}
\end{equation}
Below we will study junctions that satisfy this condition.

\subsubsection{\boldmath{$SU(N)$} SQCD with \boldmath{$F=N$} flavors}
\label{sec:multi_wall_junc_F=N}

Let us start with the case $F = N$.  
We will focus for now on the ``minimal'' junctions defined by condition \eqref{kJa_condition}.

Consider a junction of $N_\text{wall} \geqslant 3$ domain walls.  
In this case, we necessarily have $J_a = k_a$ for all walls; see eq.~\eqref{J_interval}.  
The effective theory on each domain wall, eq.~\eqref{domain_wall_theory}, is a sigma model (without TQFT factors).  
The requirement that each flavor modulus winds clockwise in exactly one wall (the ``minimal junction'' condition, see eq.~\eqref{kJa_condition}) leads us to parametrize the junction by a cyclically ordered partition of the $N$ flavor moduli labels into subsets of cardinalities $k_a$.  
The $a^\text{th}$ subset specifies which of the flavor moduli wind clockwise in the $a^\text{th}$ wall.

This breaks the underlying $U(N)$ flavor symmetry into a product $\prod U(k_a)$, and the junction moduli space becomes
\begin{equation}
	\mathcal{M}^{\text{junc, } F=N}_{\left(k_a\right)} = \frac{ U(N) }{ \prod_a U(k_a) } \,.
\label{semiclassical_junc_moduli_space_flavor}
\end{equation}
The difference from the pure SYM case is that now, when we decompactify the bulk, the flavor moduli do not disappear.  
The effective theory on the junction becomes a 2d sigma model with the target space \eqref{semiclassical_junc_moduli_space_flavor}.  
This manifold is called a flag variety of type $(k_1, \ldots, k_{N_\text{wall}})$.

From the target space \eqref{semiclassical_junc_moduli_space_flavor}, we can infer the multiplicity of the minimal junctions.  
Since this target space is evidently K\"aler, the Witten index of the sigma model is given by the Euler characteristic of the target space \cite{Witten:1982df}.
Let us assume that the Witten index counts the vacua of the sigma model, and that each vacuum of the effective theory corresponds to a particular junction in the bulk.  
Using the formula
\begin{equation}
	\chi(G / H) = \frac{\left|W_G\right|}{\left|W_H\right|} \,,
\end{equation}
where $W_G = N_G(T) / T$ is the Weyl group of $G$, and the fact that the Weyl group of the unitary group is the permutation group, $W_{U(k)} = S_k$, we obtain the junction multiplicity as the multinomial coefficient
\begin{equation}
	\nu^{\text{junc, } F=N}_{\left(k_a\right)} = \chi \left( \mathcal{M}^{\text{junc, } F=N}_{\left(k_a\right)} \right) = \frac{ N! }{ \prod_a k_a! } \,.
\label{minimal_junction_multiplicity}
\end{equation}
We can perform a simple consistency check of the multiplicity formula \eqref{minimal_junction_multiplicity} using a junction manipulation.  
Consider a 4-wall junction.  
According to eq.~\eqref{minimal_junction_multiplicity}, the multiplicity of such a junction is
\begin{equation}
	\nu^{(4-\text{wall})} = \frac{ N! }{ k_1! k_2! k_3! k_4! }
\label{junc_mult_test_1}
\end{equation}
where, of course, $k_1 + k_2 + k_3 + k_4 = N$.
Next, we can pull this 4-wall junction apart into two 3-wall junctions to form a configuration similar to the one in Fig.~\ref{fig:junc_multi}, with four outgoing walls ($k_1$, $k_2$ on the left and $k_3$, $k_4$ on the right), two 3-wall junctions, and a segment of a fifth wall stretched between them.  
Naively, the multiplicity of this configuration is given by the product
\begin{equation}
	\nu^{\text{junc, } F=N}_{\left( k_1 k_2 k_5 \right)} \cdot \nu^{\text{junc, } F=N}_{\left( k_5 k_3 k_4 \right)} \,.
\label{double_counting_junc}
\end{equation}
However, these two 3-wall junctions are not completely independent.  
The particular set of $k_5$ flavor moduli winding clockwise in the $5^\text{th}$ wall is the same near both junctions, so eq.~\eqref{double_counting_junc} effectively double-counts the number of choices for these modes.
The correct multiplicity is then given by
\begin{equation}
	\nu^{(3-\text{wall} \times 3-\text{wall})}
		= \frac{ \nu^{\text{junc, } F=N}_{\left( k_1 k_2 k_5 \right)} \cdot \nu^{\text{junc, } F=N}_{\left( k_5 k_3 k_4 \right)} }{ \nu^\text{wall}_{k_5} } \,.
\label{junc_mult_test_2}
\end{equation}
By using eqs.~\eqref{wall_multiplicity_F=N} and \eqref{minimal_junction_multiplicity}, we can verify that the multiplicity given by eq.~\eqref{junc_mult_test_2} indeed matches the result in eq.~\eqref{junc_mult_test_1}.

The case with $F = N - 1$ flavors is very similar to the $F = N$ case (since all monopoles are still confined), and we will not discuss it here in detail.

\subsubsection{\boldmath{$SU(N)$} SQCD with \boldmath{$0<F<N-1$} flavors}

When the number of quark flavors is less than $N-1$, there are unconfined monopole moduli when the bulk theory is compactified on the cylinder.

Let us again constrain ourselves to the case of the minimal junctions of $N_\text{wall}$ domain walls, see eq.~\eqref{kJa_condition}.
In this case, the set of $F$ flavor moduli is partitioned into $N_\text{wall}$ sets of cardinalities $J_a$ (some of the $J_a$ might vanish), while the set of $N-F$ unconfined monopole moduli is partitioned into $N_\text{wall}$ sets of cardinalities $k_a - J_a$ (again, some of these sets might be empty).
All these sets signify which of the moduli wind clockwise inside each particular wall.

Then, extending the above analysis, the moduli space of the junction (under the circle compactification) becomes\footnote{Our convention is that $U(k)$ for $k=0$ is a trivial group consisting only of the identity element.}
\begin{equation}
	\mathcal{M}^{\text{junc, } F}_{(k_a,J_a)} = \left[ \frac{ U(N-F) }{ \prod_a U(k_a-J_a) } \right]_\text{mon} \times \left[ \frac{ U(F) }{ \prod_a U(J_a) } \right]_\text{fl} \,.
\label{semiclassical_junc_moduli_space_F}
\end{equation}
The second factor on the r.h.s. corresponds to the flavor moduli and produces a 2d sigma model under decompactification.

The first factor in \eqref{semiclassical_junc_moduli_space_F} describes the monopole moduli; since there are no monopoles left when $L\to\infty$, this piece must be modified in the decompactification limit, in the spirit of Sec.~\ref{sec:wall_effective} here.
To comply with the pure SYM result \cite{Gaiotto:2013gwa} (see eq.~\eqref{gaiotto_junc_scft} here), we conjecture that under decompactification this monopole piece flows to a coset CFT.

As a result, we arrive at the following proposal for the effective theory on the junction:
\begin{equation}
	\left[ \frac{ U(N-F)_{N-F} }{ \prod_a U(k_a-J_a)_{N-F} } \right]^\text{2d CFT} \times \left[ \frac{ U(F) }{ \prod_a U(J_a) } \right]^\text{2d sigma model} \,.
\label{multiwall_junc_theory_N_F}
\end{equation}
In the IR, the two factors in eq.~\eqref{multiwall_junc_theory_N_F} are decoupled, as with the domain wall theory in eq.~\eqref{domain_wall_theory}.

We expect that the junction theory \eqref{multiwall_junc_theory_N_F} also possesses an enhanced $\mathcal{N}=(2,0)$ supersymmetry (two supercharges), instead of the naive minimal SUSY (one supercharge) expected for a 1/4 BPS state in an $\mathcal{N}=1$ bulk theory.
The first CFT factor in \eqref{multiwall_junc_theory_N_F} is a Kazama-Suzuki coset (cf. eq.~\eqref{gaiotto_junc_scft}) and automatically possesses enhanced SUSY, as was noted in \cite{Gaiotto:2013gwa}.
The second factor in \eqref{multiwall_junc_theory_N_F} is an NLSM with a K\"ahler target space, which automatically promotes a real supercharge to a complex supercharge.
Thus, a single real supercharge that could be naively expected on a 1/4 BPS junction in an $\mathcal{N}=1$ bulk theory gets promoted to a complex supercharge to form the $\mathcal{N}=(2,0)$ superalgebra\footnote{We note that it cannot be $\mathcal{N}=(1,1)$, because each real supercharge in the latter case, the left- and the right-moving one, would be automatically promoted to a complex supercharge, and the supersymmetry would have been enhanced to $\mathcal{N}=(2,2)$. While this is possible in principle, we believe that it is not plausible. The $\mathcal{N}=(2,0)$ superalgebra might be realized via a heterotic coupling of the junction NLSM in \eqref{multiwall_junc_theory_N_F} to the degrees of freedom on the domain wall, in the spirit similar to \cite{Shifman:2008wv}. Detailed investigation of this scenario is beyond the scope of the current study, which focuses mainly on the domain walls themselves; we leave it for future work.
}.

In our construction, the enhanced SUSY of the small-circle theory \eqref{semiclassical_junc_moduli_space_F} is naturally inherited from the enhanced SUSY on the domain walls; the latter was discussed in \cite{Ritz:2004mp} and Sec.~4.2 of \cite{Bashmakov:2018ghn}, see also Sec.~\ref{sec:walls_F=N} here.
This may help to explain why in pure SYM the junction theory has enhanced SUSY, as was noted in \cite{Gaiotto:2013gwa}, see also \cite{Rocek:2019eve} for a related discussion.

\subsection{Junctions of junctions}

\begin{figure}[h]
    \centering
    \begin{subfigure}[t]{0.49\textwidth}
        \centering
        \includegraphics[width=0.35\textwidth]{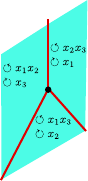}
        \caption{ $SU(3)$ with 3 flavors. The arrows indicate (counter-)clockwise winding of the flavors inside the wall}
        \label{fig:multijunction_one_wall}
    \end{subfigure}%
    ~ 
    \begin{subfigure}[t]{0.49\textwidth}
        \centering
        \includegraphics[width=0.5\textwidth]{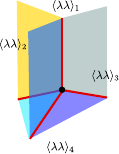}
        \caption{ $SU(4)$ with 4 flavors. Domain walls separate regions with different values of the gluino condensate}
        \label{fig:multijunction_multiple_walls}
    \end{subfigure}%
	\caption{
		Examples of ``junction of junctions'' configurations.
		Domain walls are codimension-1, wall junctions are codimension-2, while the ``junction of junctions'' is codimension-3
	}
\label{fig:multijunction}
\end{figure}

Here we also want to point out the possibility of the existence of a junction of junctions, which would be a codimension-3 object, see Fig.~\ref{fig:multijunction}.
There are various possibilities.

One possibility is a junction of several two-wall junctions within a single plane, see Fig.~\ref{fig:multijunction_one_wall}.
This configuration resembles a polygon, with multi-wall junctions as edges and junctions of junctions as vertices.
Because there are no two-wall junctions in pure SYM, we expect that this object exists only in SQCD with quark flavors.
Note that when the bulk is compactified on a circle, the domain wall is 2d and the two-wall junctions are 1d.
When the junctions can be described as kinks of the wall effective theory (see Sec.~\ref{sec:junc_2_SUN_J1=J2}), a ``junction of junctions'' corresponds to a scattering of kinks within this effective theory.

Another possibility is a junction of several multi-wall junctions, see Fig.~\ref{fig:multijunction_multiple_walls}.
This configuration resembles a polyhedron, with walls as faces, multi-wall junctions as edges, and junctions of junctions as vertices. One can expect this object to exist for any number of quark flavors, including a pure SYM bulk theory with no quark flavors.

Finally, one can also consider mixed-type intersections of the above.

When the bulk is decompactified to $\mathbb{R}^4$, the junction of junctions is a 1d object.
Recall that the usual junction is expected to have $\mathcal{N}=(2,0)$ supersymmetry on its worldsheet (two supercharges).
It might then be possible for the ``junction of junctions'' to support a supersymmetric quantum mechanics on its worldline.

\section{Endpoint phase transition on the wall}
\label{sec:phase_trans_result}

Now we would like to argue that the phase transition between the local and topological multiplicities happens at the endpoint of infinite\footnote{The ``infinite mass'' does not necessarily mean literally $1/m = 0$; rather, it means that the mass $m$ goes to the UV cutoff of the theory. For example, if the Pauli-Villars regularization scheme is used, then ``the endpoint'' is at $m = M_\text{PV}$.} quark mass, $m = \infty$.

\begin{figure}[t]
    \centering
    \includegraphics[width=0.55\textwidth]{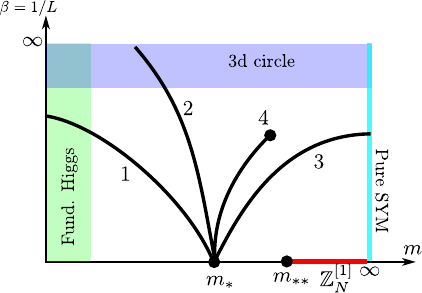}
    \caption{\footnotesize
    	Would-be phase transition lines on the plane $(m,\beta=1/L)$ where $m$ is the quark mass in the bulk SQCD, while $L$ is the size of the compact circle. 
    	Moreover, $m_*$ is the point where the local theory on the domain wall becomes topological (in the text we argue that, in fact, $m_* = \infty$).
    	To the right of the point $m_{**}$, the 1-form center symmetry emerges in the IR in a certain sense; see our discussion here on page \pageref{1-form_discussion} and in Appendix~\ref{sec:1form_quark_circle}.
    	The green-shaded region on the right corresponds to small $m$, when the gauge group is Higgsed by the fundamental quarks and the theory is at weak coupling.
    	The blue-shaded strip on the top corresponds to small circle compactification, when the gauge group is Higgsed by the 3d adjoint scalar and the 3d description is reliable.
    	}
    \label{fig:phase_diagram}
\end{figure}

\subsection{Would-be phase diagram}
\label{sec:would-be-phase-diagram}

We start by explaining the nature of this phase transition, and then discuss the corresponding phase diagram.

\subsubsection{Presence of a phase transition}

Consider SQCD with gauge group $SU(N)$ and $N$ flavors of quarks with small masses $m \ll \Lambda$.
As we saw above, flavor symmetry is spontaneously broken on the BPS domain wall.
If we IR-regularize the corresponding Goldstone modes (either by putting the theory on $\mathbb{R}^3 \times \mathbb{S}^1$ with a finite compactification scale $L$, or by introducing different masses for different quark flavors), then the wall multiplicity {\footnotesize{$\begin{pmatrix} N \\ k \end{pmatrix}$}} following from counting BPS trajectories becomes well-defined, and two-wall junctions are also possible.

However, we stress that at small $m$, the flavor symmetry is spontaneously broken on the wall irrespective of whether we IR-regularize the wall theory or not.

Now the question is what happens as we adiabatically take $m$ to be large and eventually send it to infinity.
At some point, we must recover pure SYM, where the domain wall theory is a TQFT.
When the bulk is just $\mathbb{R}^4$ and the domain wall world-volume is $\mathbb{R}^3$, this TQFT has a single gapped ground state, which corresponds to a single wall.
When the bulk is $\mathbb{R}^3 \times \mathbb{S}^1$ with a finite compactification scale $L$, the ground state degeneracy of the TQFT reproduces the formula {\footnotesize{$\begin{pmatrix} N \\ k \end{pmatrix}$}}, but contrary to the SQCD case, there is no flavor symmetry and, hence, no spontaneous breaking of the 0-form flavor symmetry.
Either way, we expect that as we increase the quark mass $m$, at some point we encounter a phase transition on the wall.

We associate this phase transition with spontaneous breaking of the 0-form flavor symmetry on the wall worldvolume: it is broken in SQCD at small $m$, while at large $m$ we decouple quarks and recover pure SYM where there is no flavor symmetry breaking.
We denote by $m_*$ the corresponding critical value of quark mass at $L = \infty$ (i.e. on $\mathbb{R}^4$), see Fig.~\ref{fig:phase_diagram} which will be discussed in detail below.

\begin{figure}[t]
    \centering
    \begin{subfigure}[t]{0.49\textwidth}
        \centering
        \includegraphics[width=0.35\textwidth]{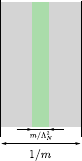}
        \caption{Domain wall at small $m$}
        \label{fig:quark_wall_weak_coupling}
    \end{subfigure}%
    ~ 
    \begin{subfigure}[t]{0.49\textwidth}
        \centering
        \includegraphics[width=0.35\textwidth]{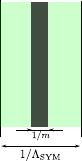}
        \caption{Domain wall at large $m$}
        \label{fig:quark_wall_strong_coupling}
    \end{subfigure}%
	\caption{
		Domain wall structure.
		\subref{fig:quark_wall_weak_coupling} At small $m$, gluons are heavy, and the wall thickness is determined by the quark mass (shown in grey); inside, there is a core of heavy gluons (represented here by a green stripe) with mass determined by the squark VEVs, see eq.~\eqref{meson_field_def} and eq.~\eqref{quark_vacua} ($\Lambda_N$ is the scale of SQCD with $N$ quark flavors).
		\subref{fig:quark_wall_strong_coupling} At large $m$, there is a thin quark core (grey stripe) and a gluon cloud around it (shown in green).
	}
\label{fig:quark_wall_struct}
\end{figure}

\subsubsection{Location of the phase transition point}

Now that we have established the existence of the phase transition, what is the precise location of the phase transition point $m_*$?
One possibility is that this phase transition happens right at the endpoint, $m_* = \infty$; we will argue here that this is exactly what happens in $\mathcal{N}=1$ SQCD.
Another possibility would be that the phase transition happens at some finite critical mass $0 < m_* < \infty$; we want to argue that this possibility is in fact excluded.
In principle, there is a third possibility with multiple phase transition points, but to the best of our knowledge, no existing knowledge about domain wall dynamics in $\mathcal{N}=1$ SQCD supports this scenario.

We are going to build our proof by contradiction.
Suppose for now that the phase transition does happen at a finite point $0 < m_* < \infty$.
Then let us compactify the bulk on $\mathbb{R}^3 \times \mathbb{S}^1$ with a circle of length $L$, and consider the plane of parameters $(m, \beta)$ where $m$ is the quark mass and $\beta = 1/L$ (we remind the reader that our compactification is not thermal but supersymmetric).
The phase transition point (associated with a 0-form symmetry) is then continued in the $(m, \beta)$ plane as a phase transition line.
Since there is only one transition point, the phase transition line cannot come back to the $m$ axis.
This leaves a few other possibilities, see Fig.~\ref{fig:phase_diagram}.

\begin{figure}[t]
    \centering
    \includegraphics[width=0.5\textwidth]{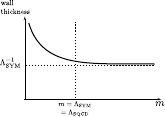}
    \caption{
    	Domain wall thickness as a function of the quark mass in the bulk (schematic).
    	Note that at the point $m = \Lambda_\text{SQCD}$ we also have $\Lambda_\text{SQCD} = \Lambda_\text{SYM}$, cf. eq.~\eqref{scales_matching_F=N}
    }
    \label{fig:plot_wall_thickness}
\end{figure}%

At small $m$, the bulk theory is under control for any $\beta$ (green-shaded region on the left of Fig.~\ref{fig:phase_diagram}), and we do not see a phase transition when going from a large to a small circle.
This excludes curve 1 from Fig.~\ref{fig:phase_diagram}.
On the other hand, for small circle compactification, the theory stays weakly coupled for any value of $m$ (blue-shaded strip on the top of Fig.~\ref{fig:phase_diagram}).
Again, we do not see a phase transition in this region, which excludes curve 2.
Strictly at $m = \infty$, we are in pure SYM, which is known to enjoy \"Unsal's adiabatic continuity.
Moreover, it would be strange if the flavor symmetry remained in this limit (even for small circle radius), since at $m = \infty$ the quarks decouple and there is no faithful flavor symmetry left.
Thus, curve 3 is also excluded.

This leaves us with curve 4.
In this case, we could start near the lower left corner of the diagram in Fig.~\ref{fig:phase_diagram} (small $m$, large $L$), go up and then right through the shaded weak coupling regions until we reach pure SYM.
In doing so, we would not encounter a phase transition, and hence we would end up with a spontaneously broken flavor symmetry in pure SYM, which is again nonsensical.
Curve 4 is then also excluded.

The argument above leaves us with only one possibility.
The phase transition point $m_*$ must be the endpoint $m = \infty$.
The phase transition line in the $(m, \beta)$ plane goes\footnote{See also Sec.~7.3 of \cite{Unsal:2021xay} for a related discussion.} directly up along the ``Pure SYM'' in Fig.~\ref{fig:phase_diagram}.

The argument above is also applicable to SQCD with fewer flavors, $F < N$.
In this case, there is a TQFT factor always present in the domain wall effective theory eq.~\eqref{domain_wall_theory}.
The argument above then applies to the ``local'' part of the wall multiplicity.

\subsubsection{Physical scenario}
\label{sec:would-be-phase-diagram:scenario}

Let us also ask ourselves what happens with the {\em low-energy theory} on the given wall {\em per se} as we move from small to large values of $m$. 
As discussed above, at least for small $m$, the wall has, in addition to the usual translational moduli, flavor moduli (or quasi-moduli if we IR regularize, cf. footnote \ref{ft:witten_IR} on p.~\pageref{ft:witten_IR}).
The corresponding flavor degrees of freedom form an NLSM on the wall worldvolume.

One might expect that at large $m$, when the matter fields are heavy in the bulk, the (quasi-)massless modes comprising the sigma model on the wall would become heavy. 
However, the following example illustrates that this is not necessarily the case; rather, the flavor modes become free and decoupled.

Consider $SU(N)$ SQCD on $\mathbb{R}^4$ with $F = N$ flavors of quarks, all having identical masses $m$.
Take a minimal $k = 1$ domain wall.
For any finite $m$, such a wall hosts a $\mathbb{CP}(N-1)$ model localized on its worldvolume $\mathbb{R}^3$.
The degrees of freedom in this 3d effective theory are the Goldstones of the broken flavor symmetry, and this 3d theory remains gapless.
On the other hand, in the limit $m \to \infty$, the bulk SQCD reduces to SYM, and we know that the domain wall theory is gapped --- it is described by a Yang-Mills plus Chern-Simons theory.
How can we reconcile the gapped and gapless nature of the effective theory?

We propose the following resolution of this puzzle.
As the bulk quark mass $m$ tends to infinity, the $\mathbb{CP}(N-1)$ NLSM coupling of the wall theory approaches zero, and the massless moduli associated with the quark wall begin to behave as free fields.
These moduli are confined to a quark core of thickness $1/m$ (illustrated as the dark grey slice in Fig.~\ref{fig:quark_wall_strong_coupling} and Fig.~\ref{fig:thick_junction}).
To probe these degrees of freedom, one needs to go to energies of the order of the heavy quark mass $m$.

Therefore, for large $m$, the flavor degrees of freedom are decoupled from the gauge (gluon) degrees of freedom at low energies; the gluon part of the wall is represented by a green cloud in Fig.~\ref{fig:quark_wall_strong_coupling} and Fig.~\ref{fig:thick_junction}.
The coupling between the quark and gluon sectors on the wall turns on only at the high scale of order $m$.
In the limit $m \to \infty$, the quark core becomes infinitesimally thin, and the NLSM becomes completely free and decoupled.
What we are left with is the effective theory on the wall governed solely by gluonic degrees of freedom, leading to the Chern-Simons description.

This phenomenon can also be interpreted from the bulk perspective.
At finite quark masses, the bulk SQCD exhibits a global $U(N)$ flavor symmetry.
The massless modes on the wall correspond to the Goldstone bosons associated with this symmetry.
When $m \to \infty$, the bulk quarks decouple, and consequently, the $U(N)$ flavor transformations become trivial (i.e., non-faithful).

\vspace{10pt}

This analysis suggests the following scenario for the domain walls (at any number of flavors in the interval $0 < F \leqslant N$).
When massive quarks are present in the theory, the corresponding sigma model for the flavor moduli on the wall (the ``NLSM'' part of the theory in eq.~\eqref{domain_wall_theory}) is present for all values of the quark mass $m$ and disappears only at the \textit{endpoint of infinite quark mass}.
This sigma model is a manifestation of the local distinguishability of different walls.

Furthermore, gauge degrees of freedom, ``frozen'' at small $m$, become light at large quark masses.
At $m > m_{**}$, the gauge degrees of freedom become noticeable at low energies and constitute the CS theory on the wall.
This CS theory coexists with the sigma model for the flavor moduli when the quark mass $m$ is larger than $m_{**}$ but finite.
At the endpoint of infinite quark mass, the flavor sigma model disappears, and the CS theory is all that remains.
Note that, while the sigma model for the flavor moduli on the wall has enhanced $\mathcal{N}=2$ supersymmetry \cite{Ritz:2004mp}, the effective theory for the gauge degrees of freedom on the wall at $m > m_{**}$ is expected to be given by $\mathcal{N}=2$ Yang-Mills with an $\mathcal{N}=1$ Chern-Simons term coupled to an adjoint chiral multiplet \cite{Acharya:2001dz}.

\subsection{Tensions of walls and junctions at large \boldmath{$m$}}
\label{sec:junc_tension_various_m}

\begin{figure}[t]
    \centering
    \includegraphics[width=0.4\textwidth]{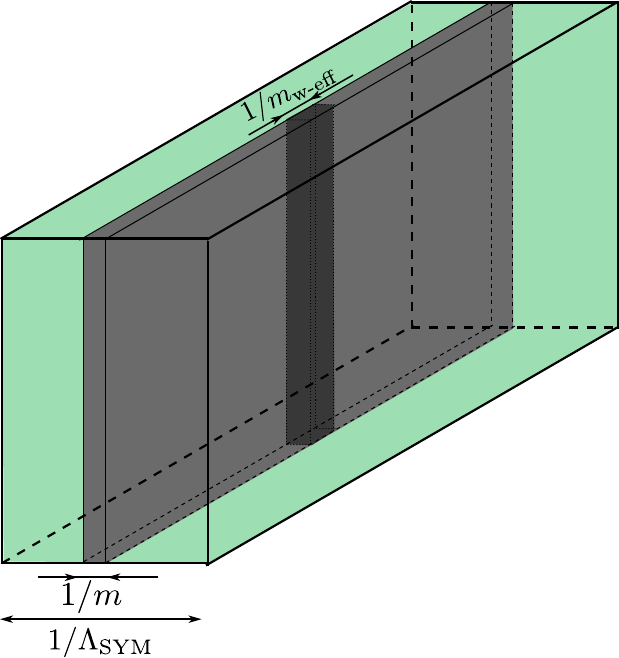}
    \caption{
    	Schematic structure of a two-wall junction for heavy quarks, $m \gg \Lambda_\text{SYM}$.
        The quark core (shaded) has a different structure inside the two walls, while the gluon clouds (green) are nearly identical.
        The transition region of the junction is localized to the core (darkened region).
        The width of the junction is determined by the mass scale of excitations on the wall worldvolume, denoted here by $m_\text{w-eff}$.
    }
    \label{fig:thick_junction}
\end{figure}

From the discussion above, it follows that as long as the quark mass $m$ is finite, an external observer can perform local experiments that distinguish between different walls.
If we IR-regularize the wall worldvolume theory, it is possible to have two-wall junctions (see Sec.~\ref{sec:junc_2} above).
Let us now discuss the fate of such junctions in the limit $m \to \infty$.

The wall structure and thickness at different values of $m$ are shown in Fig.~\ref{fig:quark_wall_struct} and Fig.~\ref{fig:plot_wall_thickness}, and for large $m$ in Fig.~\ref{fig:thick_junction}.
The width of the junction is controlled by the mass scale of excitations on the wall, denoted by $m_\text{w-eff}$ in Fig.~\ref{fig:thick_junction}.

We now argue that in the limit $m \to \infty$, the junction becomes tensionless.
On one hand, this is hinted at by the small-$m$ tension formula \eqref{junc_tension_weak_1} in the example with $N=2$ colors and $F=2$ flavors.
When decoupling some of the quarks, one should appropriately change the strong coupling scale $\Lambda$ according to \eqref{Lambda_when_integrate_out_m}.
Let us see what we obtain for the junction tension \eqref{junc_tension_weak_1} when we try to apply that formula at large masses for the two flavors, $m_1$ and $m_2$.
Using eq.~\eqref{Lambda_when_integrate_out_m}, we re-express the tension in units of $\Lambda_{F=0} \equiv \Lambda_\text{SYM}$:
\begin{equation}
	T_\text{junc} 
		= \pi\left|\frac{\left|m_1\right|-\left|m_2\right|}{\sqrt{\left|m_1 m_2\right|}}\right| \Lambda_{F=2}^2
		= \pi\left|\frac{\left|m_1\right|-\left|m_2\right|}{\left|m_1 m_2\right|}\right| \Lambda_{F=0}^3  \,.
\label{junc_tension_weak_2}
\end{equation}
If we further take $m_2 \gg m_1$, then
\begin{equation}
	T_\text{junc}
		\xrightarrow{m_2 \to \infty}
		\pi \sqrt{ \frac{\Lambda_{F=1}^5}{|m_1|} } 
		= \frac{\pi}{|m_1|} \Lambda_{F=0}^3 \,.
\label{junc_tension_weak_3}
\end{equation}
When decoupling quarks, we must hold $\Lambda_{F=0}$ fixed.
As evident from eqs.~\eqref{junc_tension_weak_2} and \eqref{junc_tension_weak_3}, in either case the junction tension vanishes in the large quark mass limit.

Of course, this argument is only a hint, since the junction tension formula \eqref{junc_tension_weak_1} is valid only at weak coupling with small quark masses.
Nevertheless, there is further evidence that the conclusion about the vanishing of the junction tension is indeed correct.

As discussed in Sec.~\ref{sec:would-be-phase-diagram:scenario}, we expect that the NLSM describing the flavor degrees of freedom on the wall worldvolume becomes free in the limit of infinitely heavy quarks.
The NLSM coupling vanishes in this limit.
It is reasonable to expect the same behavior if we compactify the bulk SQCD on a circle of finite size $L$, with the domain wall wrapping the compact dimension.
This follows from the fact that the NLSM couplings on the 3d and 2d wall worldvolumes are proportional: $g_{2d} \sim L \cdot g_{3d}$.
For fixed $L$, the vanishing of $g_{3d}$ implies the vanishing of $g_{2d}$.
Then, as follows from Sec.~\ref{sec:two_wall_junc_N=2_F=2} (see eq.~\eqref{2d_kink_mass_strong} and eq.~\eqref{kink_junc_mass_relation}), the junction tension also vanishes in the large quark mass limit.

\vspace{10pt}

This conclusion also follows naturally if we consider how much the quark core contributes to the wall tension.
Recall that the BPS wall tension is determined by the gluino condensate, via the central charge.
On one hand, in pure SYM the wall tension is
\begin{equation}
	T_\text{wall} \sim \Lambda_\text{SYM}^3 \,.
\label{T_wall_1}
\end{equation}
On the other hand, in SQCD with $F$ flavors, the quark VEVs are related to the gluino condensate by the exact Konishi relation (see eq.~\eqref{quark_vacua}).
This gives
\begin{equation}
	T_\text{wall} \sim \Lambda_F^3 \left( \frac{m}{\Lambda_F} \right)^{F/N} \,,
\label{T_wall_2}
\end{equation}
where $m$ is the quark mass scale.

When we decouple the quarks and take $m \to \infty$, we must hold $\Lambda_{F=0} \equiv \Lambda_\text{SYM}$ fixed.
Changing the scale in \eqref{T_wall_2} accordingly using \eqref{Lambda_when_integrate_out_m}, we recover the pure SYM result \eqref{T_wall_1}.
In fact, the coefficients match as well.

This teaches us that as $m$ increases, the wall tension becomes dominated by the gluonic contribution.
The energy (per unit area) contained in the quark core diminishes.
Thus, we expect that all energy scales associated with the effective theory on the quark core decrease --- including the junction tension.

\subsection{Confinement vs. heavy quarks}

An explanation is needed with regard to the point $m_{**}$ on the $(m, \beta)$ plane of Fig.~\ref{fig:phase_diagram}.
\label{1-form_discussion}
It was suggested that when the bulk is $\mathbb{R}^4$, from topological persistence one may expect that the $\mathbb{Z}_N^{[1]}$ center symmetry, explicitly broken by heavy quarks, emerges in the IR after the quark mass becomes higher than some critical value $m_{**}$ (see Fig.~\ref{fig:phase_diagram}), which would reinstate confinement in the IR. 
If we take the area law of the Wilson line as the confinement criterion, it is easy to see that the reinstatement of confinement below $m_{**}$ cannot occur. 
Indeed, the string between the probe fundamental quarks can and does break through heavy quark pair creation. 

\begin{figure}[h]
    \centering
    \includegraphics[width=0.6\textwidth]{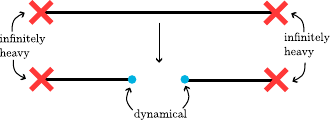}
    \caption{
    	Confining string stretched between two probe sources (red ``X'') can break by creating a pair of dynamical quark and antiquark (blue dots)
    }
    \label{fig:string_breaking}
\end{figure}%

However, one should be careful, because there are two dimensionful parameters, quark mass and the size of the Wilson loop.
Indeed, consider a process shown in Fig.~\ref{fig:string_breaking}: a string stretched between two probe (non-dynamical) sources that can break with the creation of dynamical quarks.
Let $\ell_W$ be the length of such a string.
String breaking via pair production can occur if the string is long enough,
$\ell_W \gg m/\Lambda^2$, as such a string stores enough energy to produce a $Q\bar Q$ pair of heavy quarks of mass $m$. 
Moreover, since the $Q\bar Q$ pair production is a tunneling effect, the waiting time for the pair production is 
\begin{equation}
    t_W \sim \Lambda^{-1}  \exp \left(\frac{m^2}{\Lambda^2}\right) \,.
\end{equation}
The area law for the Wilson loop is recovered provided the area
\begin{equation}
    A \ll m\Lambda^{-3}\exp \left(\frac{m^2}{\Lambda^2}\right) \,.
\end{equation}
In the opposite limit, string breaking leads to the perimeter law (for a more detailed analysis of string breaking see e.g. \cite{Monin:2008mp,Shifman:2002yi}).
The question of whether the gauge theory with heavy quarks confines or not depends on whether one first fixes a Wilson loop area and then sends the quark mass to infinity, or first fixes the quark mass $m$ and then takes an exponentially large area $A$, namely $A\gg \exp(m^2/\Lambda^2)$.

\section{Conclusions}
\label{sec:concl}

In this paper, we studied domain walls in $\mathcal{N}=1$ $SU(N)$ SQCD with $0 \leqslant F \leqslant N$ quark flavors under circle compactification.
We then argued what we can learn for the decompactified theory on $\mathbb{R}^4$.

In particular, we were interested in the domain wall multiplicity.
The multiplicity can be ``local'' (i.e. due to local physics and NLSM on the wall worldvolume) and ``topological'' (i.e. due to global effects and a TQFT on the wall worldvolume).
The different contributions are seen in the domain wall multiplicity formula in eq.~\eqref{multiplicity_cylinder_full},
\begin{equation*}
    \nu_{N,k}^\text{walls} \Big|_{\mathbb{R}^3 \times \mathbb{S}^1} = \sum_{J=\max(0,F + k - N)}^{ \min(k,F) } 
        \begin{pmatrix} N-F \\ k-J \end{pmatrix}
        \begin{pmatrix} F \\ J \end{pmatrix}
        = \begin{pmatrix} N \\ k \end{pmatrix} \,.
\end{equation*}
This formula was known before (see e.g. \cite{Bashmakov:2018ghn}), but here we rederive it by a different method.
Here, $k$ and $J$ are parameters characterizing the wall ($k$ means that the wall interpolates from some chiral vacuum $n_0$ to $n_0+k$, while $J$ is a parameter internal to the wall).

At small quark mass $m$ (at least part of the) domain wall multiplicity is local, but when we decouple quarks, we recover SYM, where all the multiplicity is topological.
One of the main results is the statement that the corresponding phase transition occurs right at the endpoint of infinite quark mass, see Sec.~\ref{sec:phase_trans_result}.

As a byproduct, a few further results are derived in the present work.
We obtain the domain wall theory in eq.~\eqref{domain_wall_theory} with the TQFT factor \eqref{TQFT_FkJ} with manifestly unbroken supersymmetry.
We also consider junctions of multiple domain walls and propose the effective theory living on the junction worldsheet in eq.~\eqref{multiwall_junc_theory_N_F},
\begin{equation*}
	\left[ \frac{ U(N-F)_{N-F} }{ \prod_a U(k_a-J_a)_{N-F} } \right]^\text{2d CFT} \times \left[ \frac{ U(F) }{ \prod_a U(J_a) } \right]^\text{2d sigma model} \,.
\end{equation*}
This extends the proposal of \cite{Gaiotto:2013gwa} from SYM to the theory with quarks.

\section*{Acknowledgments}

We thank Aleksey Cherman, Sergei Dubovsky, Thomas Dumitrescu, Jaume Gomis, Zohar Komargodski, Sahand Seifnashri, and Yue Zhao for helpful discussions. This work is supported in part by U.S. Department of Energy Grant No. de-sc0011842, and by the Simons Foundation award number 994302 (S.C.).

\appendix

\section{Instanton localization}
\label{sec:inst_loc}

This Appendix is based on Ref.~\cite{Shifman:1999mv}, Secs. 3.5.4-3.5.6, which, in turn, represents a somewhat extended version of the original derivation in 
\cite{Novikov:1985ic}.
We demonstrate that in Higgsed SQCD with one flavor (two subflavors) the instanton is localized --- in other words, we will derive Eqs.~\eqref{nine} and \eqref{inst_measure_integrated}.

SQCD with one flavor classically 
has a one-dimensional flat direction, the vacuum valley. A generic 
point from the bottom of the valley is characterized by
$$
(\phi^\alpha_f)_{\rm vac} = v\delta^\alpha_f
$$
where $v$ is an arbitrary (complex) parameter, the vacuum 
expectation value of the squark fields.\footnote{Later on, for simplicity, we will choose $v$ to be real.} Here $\alpha$ is the color 
index while $f$ is the subflavor index; $\alpha , f = 1,2$. The color 
and flavor indices get entangled, even in the topologically trivial 
sector, although in a rather trivial manner.

In the (anti)instanton background the solution takes the form
\begin{equation}
	\phi_{\dot{f}}^\alpha = v
	\frac{x^{\alpha}_{\dot{f}}}{\sqrt{x^2+\rho^2}}\,.
\label{solution_phi_inst_bkgr}
\end{equation}
Here, $x^{\alpha}_{\dot{f}} = x_\mu \varepsilon^{\alpha\beta} (\sigma^\mu)_{\beta \dot{f}}$ (our notation follows the book \cite{Shifman:2012zz}, see Sec.~10.2.1 there).

Expression \eqref{solution_phi_inst_bkgr} is approximate; it is valid only at small $\rho^2\ll (gv)^{-2}$. 
It becomes {\em exact} in the limit $\rho^2\to 0$. As we will see below, only this limit is relevant because of localization.

To generate the full superfamily, with all collective coordinates 
switched 
on, we need to apply the generalized shift operator, 
\begin{equation}
{\cal V}(x_0 ,\theta_0, \bar\beta,  \bar\zeta,\bar\omega,\rho  )
=e^{iPx_0}e^{-iQ\theta_0}e^{-i\bar{S}\bar\beta}
e^{-i\bar Q\bar \zeta}e^{i\bar M \bar\omega}e^{iD\ln\rho}\,.
\label{string1}
\end{equation}
Here 
$P,\,M,\,D$ are bosonic operators of shift, $SU(2)$ rotations and dilatation, $Q$ and $\bar Q$ are supercharges, while $\bar S$ is one of two superconformal operators.
The Grassmann moduli $\bar\beta_{\dot\alpha}$ and $\bar \zeta^{\dot\alpha}$ conjugated to $\bar{S}^{\dot\alpha}$ and 
$\bar Q_{\dot\alpha}$ are introduced. 
Supertransformations of all moduli are well-known 
\cite{Novikov:1985ic,Shifman:1999mv}
Of most interest to us are the transformations of $\bar \zeta$,
\begin{equation}
\delta \bar\zeta_{\dot\alpha}=
\bar\varepsilon_{\dot\alpha} -4 i \bar\beta_{\dot\alpha}\,
(\bar\zeta\bar\varepsilon )\, .
\label{thetatr}
\end{equation}
To linear order in $\varepsilon,\,\bar\varepsilon$, the SUSY transformation of
$\bar\zeta$ is the same as that of $\bar\theta$, but the former contains nonlinear
terms. The combination which transforms linearly, exactly as $\bar \theta$, is
\begin{equation}
(\bar\theta_0)^{\dot\alpha}=\bar\zeta^{\dot\alpha}[1-4i (\bar \beta
\bar\zeta)]\,,\qquad 
\delta(\bar\theta_0)^{\dot\alpha}=\bar\varepsilon^{\dot\alpha}\,. 
\label{thetazero}
\end{equation}
The variable $\bar\theta_0$ joins the set $\{x_0,\theta_0\}$ describing the
superinstanton center.\footnote{In the original work 
\cite{Novikov:1985ic}
the notation
$\bar\theta_0$ was used for what is called $\bar\zeta$ here. The
combination~(\ref{thetazero}) was not introduced.  }

If we commute the operators $\bar S$ and $\bar Q$ in Eq. (\ref{string1}) to arrive at a different ordering, the collective coordinate $\bar\theta_0$
in the operator ${\cal V}$, we will be able to easily define combinations of moduli which are invariant under supersymmetry transformations. 
We will denote these combinations by subscript ``inv'', namely,
\begin{equation}
{\cal V}(x_0 ,\theta_0, \bar\theta_0,\bar\beta_{\rm inv},\bar\omega_{\rm
inv},\rho_{\rm inv}
 ) =e^{iPx_0}e^{-iQ\theta_0}e^{-i\bar Q\bar
\theta_0}e^{-i\bar{S}\bar\beta_{\rm inv}} e^{i\bar M \bar\omega_{\rm
inv}}e^{iD\ln\!\rho_{\rm inv}}\;.
\label{invcoor}
\end{equation}
and
\begin{eqnarray}
&&\bar\beta_{\rm inv}=\bar\beta\,[1 + 4 i \,(\bar\beta
\bar\zeta )]=\frac{\bar\beta}{1 - 4 i \,(\bar\beta
\bar\theta_0 )}\,,\nonumber\\[0.2cm]
&&\rho^2_{\rm inv}= \rho^2 \,[1 + 4 i \,(\bar\beta 
\bar\zeta )]=\frac{\rho^2}{1 - 4 i \,(\bar\beta
\bar\theta_0 )}\,,\nonumber\\[0.2cm]
&&\left[\Omega_{\rm inv}\right]^{\dot\alpha}_{\dot\beta}\equiv \left[ e^{-i
\bar\omega_{\rm inv}}\right]^{\dot\alpha}_{\dot\beta}=\exp\left\{-4i
\left[\bar\zeta^{\dot\alpha}\,
\bar\beta_{\dot\gamma}+\frac 1 2 \delta^{\dot\alpha}_{\dot\gamma}\,
(\bar\zeta \bar\beta) \right]\right\}\Omega^{\dot\gamma}_{\dot\beta}
\,.
\label{rhoinv}
\end{eqnarray}
Here $\Omega$ is a matrix of SU(2) rotations.
With the ordering~(\ref{invcoor}) it is clear that $x_0$,
$\theta_0$, and $\bar\theta_0$ transform as $x_L$, $\theta$ and $\bar\theta$,
respectively, while the other moduli are the invariants of the supersymmetry
transformations.

Next, we pass to the invariants constructed from the coordinates
in the superspace and the moduli. 
Since the set $\{x_0,\theta_0,\bar\theta_0\}$ transforms the same way as
the superspace coordinate $\{x_L,\theta,\bar\theta\}$, such invariants are the
\begin{equation}
z_{\alpha\dot\alpha}=  (x_L-x_0)_{\alpha\dot\alpha} + 
4i\,(\theta-\theta_0)_\alpha\, (\bar\theta_0)_{\dot\alpha}\,,\quad
\theta-\theta_0\,,\quad
\bar \theta-\bar \theta_0
\,.
\label{inv2}
\end{equation}
All other invariants can be obtained by combining the sets of Eqs.~(\ref{rhoinv}) and (\ref{inv2}). For instance, the invariant combination $\tilde{x}^2/\rho^2$
where
\begin{equation}
\tilde{x}_{\alpha\dot\alpha} = (x_L - x_0 )_{\alpha\dot\alpha} 
+ 4 i \,\tilde\theta_\alpha\bar\zeta_{\dot\alpha}\, ,
\label{tildex}
\end{equation}
which frequently appears in applications, can be rewritten in the following form,
\begin{equation}
\frac{\tilde{x}^2}{\rho^2}=\frac{z^2}{\rho_{\rm inv}^2}
\,.
\end{equation}

Finally, using the above definitions, one can write the instanton-induced contribution to the ADS superpotential as follows,
\begin{eqnarray}
{\rm d}\mu_{\rm one-fl}&=& \frac{1}{2^{11}\pi^4 v^2 }\, M_{\rm PV}^5 \,\int\,\left( 
\frac{8\pi^2}{g^2}
\right)^2 \!\exp\left( -\frac{8\pi^2}{g^2}-4\pi^2|v|^2\rho^2_{\rm inv}
\right)\nonumber\\[2mm] 
&\times& {d}^4 x_0\, {d}^2\theta_0  \frac{ {\rm d}\rho^2}{\rho^2}\,
 \,
 {d}^2\bar\beta_{\rm inv}\,  {d}^2\bar\theta_0 \, .
\label{imesqcd}
\end{eqnarray}
It is obvious that the measure above is explicitly invariant under the supersymmetric transformations. Indeed, ${\rm d}\rho^2/\rho^2$ under the sign of integral is identical to ${\rm d}\rho^2_{\rm inv}/\rho^2_{\rm inv}$. Warning: we hasten to add that this ``obvious statement'' is incorrect at
the singular point $\rho^2=0$ to be discussed momentarily.

The effective superpotential is obtained by integrating over
$\rho^2,\bar{\beta}$, and $\bar{\theta}_{0}$. Let us naively replace $\int d\rho^2/\rho^2$ by $\int d\rho^2_{\rm inv}/\rho^2_{\rm inv}$ for a short while. Moreover, we 
will {\em not} perform the integrals over the “linear” moduli $d^4 x_0$ and $d^2\theta_0$ until the very end. Furthermore, we can replace the exponent in \eqref{imesqcd} 
by an arbitrary function $f(\rho^2_{\rm inv})$ provided it is smooth, falls off at infinity sufficiently fast, and is non-singular at zero.  

Since $\bar\beta_{\rm inv}$ and $\bar\theta_0$
enter the integrand of the measure only through $\rho_{\mathrm{inv}}^{2}$, at first
sight the integral might seem vanishing; indeed, changing the
variable $\rho^{2}$ to $\rho_{\mathrm{inv}}^{2}$ makes the integrand
independent of $\bar{\beta}$ and $\bar{\theta}_{0}$. As was mentioned below Eq.~\eqref{imesqcd}, naively, the integral
\begin{equation}
	\int\frac{d\rho^{2}_{\rm inv}}{\rho^{2}_{\rm inv}}\,d^{2}\bar{\beta}\,d^{2}\bar{\theta}_{0} f\left(\rho^2_{\rm inv}\right)=0
\end{equation}
vanishes because of the integration over the Grassmann moduli $d^{2}\bar{\beta}\,d^{2}\bar{\theta}_{0}$ for fixed $\rho^{2}_{\rm inv}$.

In fact, this integral 
does {\em not} vanish by virtue of the $\rho^2=0$ singularity in the measure of integration over $\rho^2$. To resolve this singularity, let us
return back from the variable $\rho^2_{\rm inv}$ to the variable $\rho^2$ and integrate over the Grassmann moduli first. 
To this end, let us note the following identity
\begin{equation}
	\int\frac{d\rho^{2}}{\rho^{2}}\,d^{2}\bar{\beta}\,d^{2}\bar{\theta}_{0} f\left(\rho^{2}(1+4i\bar{\beta}\bar{\theta}_{0})\right)
	= \int\frac{d\rho^{2}}{\rho^{2}}16\rho^{4} f''(\rho^{2})
	= 16\,f(\rho^{2}=0)
\label{nine}
\end{equation}
where two primes stand for the second derivative of $f$ over $\rho^2$.
The integration over $\rho^{2}$ was performed by integrating by
parts twice under the assumption that
$f(\rho^{2}\rightarrow\infty)=0$. It can be seen that the result
depends only on the zero-size instantons. In other words,
\begin{equation}
	\frac{d\rho^{2}}{\rho^{2}}\,d^{2}\bar{\beta}d^{2}\bar{\theta}_{0} f(\rho_{\mathrm{inv}}^{2})
	= 16\,d\rho_{\mathrm{inv}}^{2}\delta(\rho_{\mathrm{inv}}^{2}) f(\rho_{\mathrm{inv}}^{2}).
\label{ten}
\end{equation}
Equation (\ref{ten}) confirms that the one-instanton contribution is {\em saturated} at $\rho_{\mathrm{inv}}^{2}=0$, and we recover Eq.~\eqref{inst_measure_integrated}.
This implies, in turn,
that the second term in Eq.~\eqref{six} --- the instanton-induced superpotential --- can be added to the effective Lagrangian from the very beginning, without assuming that $m_0$ is small 
and before integrating out the Higgsed gauge fields. 
Eq.~\eqref{six} can be used for all values of $m_0$ except two singular points: $m_0=0$ and $m_0=\infty$, to be punctured off from the complex plane of $m_0$. Instantons are localized in Higgsed SQCD, at least for special choices of the matter sector. 

Of course, the kinetic terms cannot be established on this basis. For small $m_0$, at weak coupling,
the kinetic term should be close to canonical because the corrections are small.
At large $m_0$, when the matter field becomes very heavy, the gluon sector is soft and cannot generate strong corrections either.

\section{Running of the gauge coupling}
\label{sec:betafunc}


The gauge coupling running in $\mathcal{N}=1$ gauge theory at one loop is given by:
\begin{equation}
	E \frac{d}{d E} \frac{8 \pi^2}{g^2} = 3 C(\operatorname{adj})-C(R) \equiv b \,.
\label{RG_one_loop}
\end{equation}
For the gauge group $SU(N)$, we have $C(\operatorname{adj}) = N$ and $C(\text{fund}) = \frac{1}{2}$ (remember that each quark flavor consists of a pair of chiral and anti-chiral multiplets).
%

In the Pauli-Villars renormalization scheme, \eqref{RG_one_loop} suggests the introduction of a one-loop RG-invariant scale $\hat{\Lambda}^{b_0} = M_\text{uv}^{b_0} e^{2 \pi i \tau(M_\text{uv})}$, where $\tau$ is the complexified gauge coupling,
\begin{equation}
	\tau = \frac{4 \pi i}{ g^2 } + \frac{\theta}{2\pi} \,.
\label{tau_def}
\end{equation}
However, the exact Novikov-Shifman-Vainshtein-Zakharov beta function \cite{Novikov:1983uc} motivates an improved expression for the RG scale,
\begin{equation}
	\Lambda^{b} = M_\text{uv}^{b} \frac{1}{g^{2N}(M_\text{uv})} e^{ 2 \pi i \tau (M_\text{uv}) } \,.
\label{Lambda_def}
\end{equation}
In pure $\mathcal{N}=1$ SYM, this quantity is exactly invariant under RG flow (if the gauge group is $SU(N)$, we have $b=3N$).
With this convention, \eqref{Lambda_def} the chiral condensate takes a simple form:
\begin{equation}
	\expval{ \frac{\tr \lambda^2}{ 16 \pi^2 } } = \Lambda^3 \,.
\end{equation}
Note, however, that for gauge groups $SU(N)$, convention \eqref{Lambda_def} is suitable only for fixed-$N$ calculations.
Another definition of $\Lambda$, more common in the QCD community and also used for instanton calculations in $\mathcal{N}=1$ theories (see e.g. eq. (4.17) of \cite{Shifman:1999mv} and eq. (10.381) of \cite{Shifman:2012zz}), for the case of pure $SU(N)$ SYM, is given by
\begin{equation}
	\tilde{\Lambda}^3 = M_\text{uv}^{3} \frac{16 \pi^2}{ 3 N g^2(M_\text{uv}) } e^{ \frac{ 2 \pi i \tau (M_\text{uv}) }{N} } \,.
\end{equation}
Note the $N$-dependence in the pre-exponent.
For the sake of shorter formulas in this paper, we use the convention \eqref{Lambda_def} unless stated otherwise.

On the one-loop level, one can solve the running \eqref{RG_one_loop} as
\begin{equation}
	\tau(E)=\tau(M_\text{uv})-\frac{b}{2 \pi i} \log \frac{E}{M_\text{uv}} \,.
\label{tau_one_loop_running}
\end{equation}
Formula \eqref{tau_one_loop_running} is useful when we want to compare scales in models with different matter content.
Suppose that we have a theory $A$, and we want to decouple some flavors of matter with large mass $m$ so that below that mass scale we are left with a theory $B$.
Writing down the gauge coupling running \eqref{tau_one_loop_running} for these two theories and matching them at the scale $m$, we obtain
\begin{equation}
	\Lambda_A^{b_A} = m^{b_A - b_B } \Lambda_{B}^{b_B } 
\label{Lambda_when_integrate_out_m}
\end{equation}
where $b_A$ and $b_B$ are the beta function coefficients for the theories $A$ and $B$ respectively.
In particular, consider SQCD with $F$ pairs\footnote{The fermion component in each of them is Weyl, so that one pair comprises a legitimate Dirac flavor.} 
of (anti-)fundamental quarks $Q$ and $\tilde{Q}$, so that $b_{F} = 3N - F$.
If we decouple just one flavor, we have
\begin{equation}
	\Lambda_{F-1}^{3N-F+1} = m \Lambda_{F}^{3N-F}
\label{scales_matching_F-1}
\end{equation}
where the subscript now reminds us of the number of flavors left.
In a special case when we have $F=N$ flavors and decouple them all simultaneously, we have
\begin{equation}
	\Lambda_0^{ 3 } = m \Lambda_{N}^{ 2 }  \,.
\label{scales_matching_F=N}
\end{equation}

The 3d analog of the 4d one-instanton factor, used above, is defined in the theory with gauge group $SU(N)$ and $F$ fundamental quarks as
\begin{equation}
	\eta_F = L^N \Lambda_{F}^{3N-F}
\label{eta_def}
\end{equation}
where $L$ is the circumference of the compact dimension.
This factor enters the 3d superpotential in a product with the monopole modulus, $ \eta Y_0$.
When we integrate out one flavor with large mass $m$, from eq.~\eqref{scales_matching_F-1} we have the scale-matching rule:
\begin{equation}
	\eta_{F-1} = m \, \eta_F \,, \quad
	(Y_0)_{F-1 \text{ flavors}} = \frac{1}{m} (Y_0)_{F \text{ flavors}} \,.
\label{scales_matching_F-1_eta}
\end{equation}

\section{3d SUSY gauge theory}
\label{sec:3d_susy}

Here we briefly review some elementary facts about the 3d supersymmetric gauge theories arising from the compactification of a 4d SQCD on a cylinder $\mathbb{R}^3 \times \mathbb{S}^1$ (see e.g. \cite{Aharony:1997bx,deBoer:1997kr}).

\begin{table}
\begin{center}
\begin{tabular}{c|c|c|c}
Quantity & dim. in 4d & dim. in 3d & relation \\ 
\hline
Gauge coupling $g^2$ & 0 & 1 & $1/g_{3d}^2 = L/g_{4d}^2$ \\
Quark $Q,q$ & 1 & $\frac{1}{2}$ & $Q_{3d} = \sqrt{L} \, Q_{4d}$ \\
Meson field $M$ & 2 & 1 & $M_{3d} = L \, M_{4d}$ \\
Gauge field $A$ & 1 & 1 & $A_i^{3d} = A_{\mu=i}^{4d} $, $i=1,2,3$ \\
Adj. scalar $\sigma$ & $-$ & 1 & $\sigma = A_{\mu=4}^{4d}$ \\
K\"ahler potential $\mathcal{K}$ & 2 & 1 & $\mathcal{K}_{3d} = L \, \mathcal{K}_{4d} $ \\
Superpotential $\mathcal{W}$ & 3 & 2 & $\mathcal{W}_{3d} = L \, \mathcal{W}_{4d} $ \\
``One-instanton'' factor \eqref{eta_def} & $3N-F$ & $2N-F$ & $\eta_F = L^N \, \Lambda_F^{3N-F}$
\end{tabular}
\caption{Relation between 3d and 4d quantities and their mass dimensions (m.d.)}
\label{tab:3d-4d}
\end{center}
\end{table}

\subsection*{Vector multiplet}

Compared to 4d, a vector multiplet in three dimensions contains an extra adjoint real scalar field $\sigma$ coming from the component of the four-potential $A_\mu$ along $\mathbb{S}^1$:
\begin{equation}
    V=-i \theta \bar{\theta} \sigma-\theta \gamma^i \bar{\theta} A_i+i \bar{\theta}^2 \theta \lambda-i \theta^2 \bar{\theta} \lambda^{\dagger}+\frac{1}{2} \theta^2 \bar{\theta}^2 D
\end{equation}
where the index $i=1,2,3$ labels $\mathbb{R}^3$, while the superspace coordinates in $\theta \bar{\theta}$ in the first term are assumed to be contracted with $\sigma^0 = \delta_{\alpha \dot{\alpha}}$ inherited from 4d.
In terms of the field strength $W_\alpha = - \frac{1}{4} \bar{D}^2 e^{-V} D_\alpha e^V$, the gauge action is
\begin{equation}
    \frac{1}{g^2} \int d^2\theta \Tr W_\alpha^2 + h.c. 
\label{3d_gauge_kin_1}
\end{equation}
This also includes a kinetic term for the real scalar $\sigma$.

In the pure SYM case, the classical moduli space is a Coulomb branch parameterized by the eigenvalues $\sigma_j$ of the adjoint scalar VEV $\expval{\sigma}$.
At a generic point, the gauge group $G$ is broken down to\footnote{More precisely, the Coulomb branch is a Weyl chamber $\mathbb{R}^r/\mathrm{Weyl}(G)$. The Weyl group elements exchange eigenvalues of the matrix $\expval{\sigma}$. On the boundaries of the Weyl chamber, some eigenvalues coincide, and the gauge group is classically enhanced.} 
$U(1)^r$ where $r = \mathrm{rank}(G)$.
These Abelian gauge fields can be dualized to dimensionless scalars $\gamma_j$:
\begin{equation}
    \partial_\mu \gamma_j = \frac{2 \pi}{g^2} \epsilon_{\mu\nu\rho} F^{\nu\rho} \,, \quad 
    F_{\mu\nu} = \frac{g^2}{4\pi} \epsilon_{\mu\nu\rho} \partial^{\rho} \gamma_j \,, \quad
    j = 1, \ldots, r \,.
\end{equation}
%
The Dirac quantization condition implies that the scalars $\gamma_j$ are periodic, $\gamma_j \sim \gamma_j + 2 \pi$ (see e.g. Sec.~9.7.3 of \cite{Shifman:2012zz} and Sec.~2 of \cite{Aharony:2013dha}).

Two real scalars $\sigma_j$ and $\gamma_j$ can be combined to form a complex scalar:
\begin{equation}
    x_j = i ( \tau \sigma_j + i \gamma_j )
\label{xj_def}
\end{equation}
where the complexified gauge coupling is
\begin{equation}
	\tau = i \frac{4 \pi}{g^2} + \frac{\theta}{2\pi} \,.
\label{tau}
\end{equation}
Supersymmetry is made manifest by writing a chiral superfield whose lowest component is $x_j$ from eq. \eqref{xj_def}, also known as the linear multiplet or the scalar field strength (here we use the same letter to denote a superfield and its lowest component),
\begin{equation}
    X_j = \epsilon^{\alpha\beta} \bar{D}_\alpha D_\beta V_j        
\label{dualized_susy_scalar}
\end{equation}  
such that $D^2 X_j = \bar{D}^2 X_j = 0$ and $X_j$ is gauge invariant.
The superfield $X_j$ is called a field strength linear superfield in \cite{deBoer:1997kr}.
Since $X_j$ has a shift symmetry with respect to the imaginary part, good Coulomb branch coordinates are the exponentiated chiral superfields, roughly speaking $Y_j \sim \exp(X_j)$.

When the gauge group is SU(N), it is convenient to use $\sigma_j$, $\gamma_j$, $j = 1, \ldots, N$, subject to the constraints $\sum \sigma_j = \sum \gamma_j = 0$.
The authors of \cite{Aharony:2013dha} use the following specific choice for the Coulomb branch coordinates:
\begin{equation}
    Y_j \sim \exp( \frac{\sigma_j - \sigma_{j+1} }{g^2} + i (\gamma_j - \gamma_{j-1}) ) \,, \quad
    j = 1, \ldots , N-1 \,.
\end{equation}
Another convenient choice for the Coulomb branch coordinates \cite{Davies:2000nw} is to take $Y_j \sim \exp( \vec{\alpha}_j^* \cdot \vec{X})$, where $\vec{\alpha}_j^*$ are the co-roots of the gauge algebra, and we think of $X$ as an $r$-component vector.
The latter choice is used in Sec.~\ref{sec:semicl} here.



\subsection*{Matter multiplets}

A chiral multiplet has the same components as in 4d,
\begin{equation}
    Q = q + \theta \psi + \theta^2 F \,.
\end{equation}
The complex mass term is inherited directly from 4d,
\begin{equation}
    \int d^2\theta m \tilde{Q} Q \,.
\end{equation}
The Lagrangian of a $U(1)$ gauge theory with scalar field strength $X$ from eq.~\eqref{dualized_susy_scalar} coupled to matter can be written as (see e.g. Appendix~A of \cite{Intriligator:2013lca})
\begin{equation}
	\mathcal{L} = \int d^4\theta \left(  
		-\frac{1}{e^2} X^2-\frac{k}{4 \pi} X V-\frac{\zeta}{2 \pi} V
		+ Q^{\dagger} e^{ V} Q
	\right) 
\end{equation}
where $k$ and $\zeta$ are the possible CS level and a Fayet–Iliopoulos term respectively (they do not appear in our case).
In 3d, it is also possible to introduce a real mass term 
\begin{equation*}
    \mathcal{L}_m 
        = \theta \left( Q^\dagger e^{m_r \theta \bar{\theta}} Q + \tilde{Q} e^{ \tilde{m}_r \theta \bar{\theta}} \tilde{Q}^{\dagger} \right)
        \sim \int d^2 (\theta + \bar{\theta}) \left(m_r Q^{\dagger} Q+\tilde{m}_r \tilde{Q} \tilde{Q}^{\dagger}\right)
\end{equation*}
which is anomaly-free when $m_r = - \tilde{m}_r$.
Our interest will be only in a 3d theory that comes from compactifying a 4d $\mathcal{N}=1$ theory.
This corresponds to introducing a background gauge field with a non-trivial holonomy along the compact direction.
In this paper, we consider the case without such holonomy and assume that all the bare real masses vanish, but they can be induced by the VEV of the adjoint scalar $\sigma$ on the Coulomb branch via the interaction
\begin{equation}
    g^2 \int | \sigma Q |^2 \,.
\label{adjoint_mass_for_matter}
\end{equation}
In addition to the Coulomb branch, matter scalar VEVs can form a Higgs branch.
Because of the coupling \eqref{adjoint_mass_for_matter}, one generally needs to have either $\expval{Q}=0$ or $\expval{\sigma}=0$; mixed branches are also possible.

We note in passing that, when integrating out fundamental flavors, there can be a possible aftereffect.
Suppose we have a chiral multiplet $Q$ with a total mass $m_Q$ and charges $q_1$, $q_2$ with respect to two U(1)'s with gauge fields $A_\mu^{(1)}$ and $A_\mu^{(2)}$ (they may be actually the same).
Integrating out this chiral multiplet at one loop induces a mixed CS term $A^{(1)} \wedge d A^{(2)}$ with a coefficient \cite{Niemi:1983rq,Redlich:1983dv} (see also \cite{Aharony:1997bx,Aharony:2013dha}). 
\begin{equation}
    k = \frac{1}{2} q_1 q_2 \mathrm{sign} (m_Q) \,. 
\end{equation}
The mass $m_Q$ here is the sum of the contributions from the vector multiplets \eqref{adjoint_mass_for_matter} and the real masses (the latter vanish in our case).
In the superfield language, such a CS term is written as
\begin{equation}
    k \int d^4 \theta X^{(1)} V^{(2)}
\end{equation}
where $V$ is a vector superfield and $X$ is a dualized scalar field strength (see eq.~\eqref{dualized_susy_scalar}).
Now, in our case one flavor consists of two chiral multiplets. 
The two multiplets have opposite charges and, consequently, opposite signs of the mass terms $m_Q \sim q \sigma$; therefore, in the case of a single U(1), integrating out one such flavor does not induce the CS term.

\section{Topological order under circle compactification}
\label{sec:1form_quark_circle}

A topological order is nothing but the spontaneous breaking phase of (typically anomalous) finite higher-form symmetry.
However, unlike 0-form symmetries, a higher-form symmetry can emerge in the infrared.
Therefore, topological order can appear even if the microscopic model has no manifest higher-form symmetry.

For example, let us consider the $U(1)_k$ Chern-Simons theory, which has a $\Z_k$ 1-form symmetry (with a 't Hooft anomaly) and manifests a topological order.
On a spatial Riemann surface $\calM_g$ with genus $g$, this theory has $k^g$ degenerate vacua and no excitation states.
Such a Hilbert space is independent of any geometric detail of the spatial manifold, such as the metric and the size, which is thus dubbed topological degeneracy.

Now let us couple some charge-1 heavy matter to the theory, which explicitly destroys the 1-form symmetry.
As long as the matter mass is sufficiently large, the $\Z_k$ 1-form symmetry still emerges in the infrared.
The topological degeneracy is lost for any finite-size spatial $\calM_g$ but still appears in the large volume limit.
In addition, there are now a bunch of excitation states in the Hilbert space.
The system is still non-trivially topologically ordered, i.e., it is separated by a phase transition from the trivial phase.

The above situation is drastically different if we introduce circle compactification to reduce 3d to 2D, either as a finite-size spatial dimension or a finite temperature.
For pure Chern-Simons without matter, the 3d $\Z_k$ 1-form symmetry splits into a 2d $\Z_k$ 1-form symmetry and a 2d $\Z_k$ 0-form symmetry.
The 't Hooft anomaly of the 3d $\Z_k$ 1-form symmetry is converted into a mixed 't Hooft anomaly between the 2d $\Z_k$ 1-form symmetry and the 2d $\Z_k$ 0-form symmetry.
Therefore, we can still see topological degeneracy, which now corresponds to the spontaneous breaking of the 2d $\Z_k$ 1-form symmetry and 2d $\Z_k$ 0-form symmetry.

Nevertheless, for Chern-Simons with heavy matter, where no exact 3d 1-form symmetry is present, the topological degeneracy is completely destroyed by circle compactification.
This is because spontaneous 0-form symmetry cannot emerge in the infrared.
Now this theory just sits in the trivial phase.
Therefore, the phase boundary between a topological order and the trivial phase is removed by a finite temperature or a finite spatial size.
Hence such a phase transition is sometimes called a \textit{quantum phase transition}, which has a very different nature from the ordinary phase transition of the Landau-Ginsburg-Wilson type.

This argument shows that, as soon as we introduce matter in the fundamental representation (with arbitrary finite mass $m$), we immediately destroy the 1-form center symmetry $\mathbb{Z}_N$ of the $SU(N)$ gauge theory.

\section{\boldmath{$\lambda\lambda$} condensate from 3d}
\label{sec:lamlam_3d}


Here we would like to comment on how to compute the gluino condensate in 4d SYM from the 3d EFT superpotential.
In particular, we want to discuss how to derive eq.~(5.7) of \cite{Davies:2000nw},
\begin{equation}
    \left\langle\frac{\operatorname{tr} \lambda^2}{16 \pi^2}\right\rangle=b_0^{-1} \Lambda \frac{\partial}{\partial \Lambda}\left\langle\frac{1}{2 \pi R} \mathcal{W}_{\text {3d}}\right\rangle
\label{davies-5-7}
\end{equation}
where on the l.h.s. we have the gluino condensate in 4d, while the r.h.s. is in terms of the monopole-induced superpotential of the 3d EFT after the circle compactification.
The convention for $\Lambda$ is the one in eq.~\eqref{Lambda_def}, which in our case reads
\begin{equation}
    \Lambda^3=\mu^3 \frac{1}{g^2(\mu)} \exp \frac{2 \pi i \tau(\mu)}{c_2}
\end{equation}
where $\mu$ is a renormalization scale.

To understand eq.~\eqref{davies-5-7}, we can start with a 4d SYM with the UV Lagrangian (holomorphic part)
\begin{equation}
    \mathcal{L}_{uv} = \frac{1}{2g_{uv}^2} \int d^2 \theta \Tr(W^2) 
\end{equation}
and evolve it down to a scale $\mu$.
At one loop we obtain
\begin{equation}
    \mathcal{L}_{eff} = \frac{b_0}{16 \pi^2} \ln\frac{\mu}{\Lambda} \int d^2 \theta \Tr(W^2) \,.
\label{4d_3d_matching_gauge}
\end{equation}
Here, $W^2 = W^\alpha W_\alpha$ and $W_\alpha$ is the 4d gauge field strength superfield.
We want to simulate this 4d Lagrangian with a 3d superpotential that would give the same scaling of the effective Lagrangian, i.e., give the same value of 
$ \pdv{}{\mu} \mathcal{L}_{eff}$ or $ \pdv{}{\Lambda} \mathcal{L}_{eff}$.
The idea is to match the microscopic theory and the EFT at the compactification scale.
To this end we note that a 3d superpotential $\mathcal{W}_{3d}$ is related to its 4d version via an integral over the compact direction
\begin{equation}
    \mathcal{W}_{3d} 
        = \int_0^{2 \pi R} dx \text{``} \mathcal{W}_{4d} \text{''}
        = 2 \pi R \text{``} \mathcal{W}_{4d} \text{''} \,.
\label{4d_3d_matching_superpotential}
\end{equation}
A 4d Lagrangian would be $\int d^2\theta \text{``} \mathcal{W}_{4d} \text{''}$.
Of course, there is no superpotential in pure SYM in 4d, hence the quotation marks.
Instead of a 4d superpotential, we match the chiral field entering eq.~\eqref{4d_3d_matching_superpotential} to the chiral field in eq.~\eqref{4d_3d_matching_gauge}.
To this end we require
\begin{equation}
	\frac{b_0}{16 \pi^2} \ln\frac{\mu}{\Lambda} \Tr(W^2)  = \text{``} \mathcal{W}_{4d} \text{''}
\end{equation}
or, equivalently,
\begin{equation}
    \frac{b_0}{16 \pi^2} \Tr(W^2) = - \Lambda \pdv{}{\Lambda} \text{``} \mathcal{W}_{4d} \text{''} \,.
\end{equation}
Noting that the lowest component of $W^2$ is $- \lambda \lambda$ and using eq.~\eqref{4d_3d_matching_superpotential}, we obtain
\begin{equation}
    \frac{ \Tr \lambda \lambda }{16 \pi^2} =  \frac{1}{b_0} \cdot \frac{1}{2 \pi R} \mathcal{W}_{3d}
\end{equation}
where on the r.h.s. the superpotential is evaluated at its lowest (i.e., scalar) component.

\bibliographystyle{unsrt}

\bibliography{main} 

\end{document}